\documentclass[fleqn,usenatbib]{mnras}

\usepackage{newtxtext,newtxmath}

\usepackage[T1]{fontenc}
\usepackage{ae,aecompl}

\usepackage{graphicx}	
\usepackage{amsmath}	

\usepackage{amssymb}	
\usepackage[flushleft]{threeparttable}
\usepackage[normalem]{ulem}
\usepackage{subfig}

\def\prl{Phys. Rev. Lett.}
\def\prd{Phys. Rev. D}

\def\apj{Astrophys. J.}
\def\apjs{Astrophys. J. Suppl.}
\def\apjl{Astrophys. J. Lett.}

\def\mnras{Mon. Not. R. Astron. Soc.}

\def\part_n{\partial_\perp}
\def\eul{\mathrm{e}}
\def\ga{\,\,\raise0.14em\hbox{$>$}\kern-0.76em\lower0.28em\hbox
{$\sim$}\,\,}
\def\la{\,\,\raise0.14em\hbox{$<$}\kern-0.76em\lower0.28em\hbox
{$\sim$}\,\,}

\title[PPISNe: collapse, black-hole formation, and beyond]{Pulsational pair-instability supernovae: gravitational collapse, black-hole formation, and beyond}

\author[N.~Rahman et al.]{
N. Rahman,$^{1,2,3}$\thanks{E-mail: nrahman@mpa-garching.mpg.de}
H.-T. Janka,$^{2}$
G. Stockinger,$^{2,1}$ and
S.~E. Woosley$^{4}$
\\
$^{1}$Physik-Department, Technische Universit{\"a}t M{\"u}nchen, James-Franck-Str. 1, D-85748 Garching, Germany \\
$^{2}$Max Planck Institute for Astrophysics, Karl-Schwarzschild-Str. 1, Postfach 1317, D-85741 Garching, Germany \\
$^{3}$GSI Helmholtzzentrum f{\"u}r Schwerionenforschung, Planckstr. 1, 64291 Darmstadt, Germany \\
$^{4}$Department of Astronomy and Astrophysics, University of California, Santa Cruz, CA 95064, USA
}

\date{Accepted XXX. Received YYY; in original form ZZZ}

\pubyear{xxxx}

\begin{document}
\label{firstpage}
\pagerange{\pageref{firstpage}--\pageref{lastpage}}
\maketitle

\begin{abstract}
We investigate the final collapse of rotating and non-rotating 
pulsational pair-instability supernova progenitors with zero-age-main-sequence 
masses of 60, 80, and 115\,$\mathrm{M}_\odot$ and iron cores between 
2.37\,$\mathrm{M}_\odot$ and 2.72\,$\mathrm{M}_\odot$ by 2D hydrodynamics simulations.
Using the general relativistic \textsc{NADA-FLD} code with energy-dependent three-flavor
neutrino transport by flux-limited diffusion allows us to follow the evolution
beyond the moment when the transiently forming neutron star (NS) collapses to a
black hole (BH), which happens within 350--580\,ms after bounce in all cases.
Because of high neutrino luminosities and mean energies,
neutrino heating leads to shock revival within $\lesssim$\,250\,ms post bounce
in all cases except the rapidly rotating 60\,$\mathrm{M}_\odot$ model. 
In the latter case, centrifugal
effects support a 10\% higher NS mass but reduce the radiated neutrino luminosities
and mean energies by $\sim$20\% and $\sim$10\%, respectively, and the 
neutrino-heating rate by roughly a factor of two compared to the non-rotating 
counterpart. After BH formation, the neutrino luminosities drop steeply but continue
on a 1--2 orders of magnitude lower level for several 100\,ms because of aspherical
accretion of neutrino and shock-heated matter, before the ultimately spherical 
collapse of the outer progenitor shells suppresses the neutrino emission to 
negligible values. In all shock-reviving models BH accretion swallows the entire 
neutrino-heated matter and the explosion energies decrease from maxima around
$1.5\times 10^{51}$\,erg to zero within a few seconds latest. Nevertheless, the
shock or a sonic pulse moves outward and may trigger mass loss, which we estimate
by long-time simulations with the \textsc{Prometheus} code. We also provide 
gravitational-wave signals.
\end{abstract}




\section{Introduction}\label{sec:introduction}

Pulsational pair-instability supernovae (PPISNe) are violent pulsations,
accompanied by episodes of supernova (SN)-like mass ejection, of very massive 
stars (VMSs) with pre-SN helium core masses between about 30\,M$_\odot$ and about
65\,M$_\odot$, corresponding to zero-age-main-sequence (ZAMS) masses 
from roughly 70\,M$_\odot$ to roughly 140\,M$_\odot$. The exact ZAMS
mass range depends on the metallicity, stellar mass-loss evolution, 
reaction rates for $3\alpha$ and $^{12}$C$(\alpha,\gamma)$$^{16}$O, and 
on the stellar rotation, which shifts the boundaries of the ZAMS mass interval 
to lower values (\citealt{2002RvMP...74.1015W,2003ApJ...591..288H, 2015ASSL..412..199W,Woosley+2021,2017ApJ...836..244W,2019ApJ...878...49W,2020A&A...640L..18M,2020A&A...640A..56R,2019ApJ...887...53F,2020ApJ...902L..36F}).

PPISNe occur for pre-SN helium core masses below the mass range where the
stars are completely destroyed in a single violent pulse called pair-instability
supernova (PISN). Both phenomena are triggered in stellar cores with high
entropies, where the temperatures increase over $\sim$\,$7\times 10^8$\,K
already after core-carbon burning, enabling the onset of electron-positron 
pair production. The pair formation reduces the structural adiabatic index
below the critical limit of 4/3 for stability (or slightly greater with rotation). 
For this reason the subsequent evolution is unstable, at least transiently, a phenomenon
called ``pair-instability'' (\citealt{1964ApJS....9..201F,1967PhRvL..18..379B,1967ApJ...148..803R}). Since the equation 
of state becomes softer when the adiabatic index drops, the core contracts
and ignites nuclear burning of oxygen and/or silicon. In helium cores above
about 65\,M$_\odot$ a single, giant nuclear flash disrupts the entire star in 
a PISN, whereas for less massive helium cores the nuclear energy release is
not sufficient for this to happen. Instead, the core responds to the nuclear
energy release by vigorously expanding and then contracting in a Kelvin-Helmholtz 
phase of varying duration before becoming, once more, unstable.
This ``pulsational pair-instability'' (PPI) thus
proceeds in a series of pulsations, which can drive mass ejection 
in the PPISN events. In course of the mass shedding the star can lose its
hydrogen envelope and usually the outer layers of its helium core as well.
The multiple cycles of contraction, burning, expansion, and cooling recur
until the relic core has a mass and entropy too low for the PPI to happen.
Therefore, after the episodes of pulsing, the remnant of this active period
settles into hydrostatic equilibrium again and evolves towards ultimate core
collapse by building up an iron core though central silicon burning (for more
details of the evolution and its astrophysical implications, see, e.g.,
\citealt{2015ASSL..412..199W,2017ApJ...836..244W}).

PPISNe have also been invoked to explain SNe
with unusually high brightness (``superluminous SNe'') or unusual time
variations over long periods of visibility, e.g. iPTF14hls \citep{2017Natur.551..210A}. PPISNe were proposed as possible explanations 
of both phenomena (e.g., \citealt{2007Natur.450..390W,2017ApJ...836..244W,2018ApJ...863..105W}), because their mass loss
cycles create shells of dense circumstellar material with an enormous
diversity due to a wide range of pulse numbers, pulse strengths, and 
activity durations from the onset of the pulsing until iron-core collapse.
Also ultra-long gamma-ray bursts were suggested as observable phenomena
in connection to the gravitational collapse that follows the PPISN phase
\citep{2020A&A...640L..18M,2020A&A...641L..10M}, since 
the stars develop very massive iron cores and very extended envelopes, likely
to collapse to rapidly spinning black holes (BHs) or strongly magnetized 
neutron stars (NSs) with proto-magnetar activity and the possible creation
of collimated, jet-like outflows \citep{2017ApJ...836..244W}.

Another strong push came from the detection of gravitational-wave (GW) 
signals radiated by the inspiral and final merging of binary BHs
(e.g., \citealt{2016PhRvL.116f1102A,2019ApJ...882L..24A}), whose 
astonishingly big masses of more than 30\,M$_\odot$ match expectations 
for the evolution of low-metallicity, non-rotating stars with masses
in the PPISN regime (e.g., \citealt{2016ApJ...824L..10W,2017ApJ...836..244W,2017MNRAS.470.4739S,2018MNRAS.480.2011G,2019ApJ...878...49W,2019ApJ...887...53F}).
Very recently, \citet{2021MNRAS.503.2108P} performed 3D simulations to
determine the GW emission from the ultimate gravitational collapse of the
iron cores formed in PPISNe of Population~III progenitors with initial 
masses of 85\,M$_\odot$ and 100\,M$_\odot$. 
They also investigated the potential detectability of 
these signals, which are several times stronger than the GW amplitudes 
created during NS and BH formation in less massive progenitor stars.

Interestingly, \citet{2021MNRAS.503.2108P} witnessed shock revival by neutrino
heating prior to BH formation in their models, in line with results of 
various other studies of the collapse of massive stars with
$M_\mathrm{ZAMS}\ge 40\,$M$_\odot$ to BHs (e.g., \citealt{2018MNRAS.477L..80K,2018ApJ...852...28S,2018ApJ...855L...3O,2018ApJ...857...13P,2021ApJ...914..140P,2020MNRAS.491.2715B,2018ApJ...852L..19C,2020MNRAS.495.3751C}). The neutrino energy transfer increases for more compact NSs and 
with longer time span between shock revival and
the collapse of the transiently stable NS to a BH. For these reasons
NS equations of state (EOSs) that yield small NS radii and high threshold
masses for BH formation foster shock revival \citep{2021MNRAS.503.2108P}.
\citet{2018ApJ...852L..19C,2020MNRAS.495.3751C} followed the evolution of 
their zero-metallicity 40\,M$_\odot$ models until shock breakout
and obtained considerable mass ejection (more than 10\,M$_\odot$), 
if they employed favorable assumptions to artificially trigger an early
onset of shock expansion or to boost the neutrino energy deposition in
the postshock matter. Mass ejection occurred 
despite the fact that the neutrino-heated gas itself fell back and
was accreted by the newly formed BH, because buoyant plumes of the hot
gas pushed the outgoing shock and the shock transferred energy to the 
overlying shells by $p\mathrm{d}V$ work. 

In the project reported here we consider models from \citet{2017ApJ...836..244W} to
investigate the iron-core collapse after the PPISN 
phase of rotating and non-rotating VMSs with ZAMS masses of 60\,M$_\odot$,
80\,M$_\odot$, and 115\,M$_\odot$ and a metallicity of 10\% of the solar
value. Our goal is to address the following questions:
\begin{itemize}
\item
Does neutrino heating revive the stalled bounce shock and is the 
energy transfer sufficiently powerful to cause significant mass ejection?
Does the stellar collapse lead to BH formation and what are the BH
masses?
\item
What are the neutrino and GW signals of these stellar core-collapse
events? How do they evolve through the moment of BH formation and
afterwards when initially asymmetrically ejected matter falls back 
and is accreted by the newly formed BH?
\end{itemize}

In order to answer these questions we perform two-dimensional (axisymmetric; 2D)
neutrino-hydrodynamics simulations with the general relativistic NADA-FLD
code \citep{2019MNRAS.490.3545R}, which employs an energy-dependent flux-limited 
diffusion solver for the multi-dimensional transport of neutrinos of all 
three flavors. Using the formulation of the Einstein equations developed 
by Baumgarte, Shapiro, Shibata, and Nakamura (BSSN formalism), NADA-FLD
is able to track the hydrodynamic flow beyond the instant when the transiently
stable NS collapses to a BH. This asset permits us to determine
the neutrino and GW emission also during the period of several hundred
milliseconds that it takes most of the anisotropic initial ejecta to be 
swallowed by the BH. The neutrino emission from non-radial accretion 
flows into the BH, which are shock heated by mutual collisions before
being swallowed by the BH, is superimposed on the emission stretching 
connected to the neutrino propagation along non-radial geodesics
and a scattering echo due to neutrinos interacting with particles in the 
infalling mass flow. The geodesic and echo effects have recently been
discussed in detail by \citet{Wang+2021} and \citet{Gullin+2021}.
They were shown to smoothen the sharp (millisecond-long) cut-off of the 
neutrino signal at the NS-BH transition by fractions of a millisecond
and up to 15\,ms, respectively, but both of these effects are subdominant
compared to the ongoing emission from anisotropic fallback accretion.

In four of our five simulations we obtain initially outward shock motion
driven by the neutrino energy deposition. The only exception is our 
fastest-rotating 60\,M$_\odot$ case, where rotational deformation leads
to a cooler neutrinosphere and thus decreases
the neutrino luminosities, mean energies, and consequently the neutrino 
heating during the phase when shock revival occurs in the other models.
By mapping the NADA-FLD results to the \textsc{Prometheus} code we also
follow the outward expansion of the shock to estimate upper limits for 
the mass loss triggered by the shock breakout from the stellar surface.
In all cases the initial diagnostic energies are only 
$\sim$\,$1.5\times 10^{51}$\,erg,
which is much lower than the binding energies of the overlying stellar 
layers. Since the accretion of neutrino-heated matter by the BH reduces 
the energy available in the postshock region to several $10^{49}$\,erg,
the outgoing shock (or sonic pulse) is too weak to unbind more than a 
few solar masses in any of our models. The final BH masses are therefore
expected to be close to the gravitationally bound masses of the
progenitors at the onset of iron-core collapse.

Our paper is structured as follows. In Section~\ref{sec:setup}, we
introduce the properties of the employed progenitors and our numerical
setup. In Section~\ref{sec:definition}, we define important quantities 
that are used in the analysis of our simulations. In 
Section~\ref{sec:result_before_BH_formation}, we present results of our models for the
hydrodynamic evolution until BH formation including a comparison to
previous works, and in Section~\ref{sec:result_after_BH_formation}, we describe the
evolution after the collapse of the transiently existing NSs to BHs.
In Section~\ref{sec:neutrino_properties}, we discuss the neutrino signals and in
Section~\ref{sec:gravitational_waves}, the GW emission of our models. A summary
and conclusions follow in Section~\ref{sec:summary}.

\section{Numerical Setup and Progenitor Properties}\label{sec:setup}

In this section, we describe our numerical setup and various properties of the progenitors used in our study. Throughout the paper, the speed of light, the gravitational constant, the Boltzmann constant, and the solar mass are denoted by $c$, $G$, $k_\mathrm{b}$, $\mathrm{M}_\odot$, respectively.

\subsection{Progenitor Properties}\label{subsec:progenitor_property}
\begin{table*}
	\centering
    \caption{Pre-collapse properties of the simulated models.}
	\begin{tabular*}{\textwidth}{lcccccccccccc}
		\hline
		Model & $M_\mathrm{ZAMS}$ & $\mathrm{Rotation}$ & $M_\mathrm{prog}$ & $R_\mathrm{prog}$ & $M_\mathrm{Fe}$ & $R_\mathrm{Fe}$ & $\langle j(R_\mathrm{Fe}) \rangle$ & $\xi_{2.5}$ & $\omega_\mathrm{in}$ & $J_\mathrm{prog}$ & $J_\mathrm{Fe}$ & $a_\mathrm{prog}$ \\
        & $[\mathrm{M}_\odot]$ & & $[\mathrm{M}_\odot]$ & $[10^6\,\mathrm{km}]$ & $[\mathrm{M}_\odot]$ & $[\mathrm{km}]$ & $[10^{16}\,\mathrm{cm^2/s}]$ & & $\mathrm{rad/s}$ & $[10^{50}~\mathrm{erg\,s}]$ & $[10^{48}~\mathrm{erg\,s}]$ \\
        \hline
        C60C-NR  &  60 & no  & 41.54 & 7.32  & 2.37 & 2893 &       & 0.77 &      &      &     &      \\
        C60C     &  60 & yes & 41.54 & 7.32  & 2.37 & 2893 & 1.540 & 0.77 & 2.23 & 105  & 37  & 0.69 \\
        R80Ar-NR &  80 & no  & 47.64 & 16.36 & 2.72 & 3464 &       & 0.84 &      &      &     &      \\
        R80Ar    &  80 & yes & 47.64 & 16.36 & 2.72 & 3464 & 0.225 & 0.84 & 0.25 & 14   & 7.2 & 0.07 \\
		C115     & 115 & no  & 45.50 & 5.67  & 2.46 & 2747 &       & 0.89 &      &      &     &      \\
        \hline        
	\end{tabular*}
	\flushleft
    \textit{Notes}: $M_\mathrm{ZAMS}$ is the ZAMS mass of the progenitor, the column ``Rotation'' indicates the presence or absence of rotation, $M_\mathrm{prog}$ is the gravitationally bound baryonic mass of the pre-collapse star and $R_\mathrm{prog}$ is the corresponding radius (defined as the innermost location where the expansion velocity of stellar mass loss is above the escape velocity), $M_\mathrm{Fe}$ is the iron core mass and $R_\mathrm{Fe}$ is the corresponding radius (defined at a location where $X_{^{28}\mathrm{Si}}$ = $X_{^{54}\mathrm{Fe}}$), $\langle j(R_\mathrm{Fe}) \rangle$ is the shell-averaged specific angular momentum at the edge of the iron core, $\xi_{2.5}$ is the compactness parameter at a mass coordinate of 2.5\,$\mathrm{M}_\odot$, given by equation~\eqref{eq:compactness}, $\omega_\mathrm{in}$ is the angular frequency of the innermost grid cell of the pre-collapse model, $J_\mathrm{prog}$ is the total angular momentum of the gravitationally bound progenitor mass, $J_\mathrm{Fe}$ is the total angular momentum of the iron core, $a_\mathrm{prog}$ is the Kerr parameter of the gravitationally bound progenitor, defined by equation~\eqref{eq:kerr_rem}, all given at the onset of stellar Fe-core collapse. Further information on the progenitors can be found in \citet{2017ApJ...836..244W}. Note that the iron-core masses provided there are 2.35\,$\mathrm{M}_\odot$ for model C60C and 2.74\,$\mathrm{M}_\odot$ for Model R80Ar, both of which are very close to the values obtained with our definition.
	\label{tab:progenitor_property}
\end{table*}

\begin{figure*}
	\includegraphics[width=0.95\textwidth]{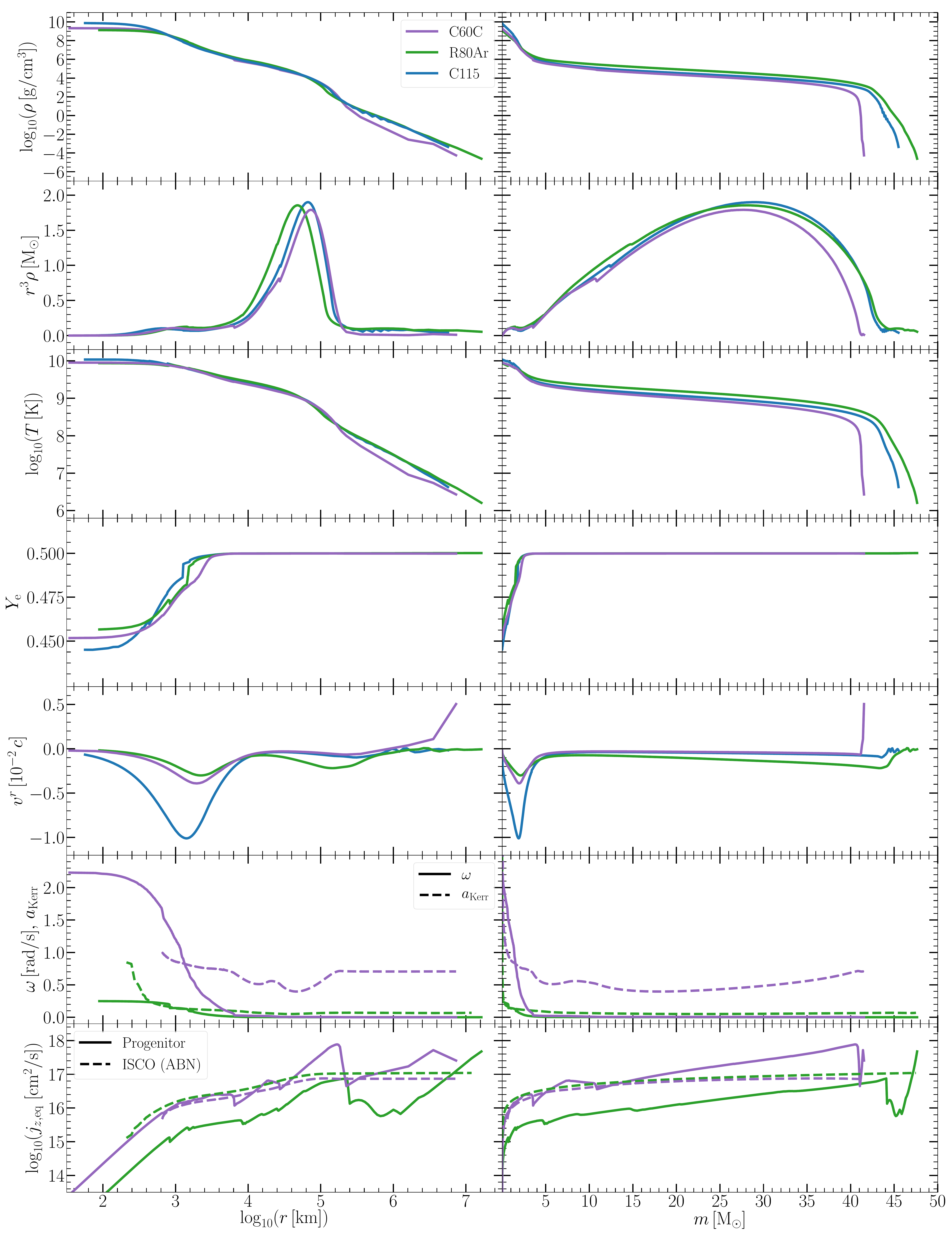}
    \vspace*{-7mm}
	\caption{Radial profiles of density, $\rho$, the product $r^3\rho$, temperature, $T$, electron fraction, $Y_\mathrm{e}$, radial velocity, $v^{r}$, angular frequency, $\omega$, Kerr parameter, $a_\mathrm{Kerr}$, given by equation\,\eqref{eq:kerr_encl}, and equatorial specific angular momentum, $j_{z,\mathrm{eq}}$ (solid lines) along with the specific angular momentum of the ISCO (dashed lines; see Section~\ref{subsec:progenitor_property} for further details) at the onset of stellar core collapse versus radius (left column) and mass coordinate (right column) for our investigated progenitors with ZAMS masses of 60\,$\mathrm{M}_\odot$ (violet line; rapidly rotating), 80\,$\mathrm{M}_\odot$ (green line; slowly rotating), and 115\,$\mathrm{M}_\odot$ (blue line; non-rotating). In the rotating models shellular rotation is assumed. The average angular momentum in each thin shell is thus 2/3 of the equatorial value.}
	\label{fig:progenitor}
\end{figure*}
\begin{figure*}
	\includegraphics[width=0.99\textwidth]{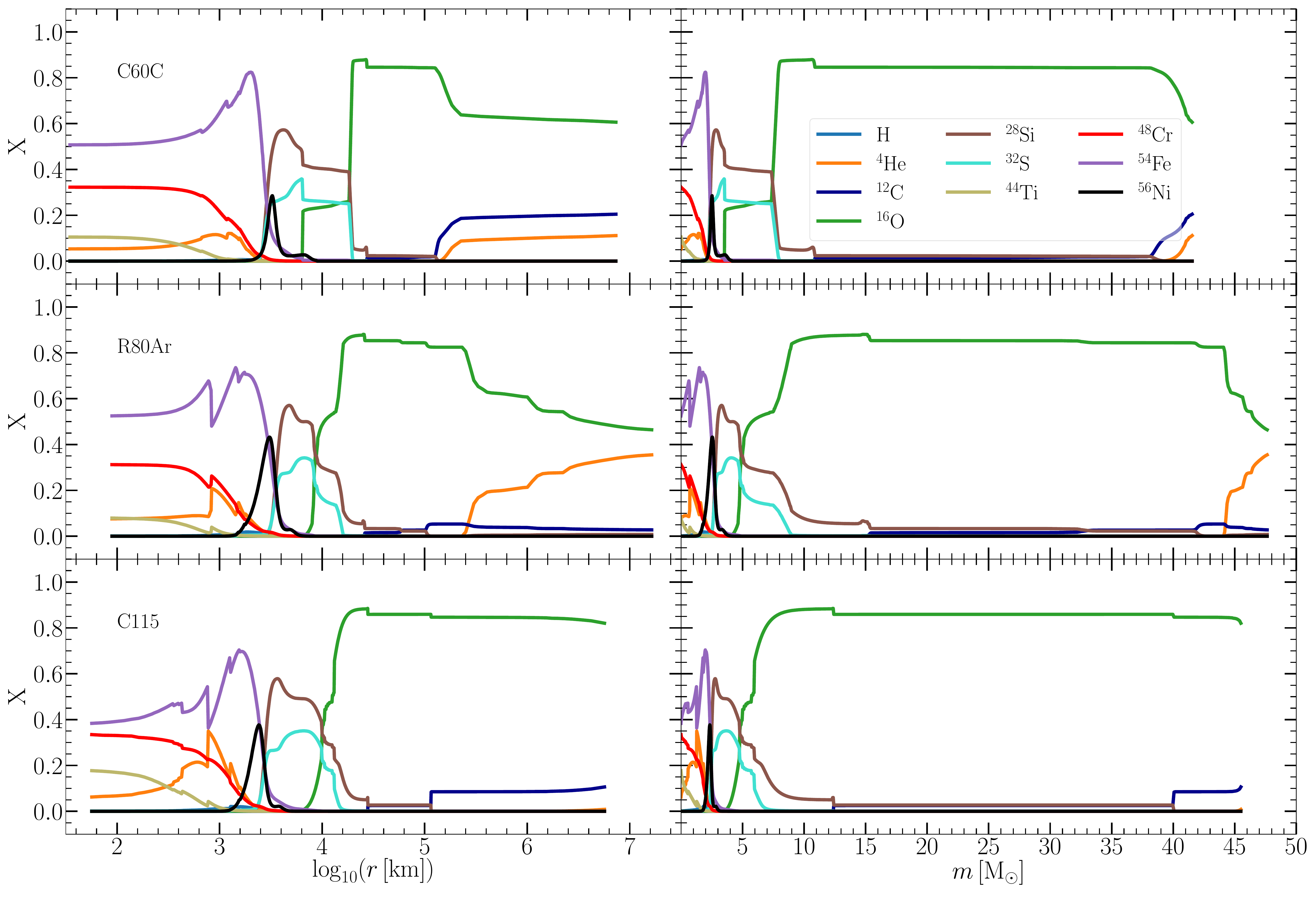}
    \vspace*{-4.5mm}
	\caption{Mass fractions of selected nuclear species for our investigated progenitors at the onset of stellar core collapse versus radius (left column) and mass coordinate (right column). Note that the H-envelopes and major parts or even all of the He-shell have been lost during stellar pulsations.}
	\label{fig:progenitor_composition}
\end{figure*}

In this section, we give a brief introduction to the VMS progenitors used in our study. We conduct core-collapse supernova (CCSN) simulations of several of these VMS progenitors from the stellar evolution calculations of \citet{2017ApJ...836..244W}. These calculations have been done using the KEPLER code (see, e.g., \citealt{2000ApJ...528..368H,2002RvMP...74.1015W,2005ApJ...626..350H,2016ApJ...821...38S}). 

We study the gravitational collapse of two rotating progenitors with ZAMS masses of 60\,$\mathrm{M}_\odot$ and 80\,$\mathrm{M}_\odot$ (Models C60C and R80Ar, respectively) and of a non-rotating progenitor with a ZAMS mass of 115\,$\mathrm{M}_\odot$ (Model C115), all evolved to the onset of gravitational instability by \citet{2017ApJ...836..244W}.\footnote{Adopting a rotationally mixed composition and structure while ignoring centrifugal forces in the progenitor modeling is admittedly physically inconsistent, but was done there in an attempt to help clarify the role of just the rotational forces.} We also investigate the 60\,$\mathrm{M}_\odot$ and 80\,$\mathrm{M}_\odot$ progenitors neglecting their rotation during the CCSN simulations (hereafter, Models C60C-NR and R80Ar-NR, respectively). All models have a ZAMS metallicity of 10\%\,$Z_{\odot}$ and include mass loss, albeit at a reduced rate. The 60\,$\mathrm{M}_\odot$ progenitor has undergone a chemically homogeneous evolution due to the efficient mixing induced by its rapid rotation (see, e.g., \citealt{2006ApJ...637..914W} for a discussion of chemically homogeneous evolution). Note that all three considered core-collapse progenitors have experienced massive mass loss in a sequence of pair-instability pulses. The data files at the onset of iron-core collapse contain the expanding shells of matter stripped off during these mass-loss episodes. We, therefore, discriminate between the unbound matter that expands faster than the local escape velocity on the one hand and the bound progenitor on the other hand. The latter is defined by all material that has started its infall (thus possesses a negative radial velocity) or has a positive radial velocity below the escape limit. Integral or average quantities of the entire pre-collapse stars are computed for this gravitationally bound mass.

Table~\ref{tab:progenitor_property} provides the ZAMS masses of our investigated progenitors, $M_\mathrm{ZAMS}$, and parameters characterizing their rotational state prior to collapse. Moreover, Table~\ref{tab:progenitor_property} lists various other properties of the progenitors at the onset of the stellar core collapse, namely, the gravitationally bound baryonic masses of the stars, $M_\mathrm{prog}$, their radii, $R_\mathrm{prog}$ (defined as the innermost location where the positive radial velocity of expanding surface matter begins to exceed the escape velocity), the iron-core masses, $M_\mathrm{Fe}$, the iron-core radii, $R_\mathrm{Fe}$ (we define the edge of the iron core as the radius where $X_{^{28}\mathrm{Si}}$ = $X_{^{54}\mathrm{Fe}}$), the angular frequencies of the innermost grid cells before collapse, $\omega_\mathrm{in}$, the specific angular momenta at the edge of the iron core, $j(R_\mathrm{Fe})$, the compactness parameters at a mass coordinate of 2.5\,$\mathrm{M}_\odot$, $\xi_{2.5}$, given by equation~\eqref{eq:compactness}, the total angular momenta of the gravitationally bound mass of the pre-collapse stars, $J_\mathrm{prog}$, and of the iron core, $J_\mathrm{Fe}$, and the Kerr parameters of the gravitationally bound pre-collapse stars, $a_\mathrm{prog}$, given by equation~\eqref{eq:kerr_rem}. 

The compactness parameter was defined by \citet{2011ApJ...730...70O} as:
\begin{eqnarray}\label{eq:compactness}
    \xi_M = \frac{M/\mathrm{M}_\odot}{R(M_\mathrm{bary} = M)/1000~\mathrm{km}},
\end{eqnarray}
where $R(M_\mathrm{bary} = M)$ is the radius where the enclosed baryonic mass, measured in solar masses, is $M$. Since the progenitor models have a constant value of the angular frequency on spherical shells, i.e., $\omega = \omega(r)$, the mass-weighted, shell-averaged value of the specific angular momentum $\langle j_z(r) \rangle$ of a radial shell of mass $M_\mathrm{shell}$ is given by:
\begin{eqnarray}\label{eq:angular_momentum_shell}
\langle j_z(r) \rangle = \frac{1}{M_\mathrm{shell}}\int_\mathrm{shell} \mathrm{d}m [\omega(r) D^2].
\end{eqnarray}
Here, $\mathrm{d}m$ is the mass element, and $D = r\sin(\theta)$ is the distance from the rotation axis ($r$, $\theta$ are the radius and the polar angle, respectively). The average $\langle j_z(r) \rangle$ gives a value of 2/3 of the equatorial specific angular momentum, $j_{z, \mathrm{eq}}$, displayed in the bottom panels of Fig.\,\ref{fig:progenitor}. The Kerr parameter of the gravitationally bound mass of the progenitor is defined as:
\begin{eqnarray}\label{eq:kerr_rem}
    a_\mathrm{prog} = \frac{J_\mathrm{prog} c}{G M_\mathrm{prog}^2},
\end{eqnarray}
and the Kerr parameter for enclosed mass $m(r)$ and associated total angular momentum $J(r)$ is given by:
\begin{eqnarray}\label{eq:kerr_encl}
    a_\mathrm{Kerr}(r) = \frac{J(r)\,c}{G m(r)^2}.
\end{eqnarray}
The listed values of $a_\mathrm{prog}$ in Table~\ref{tab:progenitor_property} will be the final values of the Kerr parameter if the whole star collapses to a black hole (BH) without any further loss of mass and angular momentum. 

The high-mass and low-metallicity progenitors considered in our study have high values of the compactness parameter $\xi_{2.5}$ compared to massive star progenitors with lower ZAMS masses ($<$60\,$\mathrm{M}_\odot$) and/or solar metallicity (see, e.g., \citealt{2014ApJ...783...10S}). Despite their high values of $\xi_{2.5} \ge 0.77$, all of our models except the rapidly rotating case of C60C will exhibit shock revival in our core-collapse simulations.  
The iron core masses of our progenitors are in the range of 2.37$-$2.72\,$\mathrm{M}_\odot$ (see Table~\ref{tab:progenitor_property}), and the ratio of the rotational energy to the gravitational energy of the pre-collapse iron core of Model C60C is 2.71$\times 10^{-3}$ and of Model R80Ar it is 5.62$\times 10^{-5}$. Fig.~\ref{fig:progenitor} displays profiles of the density, $\rho$, the quantity $r^3\rho$, the temperature, $T$, the electron fraction, $Y_\mathrm{e}$, the radial velocity, $v^{r}$, angular frequency, $\omega$, Kerr parameter of enclosed mass $m(r)$, $a_\mathrm{Kerr}$, given by equation\,\eqref{eq:kerr_encl}, and the equatorial specific angular momentum, $j_{z,\mathrm{eq}}$, versus radius (left column) and enclosed mass (right column) for our set of progenitors at the onset of core collapse. The chemical composition in terms of mass fractions of selected nuclear species is depicted in Fig.~\ref{fig:progenitor_composition}. The progenitors have already lost their outer hydrogen and helium shells and have massive oxygen shells of up to 40\,$\mathrm{M}_\odot$ with a positive gradient of $r^3 \rho$ up to $\sim$\,30\,M$_\odot$, which is likely to lead to strong shock deceleration, if shock revival should occur and the shock expands into these oxygen shells.

The rapidly rotating Model C60C formally fulfills the condition that might allow a part of the collapsing star to assemble into an accretion disk (AD) after its inner core has collapsed to a BH. In the bottom panel of Fig.~\ref{fig:progenitor}, we show the equatorial specific angular momentum, $j_{z,\mathrm{eq}}$, of the rotating models (solid lines) compared to the specific angular momentum of the innermost stable circular orbit (ISCO) of a Kerr BH (dashed lines), both plotted as functions of radius and enclosed mass. The value at the ISCO is the angular momentum a test particle needs to possess in order to be centrifugally supported on a circular orbit around the BH; in relativistic gravity, this Keplerian angular momentum as a function of radius has a local minimum at the ISCO for given values of the BH mass and spin parameter.

We use the rotating BH potential from \citet{1996ApJ...461..565A} for the calculation of the specific angular momentum at the ISCO from the properties of the progenitors, assuming mass and angular momentum conservation during the collapse. If the specific angular momentum of a progenitor at a certain mass coordinate, $m(r)$, exceeds the specific angular momentum of the ISCO corresponding to that mass coordinate $m(r)$ and the associated Kerr parameter $a(m) = 4 \pi \int_0^r \mathrm{d}r' (r')^2 \rho(r')\langle j(r') \rangle c/(Gm^2(r))$, an AD can be formed from the matter at that mass coordinate $m(r)$ \citep[see also][]{2006ApJ...637..914W}. We notice that the specific angular momentum of the C60C progenitor is marginally greater than that of the ISCO near $m(r)\sim 2$--3.5\,$\mathrm{M}_\odot$ and around $m(r)\sim 5$--11\,$\mathrm{M}_\odot$, and it exceeds the ISCO limit considerably for $m(r) \gtrsim 13$\,$\mathrm{M}_\odot$. Therefore, Model C60C has the potential to form an AD around its BH. Here, we compare the specific angular momentum in the equatorial plane with the specific angular momentum of the ISCO, because the disk is mainly formed by equatorially infalling material. Similar analysis for model R80Ar shows that in this model the rotational support against gravity is not strong enough to form an AD at any given mass coordinate, except, maybe, from a small amount of matter in the outermost layers close to the stellar surface (Fig.~\ref{fig:progenitor}, bottom panels). However, these loosely bound near-surface shells of Model R80Ar are likely to become unbound by a sonic pulse sweeping out through the star after successful neutrino-driven shock revival (see Section~\ref{sec:result_after_BH_formation}).

The stellar evolution of the rotating models employed in our study was computed with the angular frequency being constant on spheres. Correspondingly, we set up our core-collapse simulations with such an initial condition. This is in stark contrast to the assumption made by \citet{2021ApJ...919...80F}, who assumed the angular frequency to be constant on cylinders and imposed such a rotation profile on pre-collapse stellar models that had been evolved without including rotation. We point out that the latter assumption is not compatible with rotational equilibrium in the pre-collapse star. Moreover, the specific angular momentum in the equatorial plane of the models considered by \citet{2021ApJ...919...80F} is greater than in our pre-collapse stars for radii larger than $\sim$2500\,$\mathrm{km}$. Both facts combined imply that \citet{2021ApJ...919...80F} performed their core-collapse simulations with stellar models that had much more angular momentum exterior to the iron cores than our progenitors from stellar evolution calculations.

\begin{table}
  \centering
  \caption{Neutrino opacities used for the  CCSN simulations. ``$N$'' denotes nucleons and ``$A$'' and ``$A'$'' denote nuclei. The $\nu \bar{\nu}$ pair processes are taken into account only for $\nu_x$ (for $\nu_e$ and $\bar \nu_e$ the $\beta$-processes are  by far dominant).}
	\begin{tabular}{cc}
		\hline
		Reaction & Neutrino  \\
		\hline
        $\nu + A \leftrightarrow \nu + A$ & $\nu_\mathrm{e}, \bar{\nu}_\mathrm{e}, \nu_\mathrm{x}$  \\
        $\nu + N \leftrightarrow \nu + N$ & $\nu_\mathrm{e}, \bar{\nu}_\mathrm{e}, \nu_\mathrm{x}$  \\
        $\nu_\mathrm{e} + \mathrm{n} \leftrightarrow \mathrm{e}^- + \mathrm{p}$ & $\nu_\mathrm{e}$  \\
        $\nu_\mathrm{e} + A \leftrightarrow \mathrm{e}^- + A'$ & $\nu_\mathrm{e}$  \\
        $\bar{\nu}_\mathrm{e} + \mathrm{p} \leftrightarrow \mathrm{e}^+ + \mathrm{n}$ & $\bar{\nu}_\mathrm{e}$ \\
        $\nu + \bar{\nu} \leftrightarrow \mathrm{e}^- + \mathrm{e}^+$ & $\nu_\mathrm{x}$  \\
        $\nu + \bar{\nu} + N + N \leftrightarrow N + N$ & $\nu_\mathrm{x}$ \\
        \hline
	\end{tabular}
	\label{tab:neutrino_opacity}
\end{table}

\subsection{Numerical Setup}\label{subsec:numerical_setup}

The core-collapse simulations of the VMSs are conducted using the general-relativistic hydrodynamics and transport code \textsc{NADA-FLD} \citep{2014PhRvD..89h4043M,2019MNRAS.490.3545R} in two dimensions. \textsc{NADA-FLD} is a finite difference code in spherical polar coordinates. The code solves the Baumgarte-Shapiro-Shibata-Nakamura formulation (BSSN) of the Einstein equations applying a second-order partially implicit Runge-Kutta method \citep{2013PhRvD..87d4026B}. In this current study, we solve the BSSN equations assuming spherical symmetry. The source terms for the BSSN equations are evaluated using the angle-averaged hydrodynamical and transport quantities. We employ the ``$1+\log$'' condition for the lapse function \citep{1995PhRvL..75..600B} and the non-advective hyperbolic Gamma-driver for the shift vector with a damping parameter of $10^{-5}$ (see, e.g., \citealt{2003PhRvD..67h4023A} for a discussion of the gauge conditions used in this study).

We apply the generalized Valencia formalism for the hydrodynamics equations \citep{2014PhRvD..89h4043M}. A finite difference high-resolution shock-capturing method is employed to solve the hydrodynamics equations and the piecewise parabolic method (PPM) of \citet{1984JCoPh..54..174C} is used for the reconstruction of the hydrodynamical quantities at the cell interfaces from the cell-centered values. The approximate Harten-Lax-van Leer Riemann solver \citep[HLL;][]{Harten1983} is applied for the evaluation of the numerical flux at the interfaces between adjacent cells, and the time integration of the hydrodynamics equations is conducted by applying a second-order Runge-Kutta method. We employ a spherical core of 3\,km radius to avoid excessive time step restrictions imposed by the Courant-Friedrichs-Lewy condition at the center of the spherical polar grid. The tabulated Steiner, Fischer, and Hempel (SHFo) EOS \citep{2012ApJ...748...70H,2013ApJ...774...17S} is used to close the set of hydrodynamical equations. The SFHo EOS assumes nuclear statistical equilibrium (NSE) over the whole ranges of density and temperature (the density range of the table is 1.67$\times 10^3$\,$\mathrm{g/cm^3}$--3.16$\times 10^{15}$\,$\mathrm{g/cm^3}$, and the temperature range is 0.1\,MeV--158.5\,MeV).

A general relativistic multi-dimensional multi-energy group flux-limited diffusion (FLD) scheme is employed to solve the neutrino transport equations \citep{2019MNRAS.490.3545R}. The comoving frame FLD equation is integrated using a mixed implicit-explicit method. In this study, we employ the Levermore-Pomraning flux-limiter \citep{1981ApJ...248..321L}. The neutrino energy grid consists of 16 geometrically spaced points spanning from 2.5\,MeV to 500\,MeV. We evolve the FLD equations for electron neutrinos $\nu_{\mathrm{e}}$, electron anti-neutrinos $\bar{\nu}_{\mathrm{e}}$, and $\nu_{\mathrm{x}}$ representing muon and tau neutrinos and their antineutrinos. The neutrino reactions considered are shown in the Table \ref{tab:neutrino_opacity} and details about the opacities can be found in \citet{1985ApJS...58..771B} and \citet{2002A&A...396..361R} and references cited in those articles. Additionally, we include corrections due to weak magnetism and nucleon recoil for charged-current and neutral-current neutrino-nucleon interactions \citep{2002PhRvD..65d3001H}. The e$^\pm$ pair-process and nucleon-nucleon bremsstrahlung for $\nu_\mathrm{x}$ are implemented according to the treatment by \citet{2015ApJS..219...24O}.

The CCSN simulations are initialized using density, pressure, electron fraction, and velocities of the pre-collapse progenitors from \citet{2017ApJ...836..244W}, and all other thermodynamical quantities such as temperature, internal energy, etc. are evaluated using the SFHo EOS. For the rotating progenitors, we follow the description in the stellar evolution models and assume constant angular frequency on spherical mass shells rather than taking the angular momentum constant on spherical shells. If the angular momentum is assumed constant on spheres, the angular frequency along the rotational axis becomes extremely high and leads to numerical instabilities. In our initial setup, we ensured that the total angular momentum in a mass shell of the progenitor and the initialized model are equal.

A uniform grid is applied in the polar direction with an angular resolution of 1.4 degrees. The radial grid has 500 grid points. It has a uniform resolution of $\sim$0.075\,km (5\% of $G \mathrm{M}_\odot/c^2$) until $r=5$\,km, which allows us to track the evolution when the NS collapses and forms a BH, and a resolution of $\Delta r/r \sim 1$--2\% in the region 5\,km$<$\,$r$\,$<$300\,km and of $\Delta r/r \sim$2--3\% outside of $r$=300\,km. For the initial core-collapse simulations with \textsc{NADA-FLD} the outer boundary is placed at $2 \times 10^4$\,km. (See Figs.~\ref{fig:progenitor} and \ref{fig:progenitor_composition} for the corresponding locations in the stars.) Reflecting boundary conditions are applied at the center and on the polar axis. At the outer boundary, inflow conditions, Sommerfeld, and free-streaming boundary conditions are employed for the hydrodynamics, the Einstein, and the transport equations, respectively.

We continue our simulations with the \textsc{NADA-FLD} code beyond the moment of BH formation. When the infalling mass flow towards the central BH has become supersonic at a radius of $r=800$\,km, we map our models to the Newtonian multi-fluid finite volume hydrodynamics code \textsc{Prometheus} \citep{1991ApJ...367..619F} and evolve the models for a longer period of time. The \textsc{Prometheus} code solves the multi-dimensional hydrodynamics equations, employing the PPM reconstruction scheme of \citet{1984JCoPh..54..174C}; the hydrodynamical fluxes are evaluated using the Riemann solver for ideal gases from \citet{1985JCoPh..59..264C}. The advection of different nuclear species with the fluid flow is treated with the consistent multi-fluid advection method (CMA) of \citet{1999A&A...342..179P}.

For our long-time simulations with the \textsc{Prometheus} code, we apply the tabulated Helmholtz EOS of \citet{2000ApJS..126..501T}, which takes into account arbitrarily degenerate and relativistic electrons and positrons, photons, and a set of 15 nuclear species (free neutrons, free protons, and 13 alpha nuclei from $^4$He to $^{56}$Ni; for details, see \citealt{2020MNRAS.496.2039S}). At temperatures below 7$\times 10^9$\,K, nuclear reactions are taken into account by a network solver. At higher temperatures, we assume NSE with the composition being interpolated from tabulated values. Moreover, if the density of the stellar material is below a threshold density of $10^3$\,$\mathrm{g/cm^3}$, no nuclear burning is considered.

In the process of mapping, the data from the \textsc{NADA-FLD} simulations are used for the matter interior to a radius of $r=2\times 10^4$\,km (the outer boundary of the computational domain in the \textsc{NADA-FLD} simulations), whereas the progenitor data are used for the matter exterior to the mentioned radius. The simulations with the \textsc{Prometheus} code are initialized with the density, electron fraction, internal energy, and fluid velocity from the \textsc{NADA-FLD} runs. Employing these quantities ensures an optimal match of the diagnostic explosion energy between the \textsc{NADA-FLD} and \textsc{Prometheus} simulations, but we also have an eye on a good agreement of pressure and temperature values. For mapping the composition of the matter below $r=2\times 10^4$\,km, the mass fractions of neutrons, protons, light nuclei (deuterium, tritium, $^3$He), alpha particles, and heavy nuclei from the SFHo EOS used in \textsc{NADA-FLD} have to be identified with those of the 15 nuclear species considered in the \textsc{Prometheus} runs. Because light nuclei are not tracked by the latter simulations, we add their abundances to the mass fraction of alpha particles. Moreover, since the NSE treatment of the SFHo EOS represents all nuclei heavier than $^4$He by a single nucleus with mean mass number $\bar A$ and mean charge number $\bar Z$, we convert the mass fraction of this representative heavy nucleus to a nucleus in the $\alpha$-chain reaction network that has a mass number closest to the mean mass number $\bar A$. Extra neutrons and protons originating from a mismatch of $(\bar A,\bar Z)$ with $(A,Z)$ of the chosen $\alpha$-nucleus are added to the abundance of the respective nucleon. The composition of matter above $r=2\times 10^4$\,km is initialized by using data from the progenitor at the onset of the collapse.

In the long-time \textsc{Prometheus} simulations, the innermost 400\,km are excised and are replaced by a point mass and an open inner boundary condition that allows the inflowing matter to leave the grid in free fall. At the time of the mapping, the radial infall velocities at the inner boundary are supersonic; therefore, hydrodynamical quantities outside the inner boundary are not influenced by the properties of the flow passing the inner boundary. 

The outer boundary is treated in a model dependent way. For Model C60C-NR, its location is enlarged from $5.85\times 10^{5}$\,km (mass coordinate: 41.01\,$\mathrm{M}_\odot$) to $1\times 10^{9}$\,km (mass coordinate: 43.23\,$\mathrm{M}_\odot$) at 36\,s after bounce. At the same time, the inner boundary is moved from 400\,km to $4\times 10^4$\,km and, afterwards, the inner boundary is progressively moved radially outwards (to be located at 10$-$20\% of the shock radius) to increase the integration time step. While moving the inner boundary, we ensure that the infalling matter is supersonic at the inner boundary. For both Model R80Ar and Model R80Ar-NR the outer boundary is set at a radius (mass coordinate) of $9.79\times 10^{6}$\,km (47.39\,$\mathrm{M}_\odot$). Similarly, for Model C115, the outer boundary of the simulation domain is placed at 1.95$\times 10^{7}$\,km (mass coordinate: 45.50\,$\mathrm{M}_\odot$).

In the \textsc{Prometheus} simulations we do not track the evolution of the matter in the central volume but still want to account for its gravitational effects on the medium on the computational grid by a Newtonian treatment (which is valid at large distances from the center). For the transition from the \textsc{NADA-FLD} simulations to the \textsc{Prometheus} calculations, we follow \citet{2009MNRAS.399..229K} in order to determine the relevant mass to be used in the Newtonian potential. For this purpose, we determine the gravitational potential in the Newtonian limit from the lapse function, $\alpha$, as $\Phi_\mathrm{in}(r_\mathrm{in}) = (\alpha(r_\mathrm{in})^2 - 1)\,c^2 / 2$, where $\Phi_\mathrm{in}(r_\mathrm{in})$ is the gravitational potential generated by the matter interior to the inner boundary, $r_\mathrm{in}$, of the simulation domain of the Prometheus code. Setting $\Phi_\mathrm{in}(r_\mathrm{in})$ equal to the Newtonian expression for the gravitational potential,
\begin{eqnarray}\label{eq:newtonian_potential}
    \Phi_\mathrm{in}(r_\mathrm{in}) &=& -\,\frac{G M(r_\mathrm{in})}{r_\mathrm{in}}\,,
\end{eqnarray}
allows us to determine the effective gravitational mass $M(r_\mathrm{in})$ that accounts for the gravitational effects of the mass interior to the inner boundary of the simulation domain of the Prometheus code. Self-gravity of the matter on the computational grid is then accounted for by applying the superposition principle valid in the Newtonian limit, i.e. we use $M_\mathrm{total}(r) = M(r_\mathrm{in})+\Delta M(r)$, in the total Newtonian potential, where $\Delta M(r)$ is the mass between the inner grid boundary and radius $r$. This procedure ensures that no transients are created due to the different treatments of gravity in the \textsc{NADA-FLD} code and the \textsc{Prometheus} code, because the former uses a relativistic treatment of gravity and the latter uses a Newtonian treatment in this study. During the mapping, we also ensure that the total energy ($=$ internal energy $+$ kinetic energy $+$ potential energy) of the matter in the simulation domain of the \textsc{NADA-FLD} runs ($r\le 2\times10^4$\,km) is conserved.

During the \textsc{Prometheus} simulations, self-gravity is taken into account by assuming a spherical gravitational potential, employing the Poisson solver of \citet{1995CoPhC..89...45M} and being consistent with the gravity treatment by \textsc{NADA-FLD}). The excised volume around the grid center is taken into account in the evaluation of the gravitational potential by the central point mass. We stress that in our calculations of this central point mass we account for the time-dependent increase of the mass in the central volume connected to gas infall through the inner grid boundary during the \textsc{Prometheus} simulations.

In this study, the \textsc{Prometheus} simulations use the same angular resolution of 1.4 degrees as the \textsc{NADA-FLD} simulations. The radial grids employed in both simulations are also identical up to a certain radius (i.e., the radius of the shock at the time when the data are mapped from the \textsc{NADA-FLD} to the \textsc{Prometheus} simulations), and exterior to this radius the \textsc{Prometheus} code applies a geometrical grid with a resolution of $\Delta r/r\le 1\%$.

\section{Definitions}\label{sec:definition}

In this section, we introduce the definitions of different diagnostic quantities which are used for the analysis of simulation results. In numerical relativity, the mass accretion rate at a radius $r$ is defined as:
\begin{eqnarray}\label{eq:mass_accretion}
    \dot M(r) = r^2 \eul^{4 \phi} \int_\mathrm{sphere} \mathrm{d}\Omega W \rho (v^r - c\,{\beta}^r/\alpha)\,,
\end{eqnarray}
where $\phi$, $\beta^r$, $\alpha$, $\rho$, $v^r$, $W$ are the conformal factor, radial component of the shift vector, the lapse function, the rest-mass density of baryons, the radial fluid velocity and the Lorentz factor, respectively, and $\mathrm{d}\Omega = 2 \pi \mathrm{d}(\cos \theta)$ for our 2D simulations. In the above equation, the integration is carried out over the 4$\pi$-sphere. The baryonic mass of a shell is defined as (\citealt{baumgarte_shapiro_2010}):
\begin{eqnarray}\label{eq:mass}
    \Delta m(r) = \int_\mathrm{shell} \mathrm{d}V W \rho\,,
\end{eqnarray}
with $\mathrm{d}V = \eul^{6 \phi} r^2 \mathrm{d}r \mathrm{d}\Omega$. The integral spans over the width of a mass shell in the radial direction, from 0 to $\pi$ in the polar direction, and from 0 to 2$\pi$ in the azimuthal direction. Similarly, the total angular momentum of a shell along the rotation axis is calculated by (\citealt{baumgarte_shapiro_2010}):
\begin{eqnarray}\label{eq:angular_momentum}
    \Delta J(r) = \int_\mathrm{shell} \mathrm{d}V \frac{W^2 \rho h}{c^2} v_\phi\,,
\end{eqnarray}
where $h=c^2+e+P/\rho$ is the specific enthalpy, and $e$ and $P$ are the specific internal energy without particle rest masses and the pressure, respectively. The covariant azimuthal component of the fluid velocity, $v_\phi$ in equation~(\ref{eq:angular_momentum}), contains the distance from the rotation axis squared, $(r\sin\theta)^2$, and therefore possesses units of cm$^2$/s \citep{baumgarte_shapiro_2010}.

The baryonic mass in the gain layer is given by:
\begin{eqnarray}\label{eq:gain_mass}
    M_\mathrm{gain} = \int_{R_\mathrm{gain}(\theta) < r < R_\mathrm{sh}(\theta)} \mathrm{d}V W \rho\,,
\end{eqnarray}
where the integration is performed over the volume between $R_\mathrm{gain}(\theta)$ and $R_\mathrm{sh}(\theta)$, which are the angle-dependent gain radius and shock radius, respectively. The gain radius is the radius where the neutrino heating becomes dominant over neutrino cooling. The net neutrino heating rate (i.e., heating minus cooling) per unit of mass in the gain layer is evaluated by the following formula: 
\begin{eqnarray}\label{eq:gain_neutrino_heating}
    \dot q_\mathrm{gain} = \frac{1}{M_\mathrm{gain}} \int_{R_\mathrm{gain}(\theta) < r < R_\mathrm{sh}(\theta)} \mathrm{d}V \dot{q}_{\nu}\,,
\end{eqnarray}
where $\dot{q}_{\nu}$ is the same as $S_\mathrm{E}$ in \citet{2019MNRAS.490.3545R}. It measures the net neutrino heating rate per unit volume and contains a factor of $W$ accounting for relativistic effects. Similarly, the average specific angular momentum in the gain layer along the rotation axis is given by:
\begin{eqnarray}\label{eq:gain_angular_momentum}
    j_\mathrm{gain} = \frac{1}{M_\mathrm{gain}} \int_{R_\mathrm{gain}(\theta) < r < R_\mathrm{sh}(\theta)} \mathrm{d}V \frac{W^2 \rho h}{c^2} v_\phi\,.
\end{eqnarray}

We define the surface of the proto-neutron star (PNS) at a baryonic density of $10^{11}\,\mathrm{g/cm^3}$. In the Newtonian limit, the rotation period of the PNS, assuming rigid rotation, is given by:
\begin{eqnarray}\label{eq:ns_rotational_period}
T_\mathrm{ns} = \frac{2 \pi I_\mathrm{ns}}{J_\mathrm{ns}}\,,
\end{eqnarray}
where $I_\mathrm{ns}$ and $J_\mathrm{ns}$ are the moment of inertia and the total angular momentum of the PNS. Correspondingly, the rotation period of a mass shell is given by:
\begin{eqnarray}\label{eq:rotational_period}
T_\mathrm{rot} = \frac{2 \pi \Delta I}{\Delta J}\,,
\end{eqnarray}
where $\Delta I$ and $\Delta J$ are the moment of inertia and the total angular momentum of the mass shell. The mass-weighted angular average of any quantity $X$ at radius $r$ is given by:
\begin{eqnarray}\label{eq:mass_weighted_average}
\langle X \rangle(r) \equiv \frac{\eul^{6 \phi} r^2 \int_{4 \pi} \mathrm{d}\Omega X W \rho}{\eul^{6 \phi} r^2  \int_{4 \pi} \mathrm{d}\Omega W \rho}\,.
\end{eqnarray}
Similarly, the volume-weighted angular average of any quantity $X$ is given by:
\begin{eqnarray}\label{eq:volume_weighted_average}
\langle X \rangle(r) \equiv \frac{\eul^{6 \phi} r^2 \int_{4 \pi} \mathrm{d}\Omega X}{\eul^{6 \phi} r^2 \int_{4 \pi} \mathrm{d}\Omega}\,.
\end{eqnarray}
Since we are using a spherically symmetric metric, the factors $\eul^{6 \phi} r^2$ cancel each other in the numerators and denominators of equations~(\ref{eq:mass_weighted_average}) and (\ref{eq:volume_weighted_average}).

We calculate the turbulent kinetic energy density of the fluid measured by a local stationary observer by \citep{2013rehy.book.....R}:
\begin{eqnarray}\label{eq:turb_energy}
    e_\mathrm{turb} &=& \rho\,c^2\,(W_\mathrm{turb} - 1)\,, \nonumber \\
    \mathrm{with}\quad
    W_\mathrm{turb} &=& \frac{1}{\sqrt{1 - \sum_{i = r,\theta,\phi} ({\bar v^i}_\mathrm{turb}/c)^2}}\,,
\end{eqnarray}
where $\bar v^i$ is the three-velocity in the orthonormal tetrad basis (see e.g., \citealt{2012arXiv1212.4064E} for a discussion of the orthonormal tetrad basis). The turbulent velocity ${\bar v^i}_\mathrm{turb}$ is the difference between $\bar v^i$ and $\langle \bar v^i \rangle$, where $\langle \bar v^i \rangle$ is computed using equation~\eqref{eq:mass_weighted_average}. We evaluate the difference between $\bar v^i$ and $\langle \bar v^i \rangle$ using the relativistic velocity composition laws (see e.g., \citealt{2013rehy.book.....R}). 

The total energy density without the rest-mass energy is estimated using the following formula from \citet{2017MNRAS.472..491M}:
\begin{eqnarray}\label{eq:total_energy}
    e_\mathrm{tot} &=& \alpha (\rho h W^2 - P) - \rho W c^2\ + \rho \Phi_\mathrm{out}(r)\,, \nonumber \\
    \mathrm{with}\quad
    \Phi_\mathrm{out}(r) &=& - 4 \pi G \int_{r}^{\infty} \mathrm{d}r' \rho r',
\end{eqnarray}
where $\Phi_\mathrm{out}(r)$ is the Newtonian gravitational potential generated by the spherical mass shells outside a given radius $r$, where general relativistic corrections can be assumed to be negligible. The total energy in the gain layer is given by
\begin{eqnarray}\label{eq:gain_total_energy}
    E^\mathrm{tot}_\mathrm{gain} = \int_{R_\mathrm{gain}(\theta) < r < R_\mathrm{sh}(\theta)} \mathrm{d}V e_\mathrm{tot}\,.
\end{eqnarray}
Similarly, we define the diagnostic (explosion) energy and the volume-filling factor of the neutrino-heated postshock matter that possesses positive total energy and thus contributes to
$E_\mathrm{diag}$ as
\begin{eqnarray}\label{eq:gain_diagnostic_energy}
    E_\mathrm{diag} = \int_{R_\mathrm{low}(\theta) < r < R_\mathrm{sh}(\theta)} \mathrm{d}V e_\mathrm{tot} \Theta (e_\mathrm{tot})\Theta (v^{r})\,,
\end{eqnarray}
\begin{eqnarray}\label{eq:alpha_parameter}
    \alpha_\mathrm{diag} = \frac{\int_{R_\mathrm{low}(\theta) < r < R_\mathrm{sh}(\theta)} \mathrm{d}V \Theta (e_\mathrm{tot})\Theta (v^{r})}{\int_{R_\mathrm{low}(\theta) < r < R_\mathrm{sh}(\theta)} \mathrm{d}V}\,,
\end{eqnarray}
where $\Theta(x)$ is the Heaviside step function. The lower integration bound is $R_\mathrm{low} = R_\mathrm{gain}$ before BH formation, and $R_\mathrm{low} = R_\mathrm{BH}$ after the compact remnant has collapsed to a BH, with $R_\mathrm{BH}$ being the BH radius. The overburden energy is calculated using
\begin{eqnarray}\label{eq:gain_overburden_energy}
    E_\mathrm{ob} = \, \int_{R_\mathrm{sh}(\theta) < r < R_\mathrm{prog}} \mathrm{d}V e_\mathrm{tot}\,,
\end{eqnarray}
where $R_\mathrm{prog}$ is the outer radius of the gravitationally bound pre-collapse star, defined by the outermost radius where the radial velocity of outward moving gas is still smaller than the escape velocity at that radius. 

The general relativistic local rest frame neutrino luminosity, i.e., the energy loss rate in all directions, gravitationally redshifted for an observer at infinity (``lab frame''), is given by (\citealt{2010CQGra..27k4103O}):
\begin{eqnarray}\label{eq:luminosity}
L_{\nu}(r) = \alpha(r) \eul^{4\phi(r)} r^2 \int_{4 \pi} \mathrm{d}\Omega \int_{0}^{\infty} \mathrm{d}\epsilon \mathcal{H}_{\nu}(r,\theta,\epsilon)\,,
\end{eqnarray}
with $\mathcal{H}_{\nu}$ being the neutrino energy flux density in the local rest frame. Accordingly, the lab frame mean neutrino energy of the direction-integrated neutrino flux is defined as 
\begin{eqnarray}\label{eq:emean}
\langle \epsilon_{\nu} \rangle(r) = \alpha(r) \, \frac{\int_{4 \pi} \mathrm{d}\Omega \int \epsilon \mathcal{H}_{\nu}(r,\theta,\epsilon)~\epsilon^{-1} \mathrm{d}\epsilon}{\int_{4 \pi} \mathrm{d}\Omega \int \mathcal{H}_{\nu}(r,\theta,\epsilon)~\epsilon^{-1} \mathrm{d}\epsilon}\,.
\end{eqnarray}
The corresponding lab frame root-mean-square (RMS) neutrino energy of the direction-integrated neutrino flux is given by
\begin{eqnarray}\label{eq:erms}
\sqrt{\langle \epsilon_{\nu}^2 \rangle}(r) = \alpha(r) \, \sqrt{\frac{\int_{4 \pi} \mathrm{d}\Omega \int \epsilon^2 \mathcal{H}_{\nu}(r,\theta,\epsilon)\epsilon^{-1} \mathrm{d}\epsilon}{\int_{4 \pi} \mathrm{d}\Omega \int \mathcal{H}_{\nu}(r,\theta,\epsilon)\epsilon^{-1} \mathrm{d}\epsilon}}\,.
\end{eqnarray}
We define the neutrino heating efficiency in the gain layer by
\begin{eqnarray}\label{eq:heat_effi}
\eta = \frac{\dot Q_\mathrm{gain}}{L_\mathrm{\nu_e}(r=1.5\,R_\mathrm{ns})+L_\mathrm{\bar \nu_e}(r=1.5\,R_\mathrm{ns})}\,,
\end{eqnarray}
where $\dot Q_\mathrm{gain}$ = $M_\mathrm{gain}\,\dot q_\mathrm{gain}$ is the net heating rate by neutrinos in the gain layer. Here, we evaluate the neutrino luminosities not at the gain radius but at 1.5 times the neutron star radius (denoted by $R_\mathrm{ns}$ and corresponding to a baryonic density of $10^{11}\,\mathrm{g/cm^3}$), because the gain radius has a highly fluctuating time evolution, which disfavors its use in the present analysis; but on average it fulfills $R_\mathrm{gain} \sim 1.5R_\mathrm{ns}$ in our models.

The characteristic timescale of advection of matter through the gain region, which measures how long gas falling into the shock typically stays in the gain layer, can be expressed as (\citealt{2006A&A...457..281B})
\begin{eqnarray}\label{eq:timescale_advection}
\tau_\mathrm{adv} = \frac{M_\mathrm{gain}}{\dot M(r=500\,\mathrm{km})}\,.
\end{eqnarray}
The heating timescale of matter in the gain layer is given by the following formula:
\begin{eqnarray}\label{eq:timescale_heating}
\tau_\mathrm{heat} = \frac{|E^\mathrm{tot}_\mathrm{gain}|}{\dot Q_\mathrm{gain}}\,.
\end{eqnarray}
Here, $\dot Q_\mathrm{gain}$ is again the net neutrino heating rate in the gain layer. Following \citet{2018ApJ...852...28S}, we determine the Rossby number in the rotating gain layer by
\begin{eqnarray}\label{eq:rossby_number}
\mathrm{Ro} = \frac{R_\mathrm{sh}^2}{2 \tau_\mathrm{adv} j_\mathrm{gain}}\,,
\end{eqnarray}
with $R_\mathrm{sh}$ being the average shock radius.

We calculate the quadrupole amplitude of the gravitational wave (GW) emission, $A^{\mathrm{E}2}_{20}$, according to the formula (see, e.g., \citealt{2006A&A...450.1107O})
\begin{eqnarray}\label{eq:gravitational_wave_amplitude}
A^{\mathrm{E}2}_{20} =&& \frac{G}{c^4}\frac{32\pi^{3/2}}{\sqrt{15}} \int_{-1}^{+1} \mathrm{d}z \int_{0}^{\infty} \mathrm{d}r r^2 \eul^{6 \phi} W \rho \Big[ v_r v_r (3z^2-1) \nonumber \\
&& + v_{\theta} v_{\theta} (2-3z^2) - v_{\phi} v_{\phi} - 6 v_r v_{\theta} z \sqrt{1-z^2} \nonumber \\
&& -  r \partial_r \Phi (3z^2-1) + 3\partial_{\theta} \Phi z \sqrt{1-z^2} \Big]\,,
\end{eqnarray}
where $z=\cos \theta$ and we use the relativistic differential mass element $\mathrm{d}m = 2 \pi \eul^{6 \phi} W \rho r^2 \mathrm{d}r \mathrm{d}z$ (the same as in equation \eqref{eq:mass} and following \citealt{2013ApJ...766...43M}) rather than the Newtonian differential mass element $\mathrm{d}m_\mathrm{newt} = 2 \pi \rho r^2 \mathrm{d}r \mathrm{d}z$ used by \citet{2006A&A...450.1107O}. Such a relativistic correction to $A^{\mathrm{E}2}_{20}$ shifts the GW emission frequency to higher values (see, e.g., \citealt{2013ApJ...766...43M}), which motivates us to use the relativistic differential mass element in our work. Since we assume spherical symmetry in solving the BSSN equations for the GR metric, however, we are unable to fully employ the relativistic formulation of $A^{\mathrm{E}2}_{20}$ given by \citet{2013ApJ...766...43M}, which uses the relativistic stress tensor and requires the multi-dimensional evolution of the metric quantities. For the same reason of constraining ourselves to a spherically symmetric solution of the BSSN equations, we recalculate the gravitational potential in 2D by solving $\nabla^2 \Phi = 4 \pi G\,c^{-2} (\rho c^{2} + e\rho + 3P)$ and replace the monopole part of the potential by $\Phi_0 = (\alpha^2-1)\,c^2/2$ for the GW analysis. The dimensionless strain, $h_{+}$, measured by an observer at a distance $D$ and at an inclination angle $\vartheta$ relative to the symmetry axis of our 2D models is given by 
\begin{eqnarray}\label{eq:gravitational_wave_strain}
h_{+} = \frac{1}{8} \sqrt{\frac{15}{\pi}} \sin^2 \vartheta \frac{A^{\mathrm{E}2}_{20}}{D}.
\end{eqnarray}
In this work, we assume that the observer is located on the equatorial plane ($\sin^2 \vartheta = 1$). Moreover, we denote the post-bounce time by $t_\mathrm{pb}$.

\begin{table*}
	\centering
    \caption{Characteristic properties of the simulated models until BH formation.}
	\begin{tabular*}{0.72\textwidth}{lcccccccc}
		\hline
		Model & $t_\mathrm{sh-exp}$ & $t_\mathrm{BH}$ & $M_\mathrm{f,ns}$ & $T_\mathrm{f,ns}(t_\mathrm{BH})$ & $a_\mathrm{BH}(t_\mathrm{BH})$ & $R_\mathrm{sh}(t_\mathrm{BH})$ & $E_\mathrm{diag}(t_\mathrm{BH})$ & $E_\mathrm{ob}(t_\mathrm{BH})$ \\
        & $[\mathrm{s}]$ & $[\mathrm{s}]$ & [$\mathrm{M}_\odot$] & $[\mathrm{ms}]$ & &
        [$\mathrm{km}$] & [$10^{51}~\mathrm{erg}$] & [$10^{51}~\mathrm{erg}$] \\
        \hline
        C60C-NR  & 0.250 & 0.580 & 2.58 &      &       & 4552 & 1.58 & $-$4.89 \\
        C60C     &       & 0.510 & 2.84 & 1.17 & 0.725 & 130  &      &      \\        
        R80Ar-NR & 0.246 & 0.350 & 2.67 &      &       & 1560 & 1.43 & $-$10.6 \\
        R80Ar    & 0.237 & 0.350 & 2.67 & 7.90 & 0.1   & 1721 & 1.53 & $-$10.4 \\
		C115     & 0.222 & 0.400 & 2.64 &      &       & 2239 & 1.54 & $-$7.25 \\ 
        \hline        
	\end{tabular*}
	\flushleft
    \textit{Notes}: Here, $t_\mathrm{sh-exp}$ denotes the post-bounce time at the onset of shock revival (i.e., the post-bounce time when the angle-averaged shock radius reaches a value of 400\,km), $t_\mathrm{BH}$ is the post-bounce time when a BH begins to form, $M_\mathrm{f,ns}$ the baryonic mass of the PNS at the time of BH formation, $T_\mathrm{f,ns}$ the corresponding average rotation period of the PNS (assuming rigid rotation for given angular momentum; see equation\,\eqref{eq:ns_rotational_period}), and $a_\mathrm{BH}$, $R_\mathrm{sh}$, $E_\mathrm{diag}$, and $E_\mathrm{ob}$ denote the Kerr parameter of the newly born BH, shock radius, diagnostic explosion energy, and overburden energy at the time of BH formation. The time of BH formation is defined by the moment when the apparent horizon finder first detects the appearance of an event horizon.
	\label{tab:model_property}
\end{table*}

\begin{figure}
	\includegraphics[width=0.48\textwidth]{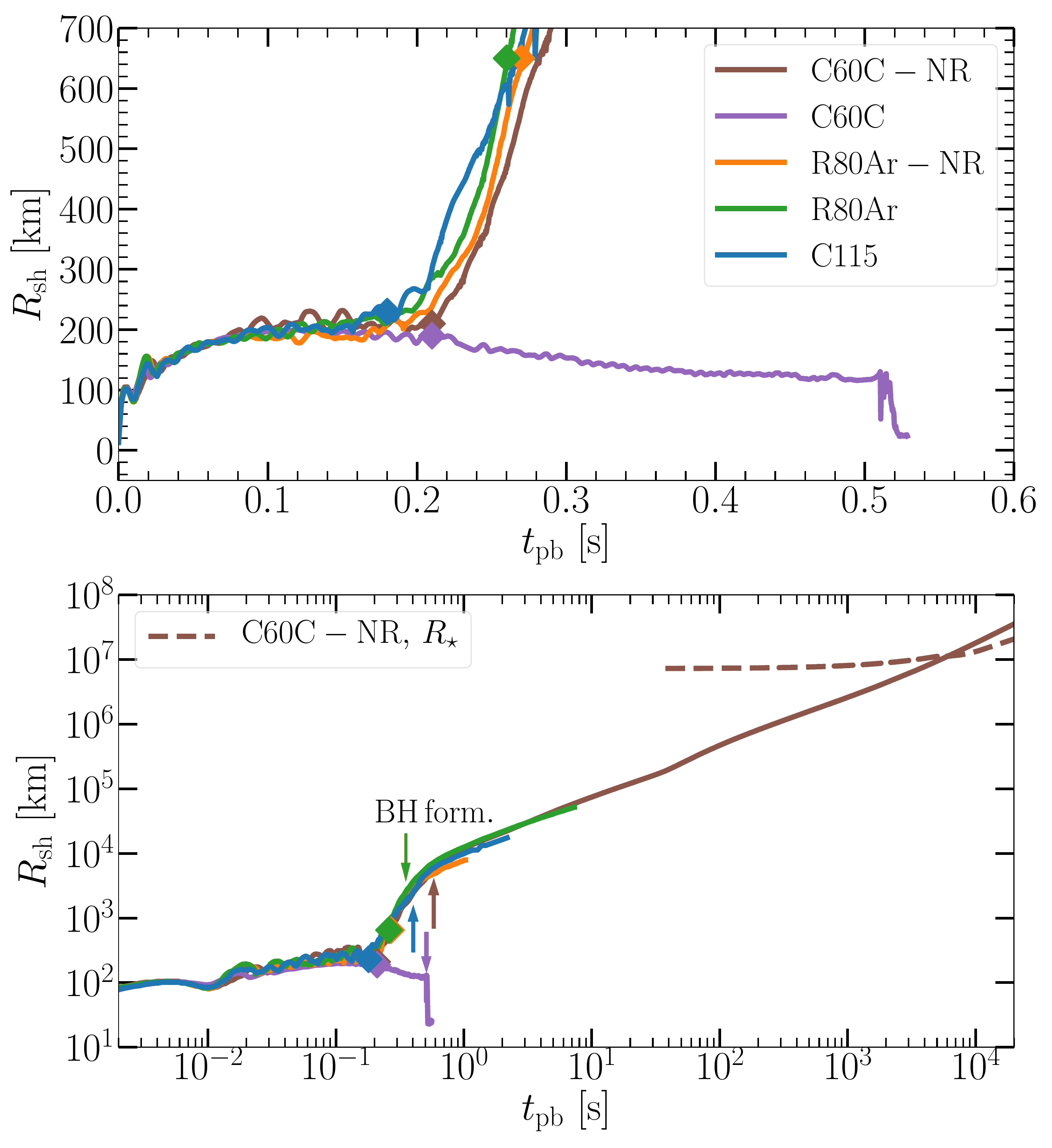}
    \vspace*{-7mm}
	\caption{Time evolution of the angle-averaged shock radius for our set of simulated models. Here, $t_\mathrm{pb}$ is the post-bounce time. Note the different timescales on the horizontal axes of the upper and lower panels. The non-rotating Models C60C-NR (brown lines), R80Ar-NR (orange lines), C115 (blue lines), and the slowly rotating Model R80Ar (green line) experience shock expansion around 200\,ms after bounce. However, only in Model C60C-NR, the shock survives as a shock wave until its breakout from the stellar surface (indicated by the brown dashed line), whereas in the other models at some point the outgoing shock converts to a sonic pulse. In contrast, the rapidly rotating Model C60C (violet line) does not show shock expansion. The moments when the interface between iron core and Si shell falls through the shock in our models are marked by the diamond shaped markers. Vertical arrows in the lower panel mark the moments of BH formation in the different simulations.}
	\label{fig:shock_radius}
\end{figure}
\begin{figure}
	\includegraphics[width=0.48\textwidth]{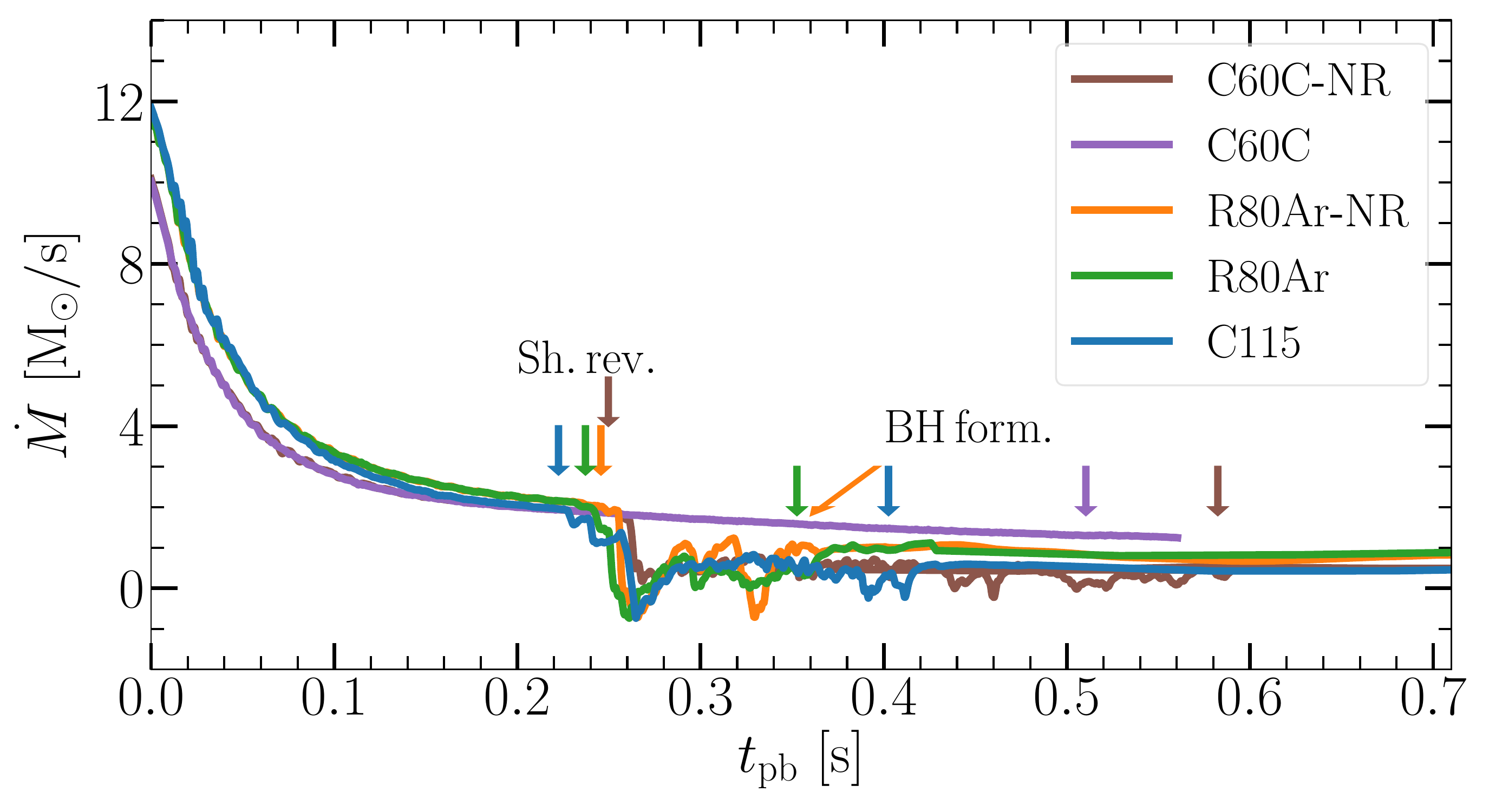}
    \vspace*{-8mm}
	\caption{Time evolution of the mass accretion rate at a radius of 500\,km for our set of simulated models. The line colors are chosen to be the same as in Fig.~\ref{fig:shock_radius}. The left cluster of arrows marks the instants of shock revival, the five arrows at later times indicate the moments of BH formation in all simulations.}    
	\label{fig:mass_accretion}
\end{figure}
\begin{figure}
	\includegraphics[width=0.48\textwidth]{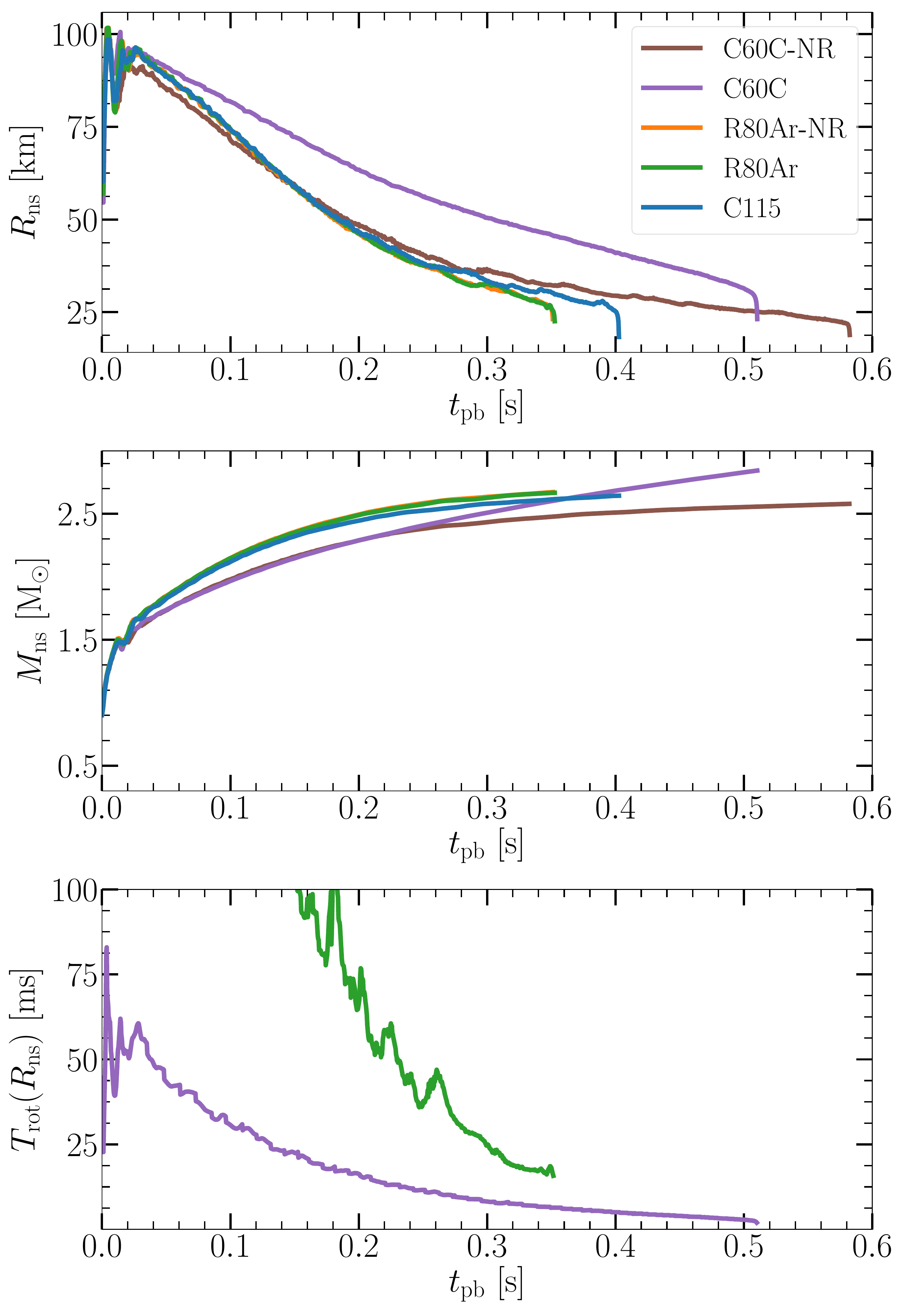}
    \vspace*{-7mm}
	\caption{Time evolution of the angle-averaged PNS radius (top panel), PNS baryonic mass (middle panel), and rotational period at the PNS surface according to equation~\eqref{eq:rotational_period} (bottom panel) for all of our models. The PNS surface is defined at a baryonic density of $10^{11}\,\mathrm{g/cm^3}$. The line colors for the different models are the same as in Fig.~\ref{fig:shock_radius}. Models C60C-NR (brown line), C60C (violet line), R80Ar-NR (orange line), R80Ar (green line), and C115 (blue line) form BHs at 580, 510, 350, 350, and 400\,ms after core bounce, respectively, with the initial BH masses of 2.58 2.84, 2.67, 2.67, 2.64\,$\mathrm{M}_\odot$, respectively.}    
	\label{fig:NS_property}
\end{figure}
\begin{figure}
	\includegraphics[width=0.48\textwidth]{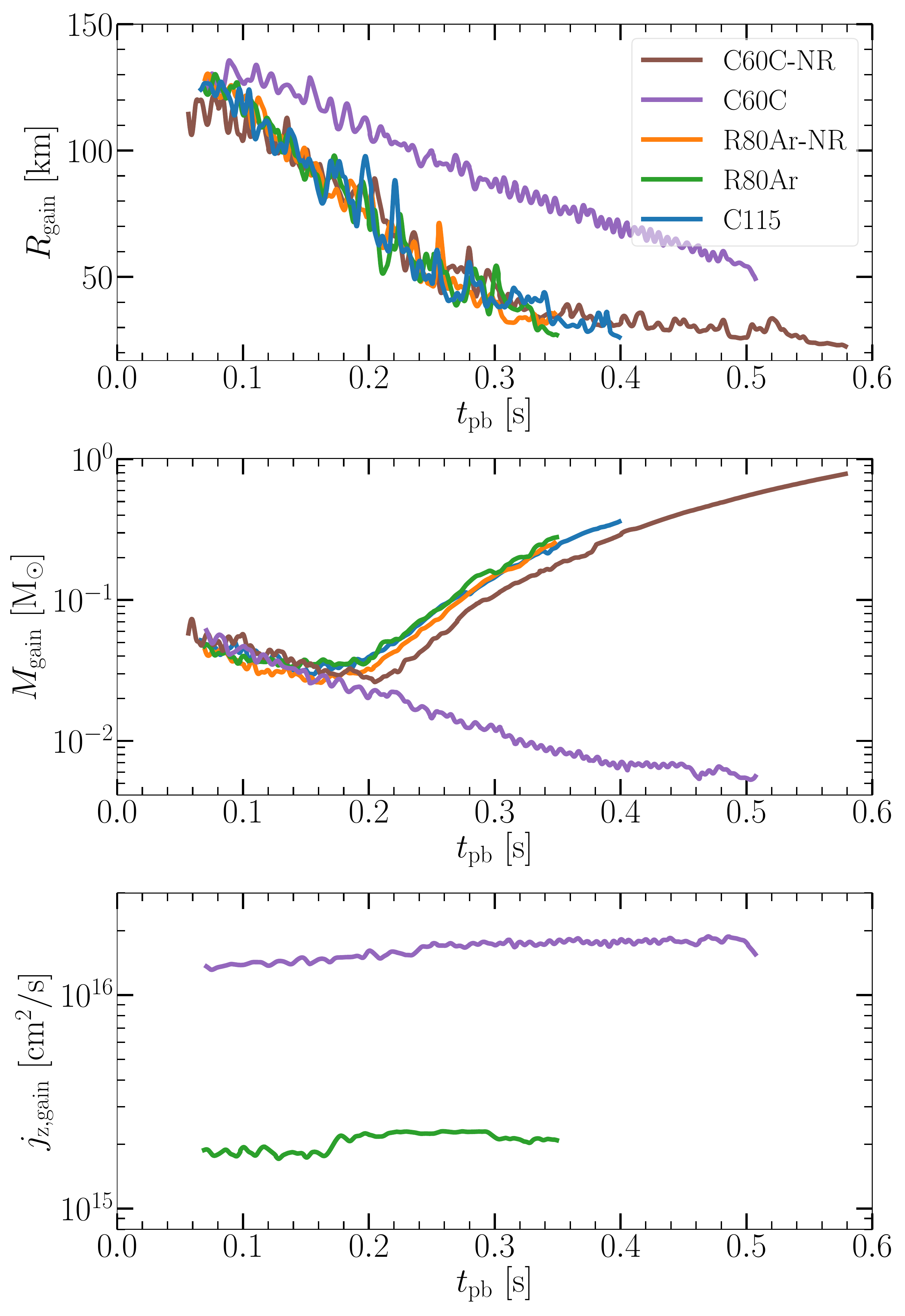}
	\vspace*{-7mm}
	\caption{Time evolution of the gain radius (top panel), mass in the gain layer (according to equation~\eqref{eq:gain_mass}; middle panel), and average specific angular momentum in the gain layer according to equation~\eqref{eq:gain_angular_momentum} (bottom panel) for our models. The line colors for the models are the same as in Fig.~\ref{fig:shock_radius}.}
	\label{fig:gain_property}
\end{figure}
\begin{figure}
	\includegraphics[width=0.48\textwidth]{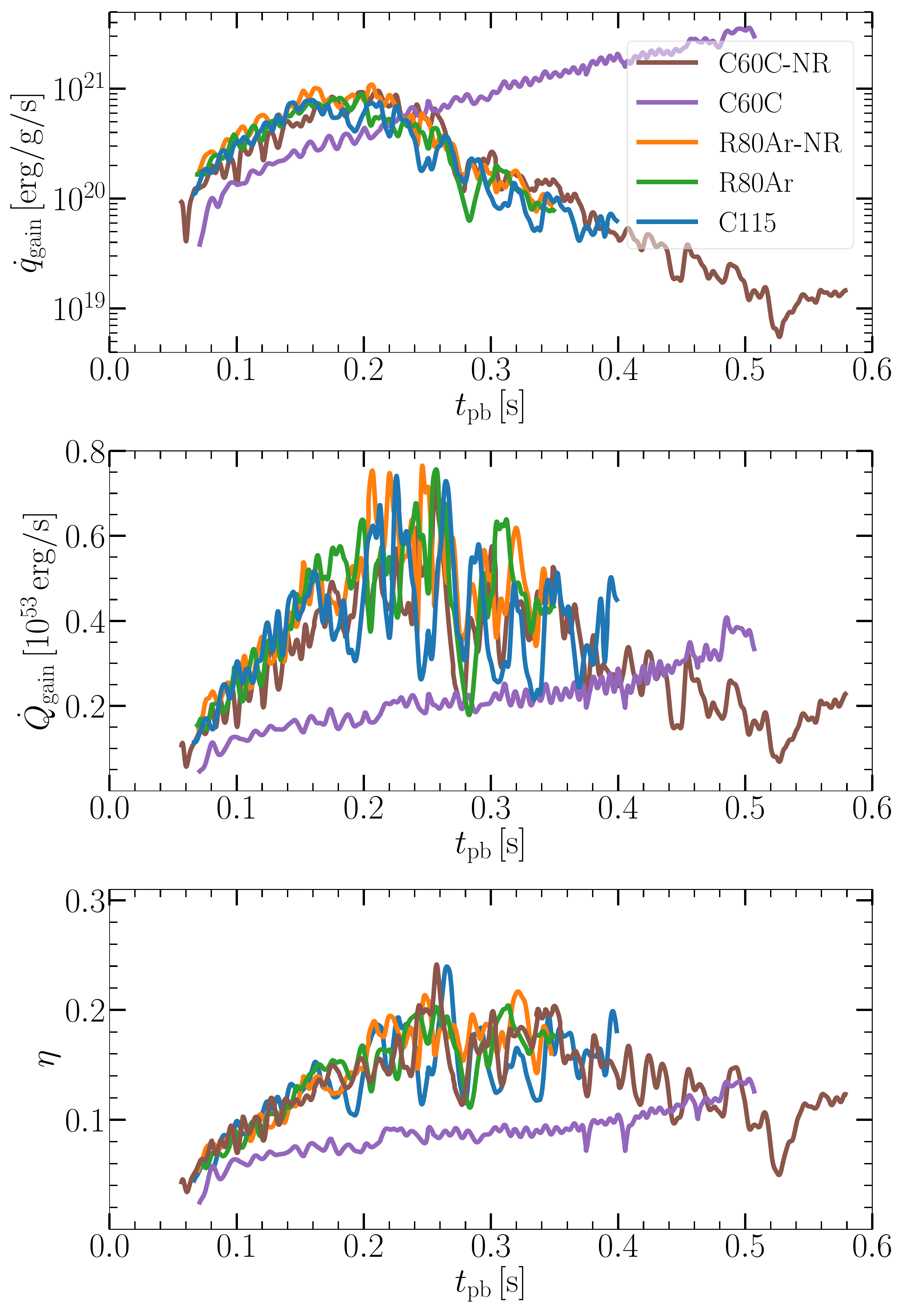}
	\vspace*{-7mm}
	\caption{Time evolution of the net neutrino heating rate per unit mass (top panel), given by equation~\eqref{eq:gain_neutrino_heating}, the total net heating rate in the gain layer (middle panel), and the heating efficiency in the gain region (bottom panel), given by equation~\eqref{eq:heat_effi} for our set of models. The models with shock revival (Models C115, R80Ar-NR, R80Ar, and C60C-NR) have higher neutrino heating efficiency compared to Model C60C without shock revival. The line colors for the different models are the same as in Fig.~\ref{fig:shock_radius}}.
	\label{fig:gain_heating}
\end{figure}
\begin{figure}
	\includegraphics[width=0.48\textwidth]{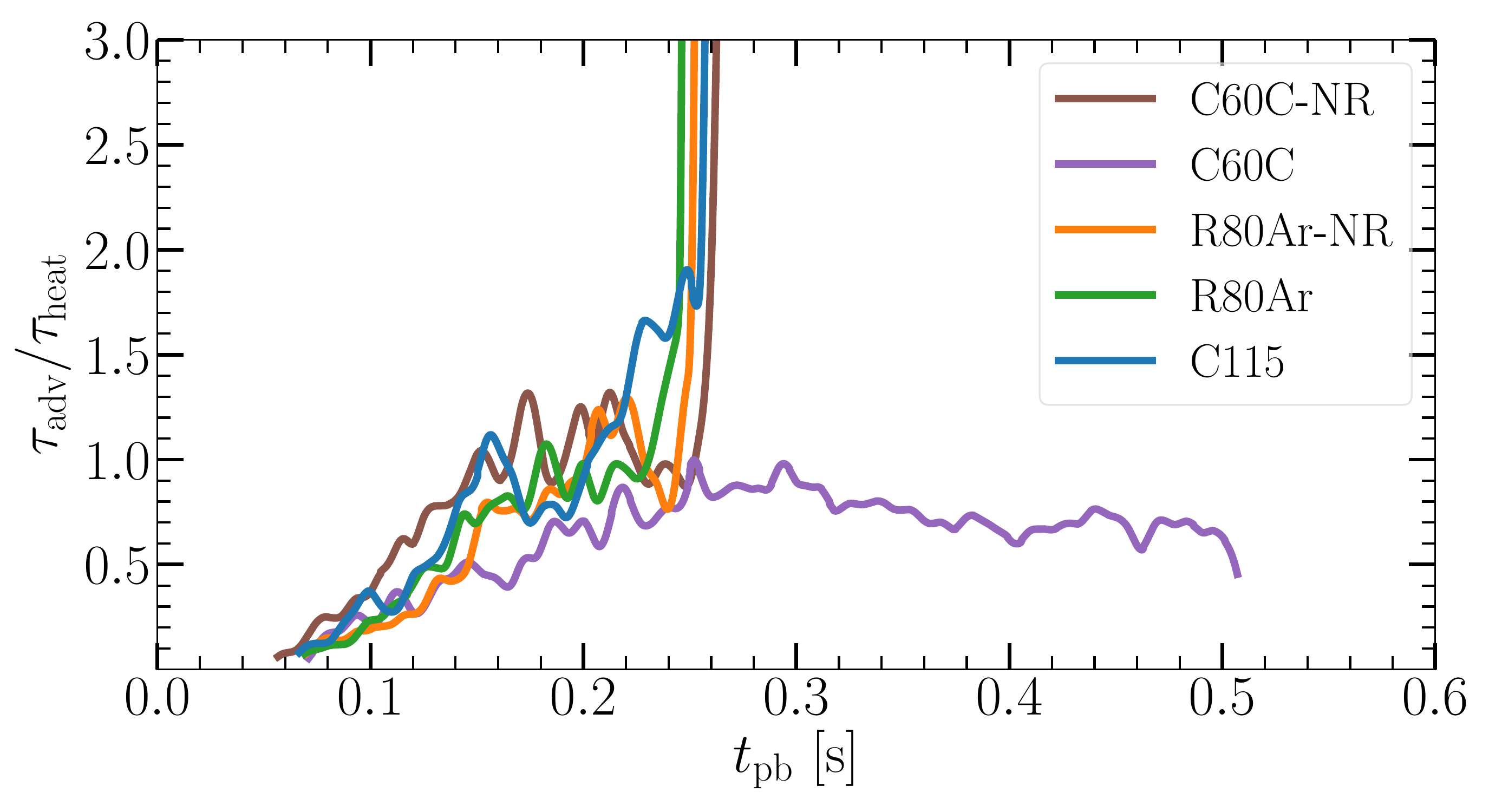}
	\vspace*{-7mm}
	\caption{Ratio of advection timescale $\tau_\mathrm{adv}$, given by equation~\eqref{eq:timescale_advection}, to heating timescale $\tau_\mathrm{heat}$, given by equation~\eqref{eq:timescale_heating}, in the gain region for our models. The ratios rise steeply at the onset of shock revival at around 220--250\,ms after bounce for Models C115 (blue line), R80Ar-NR (orange line), R80Ar (green lines), and C60C-NR (brown line).}
	\label{fig:theat}
\end{figure}
\begin{figure*}
	\includegraphics[width=\textwidth]{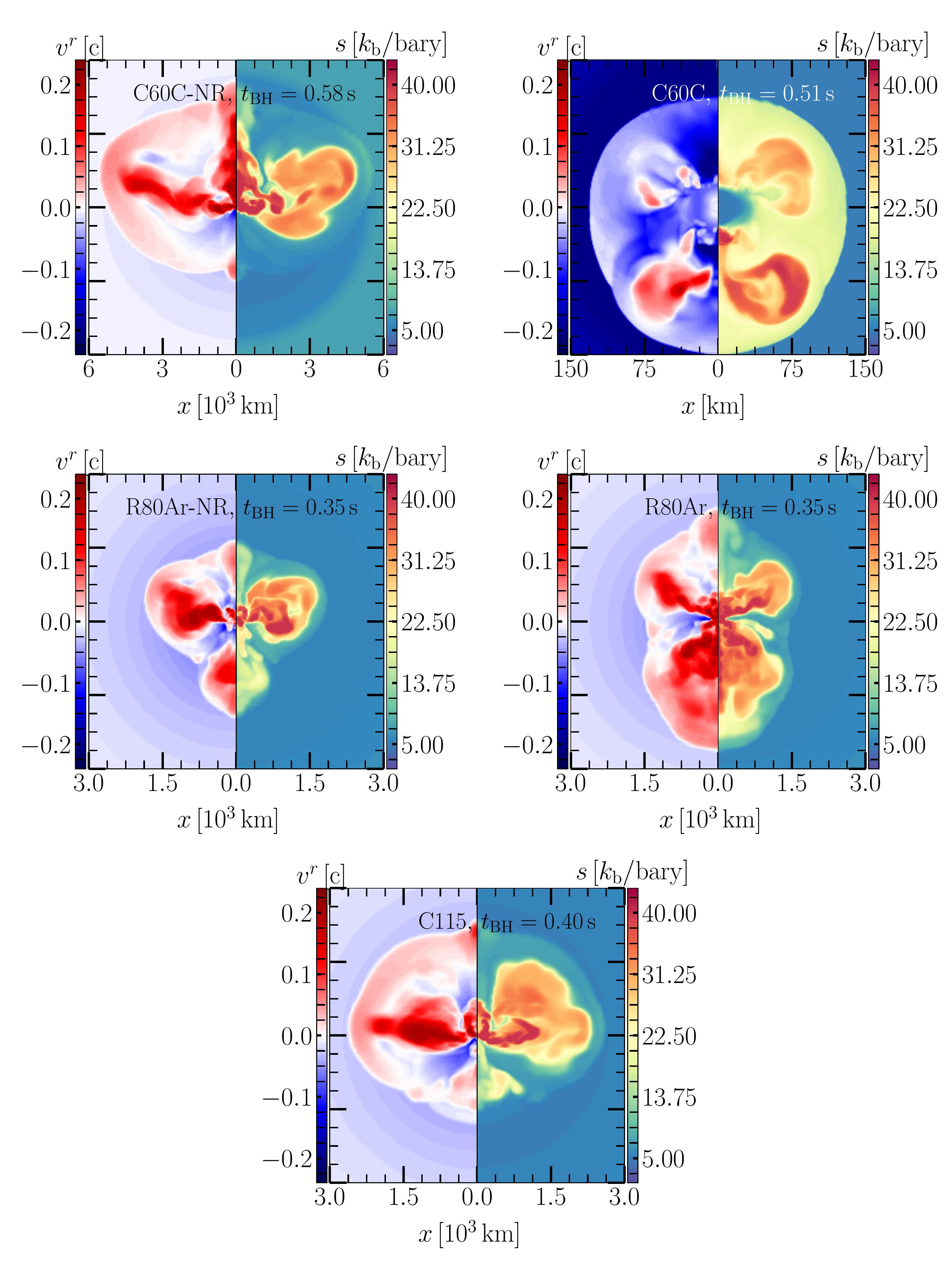}
	\vspace*{-14mm}
	\caption{Snapshots of radial velocity, $v^{r}$ (left half of panels), and gas entropy per baryon, $s$ (right half of panels), at the instants of BH formation in our models. In Model C60C-NR, the shock has expanded to the largest radius at the onset of the BH formation compared to the other models.}
	\label{fig:contour_all_model_BH}
\end{figure*}
\begin{figure}
	\includegraphics[width=0.48\textwidth]{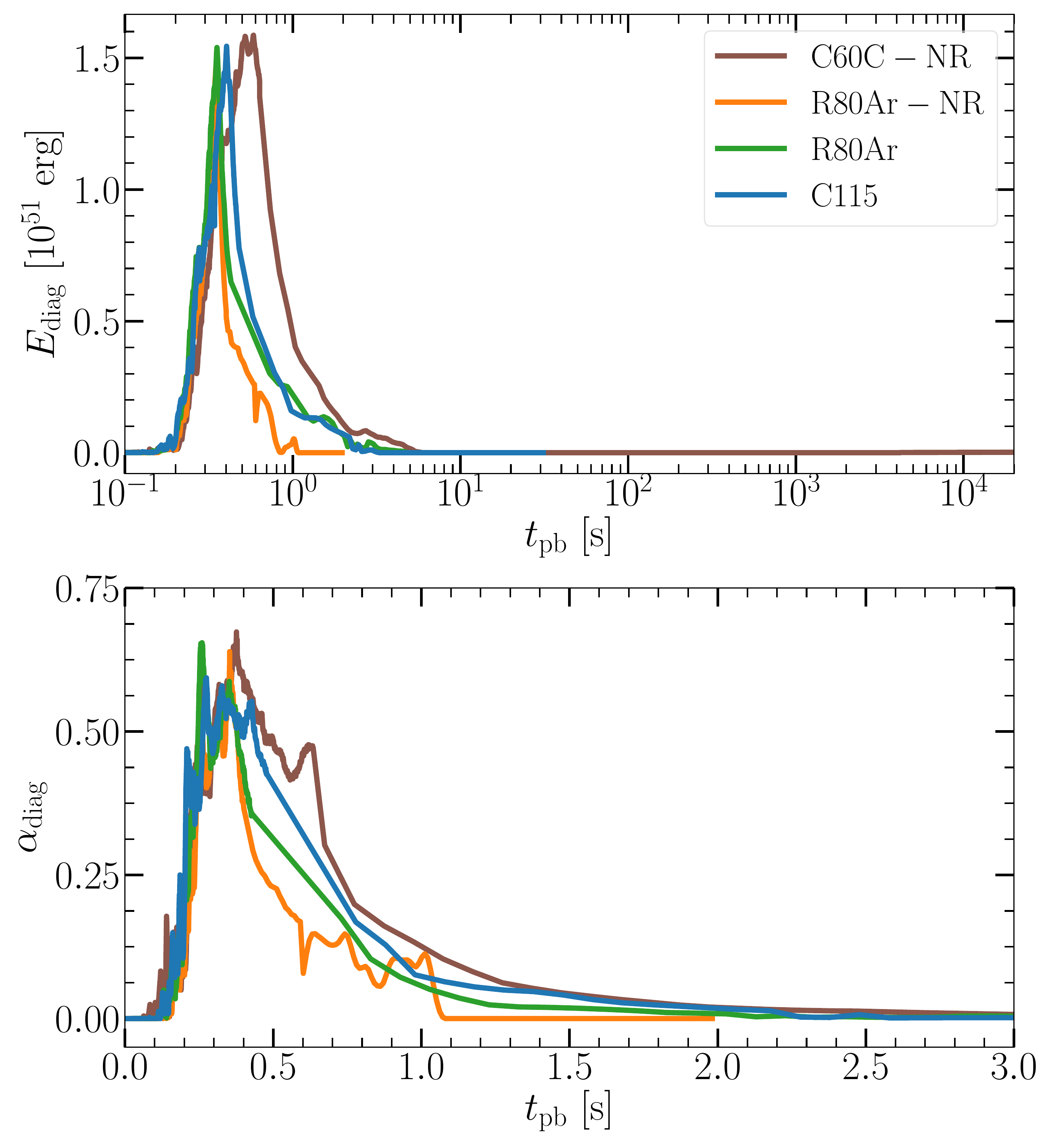}
	\vspace*{-7mm}
	\caption{Time evolution of the diagnostic energy of neutrino-heated postshock matter (equation~\eqref{eq:gain_diagnostic_energy}; top panel) and corresponding volume-filling factor defined in equation~\eqref{eq:alpha_parameter} (bottom panel) for our models. Note that both panels show largely different timescales, in the upper panel with a log scale, in the lower panel with a linear scale. The diagnostic energies decline steeply after the BH formation at 580, 350, 350, 400\,ms post bounce in Models C60C-NR (brown line), R80Ar-NR (orange line), R80Ar (green line), and C115 (blue line), respectively.}
	\label{fig:shocked_layer_properties}
\end{figure}

\section{Dynamical evolution before Black Hole Formation}\label{sec:result_before_BH_formation}

In this section, we will discuss the results of our CCSN simulations. In Fig.~\ref{fig:shock_radius}, the post-bounce time evolution of the angle-averaged shock radii in our models is shown. In all models, the shock experiences several quasi-periodic oscillations during the first tens of milliseconds after bounce, followed by a period of overall expansion. The early phases of shock contraction and expansion create negative entropy gradients and lead to prompt postshock convection. At about 100 ms post bounce, the shock expansion in all models slows down or stagnates at around 200\,km. The shock starts to expand again at about 250, 246, 237, and 222\,ms after bounce in Models C60C-NR, R80Ar-NR, R80Ar, and C115, respectively. Hereafter, we will call these models the ``shock-reviving models''. On the other hand, the stalled shock is not revived in our rapidly rotating Model C60C.

\subsection{Shock-reviving models}\label{subsec:shock-revival}

In Fig.~\ref{fig:mass_accretion}, we show the time evolution of the mass accretion rate at a radial distance of 500\,km for different models. In the early 200\,ms post-bounce, we observe that the mass accretion rates are higher in our models based on the progenitors with ZAMS masses of 80\,$\mathrm{M}_\odot$ and 115\,$\mathrm{M}_\odot$ compared to the models based on the progenitor with a ZAMS mass of 60\,$\mathrm{M}_\odot$, because the former progenitors have higher compactness values. The mass accretion rate in all of our models remains higher than\,$\sim$2\,$\mathrm{M}_\odot/\mathrm{s}$ until about 250\,ms after bounce. In Model C60C without shock revival, it exceeds\,$\sim$1.4\,$\mathrm{M_\odot/s}$ during the entire simulated evolution (see Fig.~\ref{fig:mass_accretion}). In contrast, in Models C60C-NR, R80Ar-NR, R80Ar, and C115, the mass accretion rates drop considerably after the shock expansion sets in, even before eventual BH formation, because mass infall to the compact remnant is hindered and reduced by expanding plumes of high-entropy, neutrino-heated matter. Even negative values of the (direction integrated) mass accretion rates occur temporarily, signalling the dominance of expanding mass over infalling gas at 500 km. This effect is particularly long-lasting in Model C60C-NR, which develops the highest diagnostic energy of the expanding postshock matter of all models (see Table~\ref{tab:model_property}).

During the first tens of milliseconds post bounce, the angle-averaged PNS radius in all of our simulations exhibits large-amplitude variations (Fig.~\ref{fig:NS_property}, upper panel), which correlate with the shock expansion and contraction episodes during this phase. Subsequently, the PNS contracts monotonically until BH formation. Because of the high mass accretion rates, the PNS mass in all cases grows steeply and reaches more than 2.5\,$\mathrm{M}_\odot$ within only a few 100\,ms after bounce (Fig.~\ref{fig:NS_property}, middle panel). The angular momentum associated with the accreted matter leads to a rapid spin-up of the near-surface layers in the two rotating models and a steep decline of the corresponding rotation periods (Fig.~\ref{fig:NS_property}, bottom panel). Because of its rapid rotation the PNS in Model C60C is centrifugally deformed (oblate) and therefore has a larger average radius than its counterpart in the non-rotating Model C60C-NR and the PNSs in all other models (see top panel of Fig.~\ref{fig:NS_property}).

The gain layers, i.e., the regions between gain radius and shock front, where net neutrino heating takes place, form at around 70\,ms post bounce. The top panel of Fig.~\ref{fig:gain_property} shows that the gain radius in all of our models appears at an initial distance of about 120--130 km, the middle panel of Fig.~\ref{fig:gain_property} displays the corresponding time evolution of the mass in the gain layer, and the bottom panel the average specific angular momentum in the gain layer for the two rotating Models C60C and R80Ar. In Fig.~\ref{fig:gain_heating}, we provide the net neutrino heating rate per unit mass in the gain layer, given by equation\,\eqref{eq:gain_neutrino_heating}. In the shock-reviving models it rises to peak values of up to $10^{21}$\,erg/g/s within the first 150\,ms after bounce and remains around such high values until shock expansion sets in between 220\,ms and 250\,ms after bounce. The total heating rate in the gain layer, $\dot Q^{+}_\mathrm{gain}$, scales with the relevant parameters according to (see, e.g., \citealt{2012ApJ...756...84M}):
\begin{eqnarray}\label{eq:qheat_rel}
    \dot Q^{+}_\mathrm{gain} \propto  \frac{L_{\nu} \langle \epsilon_{\nu}^2 \rangle M_\mathrm{gain}}{R_\mathrm{gain}^2},
\end{eqnarray}
where $L_{\nu}$ and $\langle \epsilon_{\nu}^2 \rangle$ are the luminosity and the mean squared energy of electron neutrinos and antineutrinos, respectively, and $R_\mathrm{gain}$ and $M_\mathrm{gain}$ are the gain radius and the mass in the gain layer, respectively. Since the progenitors have massive iron cores (see Table~\ref{tab:progenitor_property}) and the shocks stagnate at large radii of $\sim$200\,km, the gain layers in all models contain large masses of around $(3-4)\times 10^{-2}$\,$\mathrm{M}_\odot$ (see the middle panel of Fig.\,\ref{fig:gain_property}) before 200\,ms post bounce. Despite declining $M_\mathrm{gain}$ until about 200\,ms after bounce, the shrinking gain radius (top panel of Fig.~\ref{fig:gain_property}) in combination with growing neutrino luminosities and mean squared energies (see Section~\ref{sec:neutrino_properties}) leads to a steep rise of the net heating rate in the gain layer for all shock-reviving models, in agreement with equation~\eqref{eq:qheat_rel}. The specific net neutrino heating rate, $\dot q_\mathrm{gain} = \dot Q_\mathrm{gain}/M_\mathrm{gain}$ (Fig.~\ref{fig:gain_heating}, top panel), exhibits a correspondingly steeper increase by more than an order of magnitude. A considerably larger gain radius (Fig.~\ref{fig:gain_property}) as well as lower neutrino luminosities and lower mean squared energies (see Section~\ref{sec:neutrino_properties}) in the rotating Model C60C result in weaker gain-layer heating in this case, for which reason shock revival does not occur in Model C60C.

In the bottom panel of Fig.~\ref{fig:gain_heating}, the heating efficiency in the gain layer, defined according to equation~\eqref{eq:heat_effi}, is shown for our models. We observe a heating efficiency as high as 10--20\% in the shock reviving models before the onset of shock expansion. Due to the high values of the net neutrino heating rate, the heating timescale in the gain layer, $\tau_\mathrm{heat}$, evaluated by equation\,\eqref{eq:timescale_heating}, eventually becomes shorter than the advection timescale of matter passing through the gain layer, $\tau_\mathrm{adv}$, estimated by equation\,\eqref{eq:timescale_advection}, in all shock reviving models (Fig.~\ref{fig:theat}). As a result, the shocks start to expand and the mean shock radii reach 400\,km in between post-bounce times of 222 and 250\,ms in the shock reviving models (see Table~\ref{tab:model_property}).

In Models C60C-NR and C115 the interface between iron core and Si shell (located at 2.37\,$\mathrm{M}_\odot$ and 2.46\,$\mathrm{M}_\odot$; see Table~\ref{tab:progenitor_property}) falls through the shock before the shock expansion sets in (see Fig.~\ref{fig:shock_radius}). In contrast, in Models R80Ar-NR and R80Ar the corresponding interface (at a mass shell of 2.72\,$\mathrm{M}_\odot$) crosses the shock only after the shock revival (see Fig.~\ref{fig:shock_radius}).

Since recombination of free nucleons to $\alpha$-particles sets in when the shock reaches $\sim$250\,km, the additional release of energy accelerates the shock to a velocity of about 15000\,km/s within only 50\,ms after the onset of shock expansion in all cases with shock revival. Before the onset of shock expansion, we witness violent bipolar shock oscillations, which are a typical 2D phenomenon favored by the constraining assumption of axisymmetry. The large-amplitude nonspherical mass and shock motions can be a consequence either of strong postshock convection or of the standing accretion shock instability (SASI; \citealt{2003ApJ...584..971B}; \citealt{2007ApJ...654.1006F}). A highly time-dependent pattern of accretion downflows and buoyant plumes of neutrino-heated matter fills the volume between the PNS and the shock. Correlated with the shock oscillations, bubbles form during shock expansion, disappear when the shock contracts, and new bubbles form again during the next expansion phase, continuously rearranging the pattern of downflows and high-entropy plumes. The geometry that establishes when shock expansion sets is quite stochastic with different morphologies (see Fig.~\ref{fig:contour_all_model_BH}).

The shock expansion begins highly asymmetrically with accretion of postshock matter continuing through several large downdrafts. Simultaneously, neutrino-heated matter rises in prominent plumes and pushes the shock outward. In Models C60C-NR, R80Ar-NR, and C115 the strongest plumes are near the equatorial plane. More narrow outflows can also be present at both poles while accretion happens off-axis (Model R80Ar-NR), or there can be one polar accretion flow and an outflow on the opposite side (Model C60C-NR), or the polar downflows and outflows can even exhibit time-variable behavior (Model C115).  All of these three models show an oblate deformation of the expanding shock. In contrast, in the slowly rotating Model R80Ar the shock revival occurs with a large prolate asymmetry because of accretion downdrafts close to the equatorial plane and outflows in the polar directions (Fig.~\ref{fig:contour_all_model_BH}). In the rapidly rotating, nonexploding Model C60C we witness a butterfly-like pattern that is characterized by polar and equatorial downflows separated by plumes at intermediate latitudes with $\sim$45$^\circ$ inclination.

In all shock reviving models the mass accretion rate, measured at 500\,km, drops after the shock expansion sets in, but accretion (now behind the expanding shock) continues at a lower rate, as we can see in Fig.\,\ref{fig:mass_accretion}. As a result, the neutrino luminosities and therefore the net neutrino heating rate of the postshock matter drop after shock revival. The trend is amplified in the net heating rate per unit mass (Fig.\,\ref{fig:gain_heating}) because of the rapidly growing mass in the postshock layer (Fig.\,\ref{fig:gain_property}, middle panel). On the contrary, the net neutrino heating rate per unit mass (Fig.\,\ref{fig:gain_heating}), as well as the neutrino luminosities and mean energies (see Section\,\ref{sec:neutrino_properties}) grow with time in the non-exploding Model C60C because of the continuously higher mass accretion rate compared to the models with shock expansion.

The maximum gravitational mass supported by the SFHo EOS at zero temperature for nonrotating neutron stars is around 2.059 $\mathrm{\mathrm{M}_\odot}$ \citep{2013ApJ...774...17S}. Thermal pressure in the hot PNS and the centrifugal support due to rapid differential rotation can increase the mass limit above this value. Because of continued accretion, the baryonic PNS mass increases monotonically in all models (Fig.\,\ref{fig:NS_property}) and eventually a BH forms in all models. The final baryonic mass of the PNS, $M_\mathrm{f,ns}$, and the time of BH formation, $t_\mathrm{BH}$, are listed in Table~\ref{tab:model_property}. The slow rotation of Model R80Ar does not affect the BH formation time and the final mass of the PNS, since Model R80Ar and its non-rotating counterpart, Model R80Ar-NR, have the same values of these parameters. However, $M_\mathrm{f,ns}$ is higher by nearly 0.3\,$\mathrm{M}_\odot$ in the rapidly rotating Model C60C compared to its non-rotating counterpart C60C-NR, as Model C60C gets additional centrifugal support against gravity. The final rotation period of the PNS in Model C60C, assuming rigid rotation (equation\,\eqref{eq:ns_rotational_period}), is 1.17\,ms, corresponding to an initial Kerr parameter of 0.725 at the time of BH formation, whereas the corresponding values for Model R80Ar are 7.90\,ms and 0.1, respectively (Table\,\ref{tab:model_property}).

The top panel of Fig.\,\ref{fig:shocked_layer_properties} displays the diagnostic (explosion) energy of neutrino-heated postshock material (equation\,\eqref{eq:gain_diagnostic_energy}) for all models with shock revival. It increases steadily right after the onset of shock expansion and reaches about $1.58\times 10^{51}$\,erg in Model C60C-NR at the time of BH formation ($t_\mathrm{BH}$), whereas the overburden energy (equation\,\eqref{eq:gain_overburden_energy}) is around $-4.89\times 10^{51}$\,erg at the same time. Similarly, the magnitude of the overburden energy at the time of BH formation is still considerably greater than the diagnostic energy in all other models with shock revival. The corresponding values are $1.43\times 10^{51}$\,erg, $1.53\times 10^{51}$\,erg, and $1.54\times 10^{51}$\,erg for the diagnostic energy and $-1.06\times 10^{52}$\,erg, $-1.04\times 10^{52}$\,erg, and $-7.25\times 10^{51}$\,erg for the overburden energy in Models R80Ar-NR, R80Ar, and C115, respectively (Table~\ref{tab:model_property}).

The diagnostic energies reach maxima at $t_\mathrm{BH}$ and start to decline after BH formation because the neutrino heating plummets at this moment, neutrino-heated matter begins to fall back to be swallowed by the BH, and the expanding shock sweeps up gravitationally bound matter from the overlying star. Therefore, it is necessary to continue the simulations after shock revival beyond the time of BH formation to see whether the diagnostic energy of the neutrino-heated material is sufficient to unbind some of the gravitationally bound matter with its negative overburden energy ahead of the shock (see Section\,\ref{sec:result_after_BH_formation}). The onset of shock expansion (``shock revival'') is therefore no conclusive criterion for a successful explosion and ultimate shock breakout from the stellar surface, see, e.g., \citet{2013ApJ...767L...6B}; \citet{2018ApJ...852L..19C}; \citet{2020MNRAS.495.3751C}, for a detailed discussion of the relevance of the overburden energy and the dynamics of shock revival in BH forming stellar collapse events.

\begin{figure}
	\includegraphics[width=0.48\textwidth]{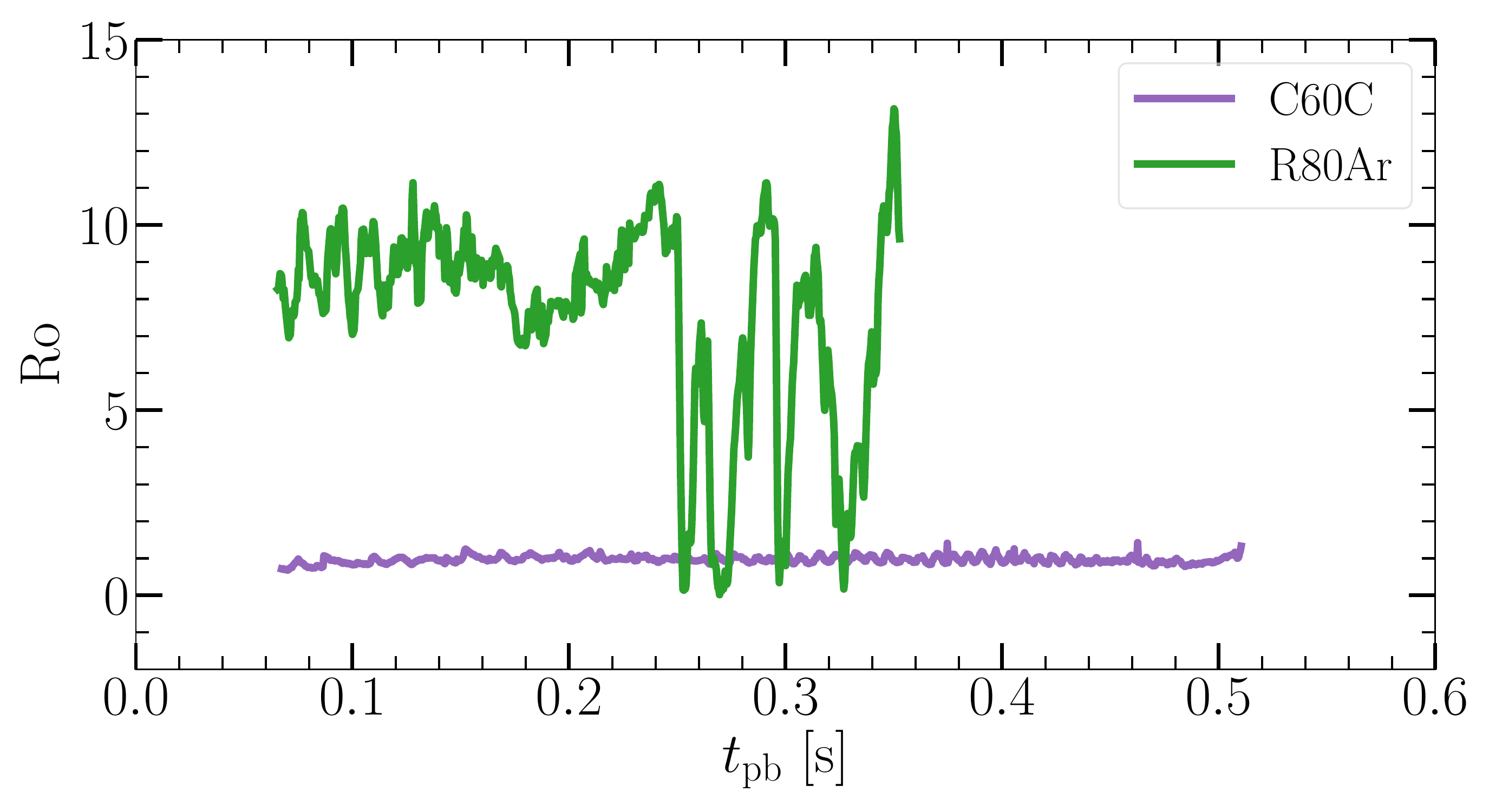}
	\vspace*{-7mm}
	\caption{Time evolution of the Rossby number, given by equation~\eqref{eq:rossby_number}, in the gain layer of our two rotating models. The Rossby number for the slowly rotating Model R80Ar-NR (green line) remains around 10 until 250\,ms post bounce, i.e., before the moment of shock revival. In contrast, in the rapidly rotating Model C60C (violet line) its value is around one.}	
	\label{fig:gain_rossby}
\end{figure}
\begin{figure}
	\includegraphics[width=0.48\textwidth]{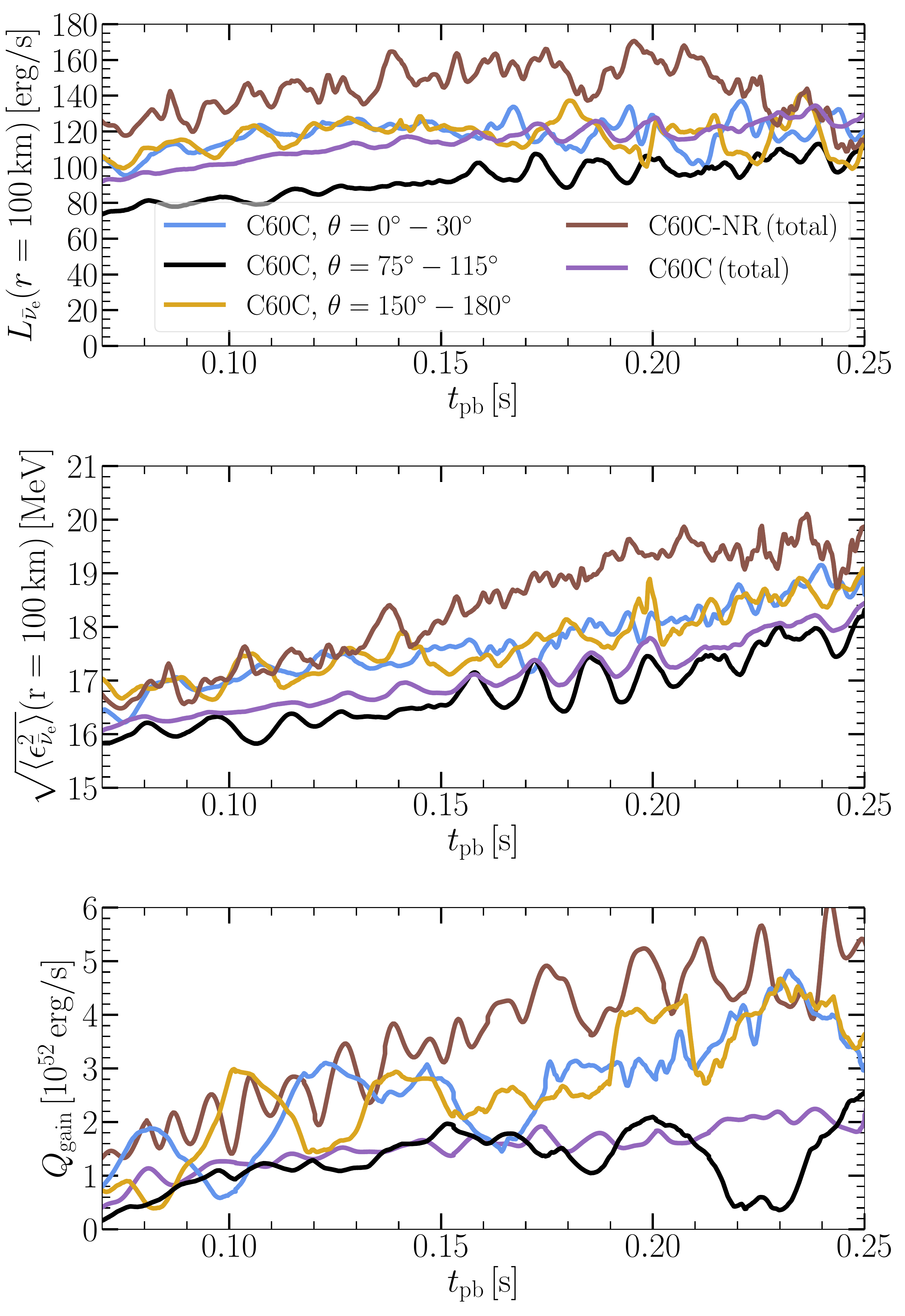}
	\vspace*{-7mm}
	\caption{Time evolution of the electron antineutrino luminosities (top panel) and RMS energies (middle panel) at a radius of 100\,$\mathrm{km}$ and of the net neutrino heating rates in the gain layer (bottom panel). The isotropic-equivalent values of these neutrino properties, evaluated in an equatorial wedge ($75^{\circ}\le \theta \le 115^{\circ}$), a north-polar wedge ($0^{\circ} \le \theta \le 30^{\circ}$), and a south-polar wedge ($150^{\circ}\le \theta \le 180^{\circ}$), are shown by black, blue, and yellow lines, respectively, for the fast rotating Model C60C. Additionally, electron antineutrino luminosities, RMS energies, and net neutrino heating rates in the gain layer, integrated over all directions, are displayed for Model C60C (violet lines) and its non-rotating counterpart, Model C60C-NR (brown lines).}
	\label{fig:c60c_neutrino_pole_eqt}
\end{figure}

\subsection{Impact of rotation}\label{subsec:impact_of_rotation}

In this section, we discuss the impact of rotation on the pre-BH-formation dynamics. The Rossby number, $\mathrm{Ro}$, evaluated using equation\,\eqref{eq:rossby_number}, is around 7--11 in the gain layer of the rotating Model R80Ar before shock revival (see Fig.~\ref{fig:gain_rossby}). As a consequence, we do not see any strong impact of rotation on the post-bounce dynamics of this model, since $\mathrm{Ro}$\,$\gg$\,1 means that the rotation period is much longer than the timescale of convective turnover in the gain layer. Thus the rotating Model R80Ar and the non-rotating Model R80Ar-NR have similar time evolution. Moreover, the rotation in Model R80Ar is not strong enough to cause large deformation of the PNS.

In contrast, the rapidly rotating Model C60C has a smaller Rossby number in the gain layer, $\mathrm{Ro}$\,$\sim$\,1 (see Fig.~\ref{fig:gain_rossby}). As a result, we notice significant differences between Model C60C and its non-rotating counterpart, Model C60C-NR, since Model C60C-NR experiences shock revival and, on the contrary, Model C60C does not. In Model C60C, the mean PNS radius (see Fig.~\ref{fig:NS_property}) and the mean gain radius (see Fig.~\ref{fig:gain_property}) are larger compared to Model C60C-NR, because the rapidly rotating model is stabilized and deformed to an oblate shape by centrifugal forces. Moreover, as we will see im Section~\ref{sec:neutrino_properties}, the nonrotating case has higher total neutrino luminosities and RMS energies before the onset of shock revival (at about 250\,ms post bounce) in Model C60C-NR. However, the mass accretion rates and the PNS masses remain similar in these two models before a post-bounce time of 250\,ms (Figs.\,\ref{fig:mass_accretion} and \ref{fig:NS_property}). As a consequence, Model C60C-NR has a larger net neutrino heating rate (by absolute value as well as per unit of mass) in the gain layer (see Fig.~\ref{fig:gain_heating} and also equation~\eqref{eq:qheat_rel}) than its rotating counterpart, Model C60C, at $0<t_\mathrm{pb}<250$\,ms. This fosters the expansion of the stagnating shock in Model C60C-NR. In summary, rotation in Model C60C quenches the chance of shock revival.

In the rapidly rotating case of C60C, the PNS has an oblate shape (the quadruple component of the PNS radius is around 10--20\% of its mean value in the time interval of $0<t_\mathrm{pb}<250$\,ms): the PNS has a smaller radius at the poles than at the equator. Consequently, the neutrinospheric layer at the poles is hotter compared to the neutrinosphere near the equator. For this reason the neutrino energy fluxes and RMS energies at the poles are higher than those near the equatorial plane and thus the net neutrino heating rate in the gain layer near the poles is also higher as shown in Fig~\ref{fig:c60c_neutrino_pole_eqt}. The top panel of Fig~\ref{fig:c60c_neutrino_pole_eqt} depicts isotropic-equivalent values of the electron antineutrino luminosities for different angular wedges in Model C60C. Specifically, the $\bar\nu_e$ energy fluxes have been angle-averaged over the wedges of $0^{\circ}\le \theta \le 30^{\circ}$ (north pole, blue line), $75^{\circ} \le \theta \le 115^{\circ}$ (equator, black line), and $50^{\circ}\le \theta\le 180^{\circ}$ (south pole, yellow line), evaluated at r = 100 km for an observer at rest at infinity and multiplied by a factor 4$\pi$ according to the following equation:
\begin{eqnarray}\label{eq:area_average_luminosity}
L(r, \theta_\mathrm{i}, \theta_\mathrm{f}) \equiv 4 \pi \alpha \eul^{4\phi} r^2 \frac{\int_{\theta_\mathrm{i}}^{\theta_\mathrm{f}} \mathrm{d}\theta \sin(\theta) \mathcal{H}_\nu(r,\theta)}{\int_{\theta_\mathrm{i}}^{\theta_\mathrm{f}} \mathrm{d}\theta \sin(\theta)}\,,
\end{eqnarray}
where $\mathcal{H}_\nu(r,\theta)$ is the energy-integrated neutrino energy flux in the local rest frame and $\theta_\mathrm{i}$ and $\theta_\mathrm{f}$ are the bracketing angles of the angular wedge. Additionally, the top panel of Fig~\ref{fig:c60c_neutrino_pole_eqt} also displays the total (i.e., for all directions) electron antineutrino luminosities for Models C60C (violet line) and C60C-NR (brown line) at $r$=100\,km. Likewise, the middle panel of Fig~\ref{fig:c60c_neutrino_pole_eqt} displays the time evolution of angle-averaged RMS energies of electron antineutrinos in the polar and equatorial wedges of Model C60C along with the values of the RMS energies angle averaged over the whole surface area at a radius of 100\,km for the rapidly rotating Model C60C as well as its nonrotating counterpart. Similarly, the bottom panel shows the isotropic-equivalent values of the net neutrino-heating rates in the mentioned angular wedges (i.e., the net heating rates per unit mass multiplied by the total mass of the gain layer) as well as the integrated values for the entire gain layer. The angular dependence of the neutrino properties observed in the rapidly rotating Model C60C complies with the findings of \citet{2009ApJ...694..664M}. These authors conducted a CCSN simulation (their Model M15LS-rot) of a rotating 15\,$\mathrm{M}_\odot$ progenitor from \citet{2005ApJ...626..350H} and they also observed higher neutrino luminosities and energies in the polar regions than near the equatorial plane (see their Fig. 15).

Infalling matter near the poles of the fast-rotating Model C60C has relatively little angular momentum. For this reason one might suspect that in Model C60C the isotropic-equivalent values of the net neutrino-heating rates in the polar wedges of the gain layer are similar to the net neutrino-heating rate in the entire gain layer of its non-rotating counterpart, Model C60C-NR. Therefore, one might expect that the former model could experience shock revival at the poles. Indeed, the PNS radii at the poles of Models C60C and C60C-NR have similar values at a given post-bounce time. However, no polar shock revival is observed in Model C60C, because the neutrino luminosities and the RMS energies in Model C60C are lower than those of Model C60C-NR also around the poles (see Fig~\ref{fig:c60c_neutrino_pole_eqt}). This result can be understood by the fact that matter with higher angular momentum, while falling towards the polar caps of the PNS, experiences strong centrifugal forces and gets deflected towards the equatorial region. Thus, the rapid rotation in Model C60C prevents the accretion of  matter onto the polar caps of the PNS and the associated strong compressional heating. As a result, the neutrinospheric layer near the poles in Model C60C is cooler than in Model C60C-NR, for which reason the polar neutrino luminosities and RMS energies are correspondingly lower. These lower luminosities and RMS energies imply less efficient neutrino heating of the matter in the gain layer around the poles and consequently successful shock expansion is not witnessed in the polar regions of Model C60C.

In line with this finding, \citet{2018ApJ...852...28S} reported a successful explosion of a non-rotating model (their Model m15\_2D\_norot\_1.4deg) of a 15\,$\mathrm{M}_\odot$ progenitor from \citet{2005ApJ...626..350H}, whereas its fast-rotating counterpart (Model m15\_2D\_artrot\_1.4deg) failed to explode on a timescale of $\sim$450\,ms after bounce. This agreement is plausible since our Model C60C and their Model m15\_2D\_artrot\_1.4deg have similar values of the specific angular momentum of some $10^{16}$\,$\mathrm{cm}^2/\mathrm{s}$ near the equatorial plane in the region between radii of 1000\,km and 10000\,km at the onset of stellar core collapse. However, in 3D the fast-rotating Model m15\_3D\_artrot\_2deg of \citet{2018ApJ...852...28S} experiences shock revival due to strong spiral SASI activity that pushes the shock outward in the equatorial direction. Therefore, the absence of shock revival in Model C60C might be an artifact of our dimensional constraint to 2D simulations. Hence, 3D simulations will be needed to reliably assess the possibility of shock revival in the rapidly rotating case of Model C60C.

After the shock revival in our Model C60C-NR, the mass accretion rate drops to a smaller value, whereas the mass accretion continues at a high rate in Model C60C (Fig.\,\ref{fig:mass_accretion}). Therefore the BH formation occurs about 70\,ms earlier in Model C60C than in Model C60C-NR. Interestingly, the mass of the PNS at the time it collapses to a BH is bigger by\,$\sim$0.26\,$\mathrm{M}_\odot$ in Model C60C (Table\,\ref{tab:model_property} and Fig.\,\ref{fig:NS_property}, middle panel) because of its rotational support.

In summary, Model R80Ar behaves similarly to its non-rotating counterpart, Model R80Ar-NR, because its slow rotation has little impact on the post-bounce dynamics. On the contrary, the fast rotation in Model C60C has a significant influence on the post-bounce dynamics and suppresses the shock revival.

\subsection{Comparison with previous works}\label{subsec:comparison_with_previous_works}

Shock expansion in collapsing progenitors with high ZAMS masses concomitant with BH formation was also witnessed by \citet{2018ApJ...852L..19C,2018MNRAS.477L..80K, 2018ApJ...857...13P,2018ApJ...852...28S,2018ApJ...855L...3O,2020MNRAS.496.2000C,2021ApJ...914..140P,2020MNRAS.491.2715B}, and \citet{2021MNRAS.503.2108P}. \citet{2021MNRAS.503.2108P} presented 3D CCSN simulations of very massive, metal-free Pop-III progenitor stars with ZAMS masses of 85\,$\mathrm{M}_\odot$ and 100\,$\mathrm{M}_\odot$. The progenitor models had been evolved until the onset of the core collapse using the stellar evolution code KEPLER \citep{1978ApJ...225.1021W,2002ApJ...576..323R}. They employed the nuclear EOS of \citet{1991NuPhA.535..331L} with a bulk incompressibility modulus of $K$=220\,MeV (LS220) as well as the SFHo and SFHx EOSs \citep{2013ApJ...774...17S} for their CCSN simulations. \citet{2018ApJ...852L..19C} studied the core collapse of a metal-free progenitor with a ZAMS mass of 40\,$\mathrm{M}_\odot$ from \citet{2010ApJ...724..341H}. \citet{2018MNRAS.477L..80K} considered a zero-metallicity 70\,$\mathrm{M}_\odot$ progenitor from \citet{2014ApJ...794...40T} and a solar-metallicity 40\,$\mathrm{M}_\odot$ progenitor from \citet{2007PhR...442..269W} in three dimensions with the LS220 EOS. \citet{2018ApJ...857...13P} conducted CCSN simulations in axisymmetry for the same 40\,$\mathrm{M}_\odot$ progenitor used by \citet{2018MNRAS.477L..80K}, however employing different nuclear EOSs, namely LS220 (the same EOS as used by \citealt{2018ApJ...852L..19C,2021MNRAS.503.2108P} and \citealt{2018MNRAS.477L..80K}), SFHo (this EOS is used in our study as well as by \citealt{2021MNRAS.503.2108P}), DD2 \citep{2014EPJA...50...46F}, and BHB$\Lambda\phi$ \citep{2014ApJS..214...22B}.

\citet{2018ApJ...852L..19C} fostered shock revival in their 40\,$\mathrm{M}_\odot$ simulation by artificially increasing the strangeness contribution to the axial-vector coupling for neutral-current neutrino-nucleon scattering, using a coupling constant of $g_\mathrm{A,s} = -0.2$, which is on the extreme side compared to experimental and theoretical constraints (see \citealt{2016PhRvC..93e2801H}). At the time of BH formation the shock location was at about 4000\,km and the expanding postshock matter had reached a diagnostic energy of 2.09\,$\times 10^{51} \mathrm{erg}$, which barely equaled the binding energy of the matter ahead of the shock with a value of 2.1\,$\times 10^{51} \mathrm{erg}$. \citet{2018ApJ...852L..19C} continued their simulation beyond the BH formation (replacing the BH by an outflow boundary), and despite the odds mentioned in Section~\ref{subsec:shock-revival} they obtained shock breakout from the surface of the star. 

\citet{2018MNRAS.477L..80K} conducted their simulations until the birth of the BH. For their metal-free 70\,$\mathrm{M}_\odot$ model they observed shock expansion before the BH formation, at which time the maximum shock radius was around 380\,km. In contrast, they did not obtain any shock revival in the solar-metallicity 40\,$\mathrm{M}_\odot$ model, in agreement with the 3D results referenced by \citet{2018ApJ...852...28S} for the same progenitor.

\citet{2018ApJ...857...13P} found shock revival also for this 40\,$\mathrm{M}_\odot$ progenitor, however in a 2D simulation with the DD2 EOS and quite late at 1.27\,s after core bounce. They reported continued shock expansion until the end of their simulation at 1.3\,s after bounce, at which time the shock had reached an angle-averaged radius of 1000\,km. No BH had formed until this moment. However, since the mass of the PNS exceeded the maximum mass of a NS at zero temperature supported by the DD2 EOS, the authors expected a BH to form when the PNS cools down. \citet{2018ApJ...857...13P} obtained shock expansion in their 2D simulation with the LS220 EOS, too, where the shock was revived just before the PNS collapsed to a BH. This result is in agreement with a 40\,$\mathrm{M}_\odot$ 2D calculation mentioned by \citet{2018ApJ...852...28S}, but it is in contrast to the 3D simulations for the 40\,$\mathrm{M}_\odot$ progenitor by \citet{2018MNRAS.477L..80K} and \citet{2018ApJ...852...28S}. Since \citet{2018ApJ...857...13P} did not witness shock revival with the SFHo and BHB$\Lambda\phi$ EOSs, they concluded that both this effect and the time of BH formation depend sensitively on the nuclear EOS. In addition, it seems that the question of shock revival or not can also depend on the dimension of the simulation.

In \citet{2021MNRAS.503.2108P}, the 85\,$\mathrm{M}_\odot$ simulations exhibited shock revival with all employed EOSs. In contrast, no shock expansion was obtained for the 100\,$\mathrm{M}_\odot$ progenitor in their CCSN calculations. They stopped all of their simulations at the onset of BH formation. At this instant, the 85\,$\mathrm{M}_\odot$ models had shock radii of 4451\,km, 2103\,km, and 1504\,km and diagnostic energies of $2.7\times 10^{51}$\,erg, $1.25\times 10^{51}$\,erg, and $0.7\times 10^{51}$\,erg for the SFHx, SFHo, and LS220 EOS, respectively. The shock radius and overburden energy in their model with SFHx are comparable with our values for Model C60C-NR. However, the diagnostic energy in this model of \citet{2021MNRAS.503.2108P} is greater than in our Model C60C-NR at the onset of the BH formation. This implies that the simulation with the SFHx EOS presented by \citet{2021MNRAS.503.2108P} is more likely to lead to mass ejection in connection to BH formation than any of our models.

\begin{table*}
	\centering
    \caption{Characteristic properties of our simulated models at the time of mapping from the \textsc{NADA-FLD} code to the \textsc{Prometheus} code and energy loss in neutrinos during
    the \textsc{NADA-FLD} simulations.}
	\begin{tabular*}{0.85\textwidth}{lccccccccc}
		\hline \vspace*{-3mm} \\
		Model & $t_\mathrm{BH}$ & $t_\mathrm{map}$ & $M_\mathrm{BH}(t_\mathrm{map})$ & $R_\mathrm{sh}(t_\mathrm{map})$ & $E_\mathrm{diag}(t_\mathrm{map})$ & $E_\mathrm{ob}(t_\mathrm{map})$ & $\Delta E_\nu(t_\mathrm{BH})$ & $t_\mathrm{end}^\mathrm{NADA}$ & $\Delta E_\nu(t_\mathrm{end}^\mathrm{NADA})$ \\
		\vspace*{-3mm} \\
        & $[\mathrm{s}]$ & $[\mathrm{s}]$ & [$\mathrm{M}_\odot$] & [$\mathrm{km}$] & [$10^{51}~\mathrm{erg}$] & [$10^{51}~\mathrm{erg}$] & [$\mathrm{M}_\odot c^2$] & $[\mathrm{s}]$ & [$\mathrm{M}_\odot c^2$] \\
        \hline
        C60C-NR  & 0.580 & 0.635 & 2.61 & 5392 & 1.41 & $-$4.60 & 0.128 & 1.06 & 0.130 \\
        C60C     & 0.510 & 0.560 & 2.91 &      &      &       & 0.116 & 0.56 & 0.117 \\        
        R80Ar-NR & 0.350 & 0.401 & 2.75 & 2346 & 0.53 & $-$9.85 & 0.099 & 0.70 & 0.100 \\
        R80Ar    & 0.350 & 0.428 & 2.74 & 2875 & 0.65 & $-$9.39 & 0.100 & 0.89 & 0.101 \\
		C115     & 0.400 & 0.478 & 2.70 & 3332 & 0.91 & $-$6.47 & 0.104 & 1.13 & 0.109 \\
        \hline        
	\end{tabular*}
	\flushleft
    \textit{Notes}: $t_\mathrm{BH}$ is the post-bounce time when the BH forms, $t_\mathrm{map}$ the post-bounce time of the mapping, $M_\mathrm{BH}$ the baryonic BH mass, $R_\mathrm{sh}$ the average shock radius, $E_\mathrm{diag}$ the diagnostic energy, $E_\mathrm{ob}$ the overburden energy at the time of mapping. $\Delta E_\nu(t_\mathrm{BH})$ is the energy drain due to neutrino escape until the time of BH formation, $t_\mathrm{end}^\mathrm{NADA}$ the time when the \textsc{NADA-FLD} simulation is stopped, and $\Delta E_\nu(t_\mathrm{end}^\mathrm{NADA})$ the total energy loss in neutrinos until the end of the \textsc{NADA-FLD} simulation. Note that the rapidly rotating Model C60C did not develop shock revival and its simulation was not continued with the \textsc{Prometheus} code. The listed values of $t_\mathrm{map}$ and $M_\mathrm{BH}$ are therefore those at the end of the \textsc{NADA-FLD} run. For the calculation of the neutrino signals, the \textsc{NADA-FLD} simulations are carried out beyond the mapping times $t_\mathrm{map}$ listed here.
	\label{tab:model_property_2}
\end{table*}

\section{Dynamical evolution after black hole formation}\label{sec:result_after_BH_formation}

In this section, we present the results of our 2D simulations after BH formation, in particular the evolution of the revived shock and of the neutrino-heated, initially expanding matter in high-entropy plumes. The initial 50--730\,ms after BH formation are simulated with the \textsc{NADA-FLD} code. When the radial infall downstream of the shock has become supersonic at a radius of around 800\,km, the models are mapped to the \textsc{Prometheus} code. We run the \textsc{Prometheus} simulations either until the moment when the shock reaches the stellar surface or shortly after it has converted to a weak sonic pulse and cannot be tracked well any longer. For the calculation of the neutrino signals, the \textsc{NADA-FLD} radiation hydrodynamic simulations are carried out beyond the mapping times for the \textsc{Prometheus} simulations listed in Table~\ref{tab:model_property_2}.

\begin{figure}
    \includegraphics[width=0.48\textwidth]{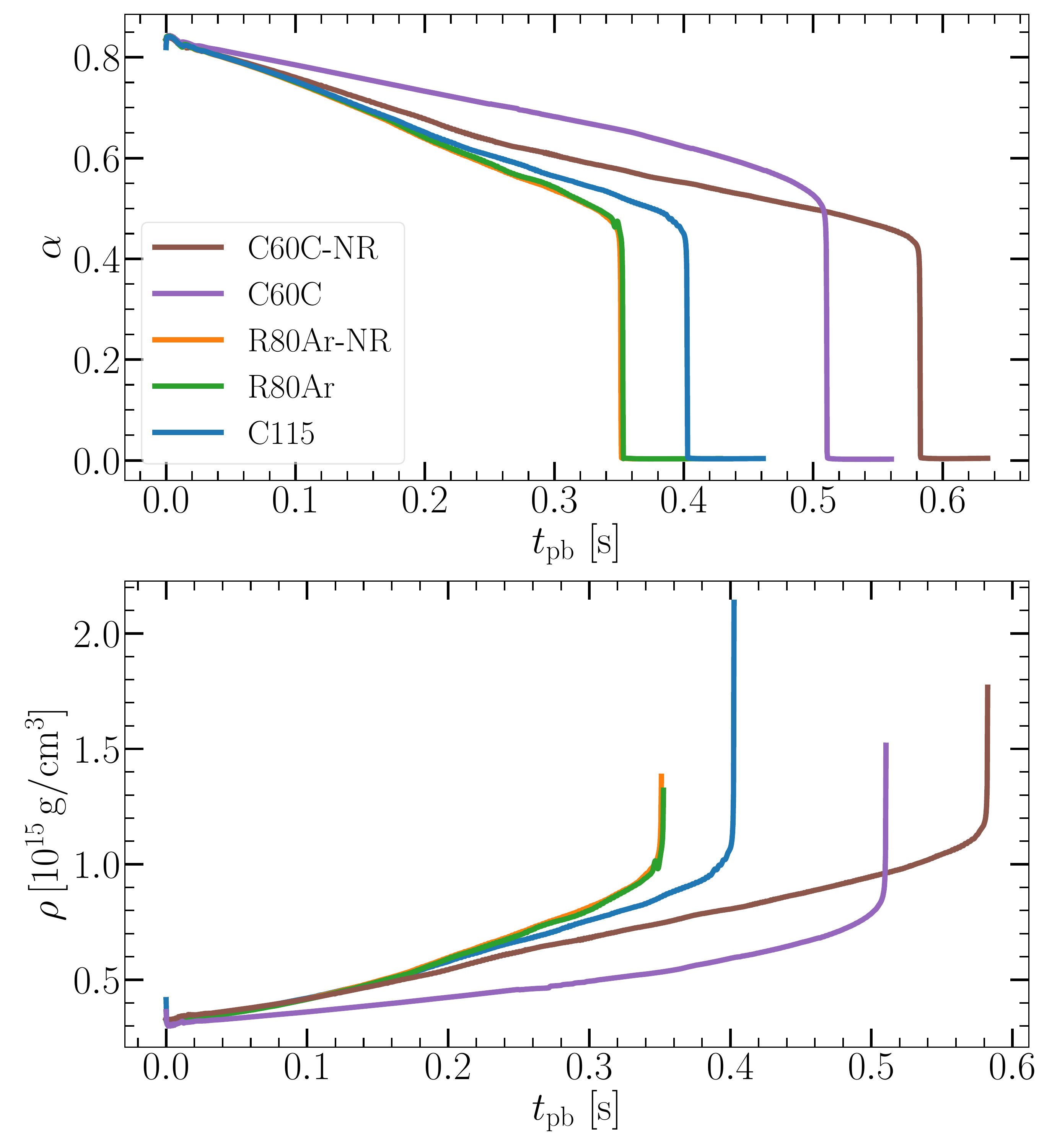}
    \vspace*{-7mm}
    \caption{Time evolution of the central lapse function (top panel) and the central baryonic density (bottom panel) for our set of models. At the time of BH formation the value of the central lapse function drops steeply towards zero and the central density rises sharply. The color scheme for the different models is the same as in Fig.~\ref{fig:shock_radius}.}
    \label{fig:central_density_lapse}
\end{figure}

\subsection{The first 50--730\,ms after BH formation}
\label{subsec:result_after_BH_formation_NADA}

The BH formation time is defined as the instant when an apparent horizon is detected (see, e.g., \citealt{Alcubierre:1138167} for a discussion of our apparent horizon finder). The times when the BHs are formed in our different models are tabulated in Tables~\ref{tab:model_property} and~\ref{tab:model_property_2}. Figure~\ref{fig:central_density_lapse} displays the central lapse function and the central density against the post-bounce time for our set of models. The values of the central lapse function drop steeply towards zero and the central densities increase sharply, signaling the emergence of the BHs. As we use a moving puncture gauge condition \citep{1995PhRvL..75..600B,2003PhRvD..67h4023A}, the values of the lapse function remain close to zero after BH formation.

\begin{figure}
    \includegraphics[width=0.48\textwidth]{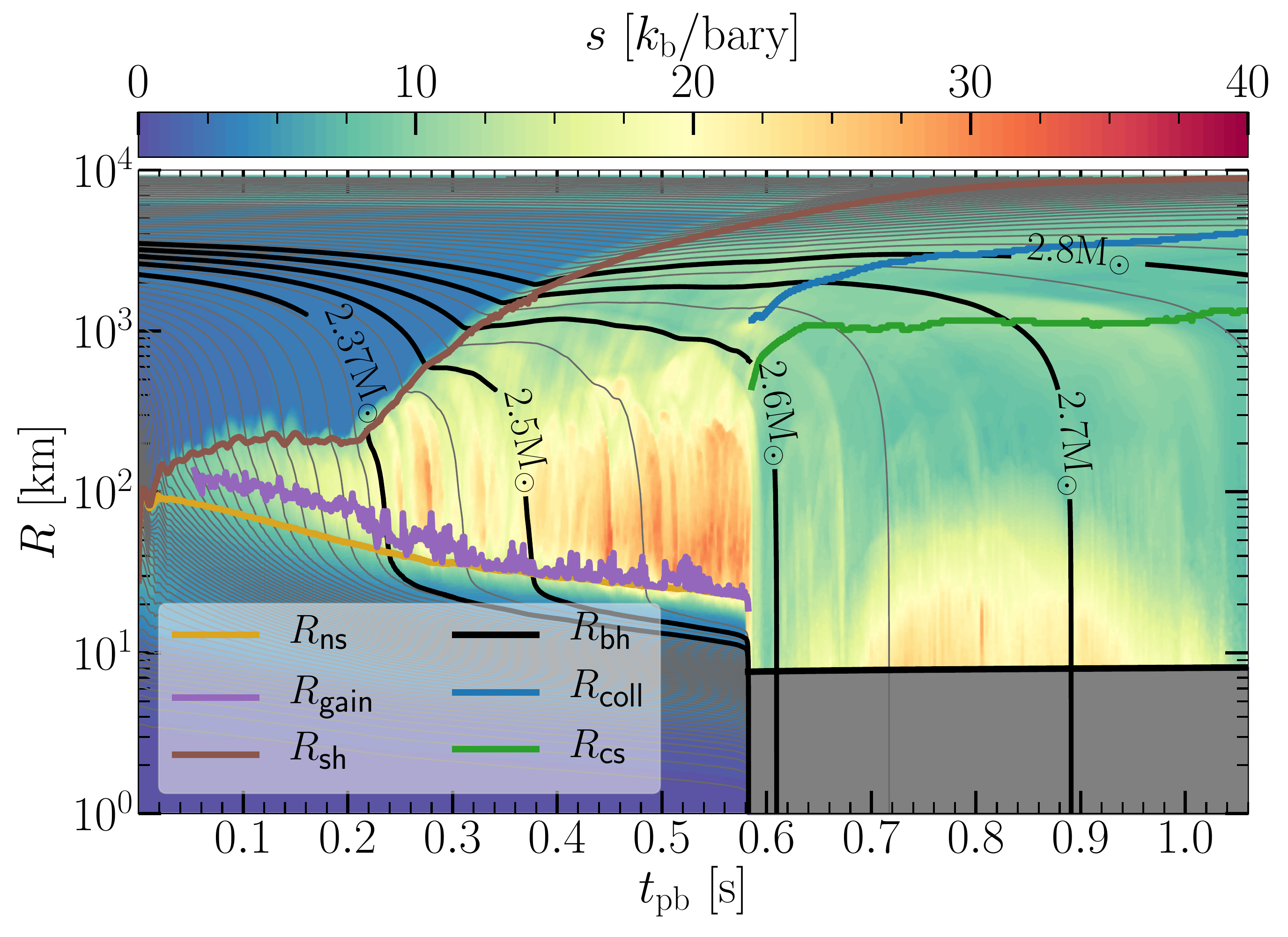}
    \includegraphics[width=0.48\textwidth]{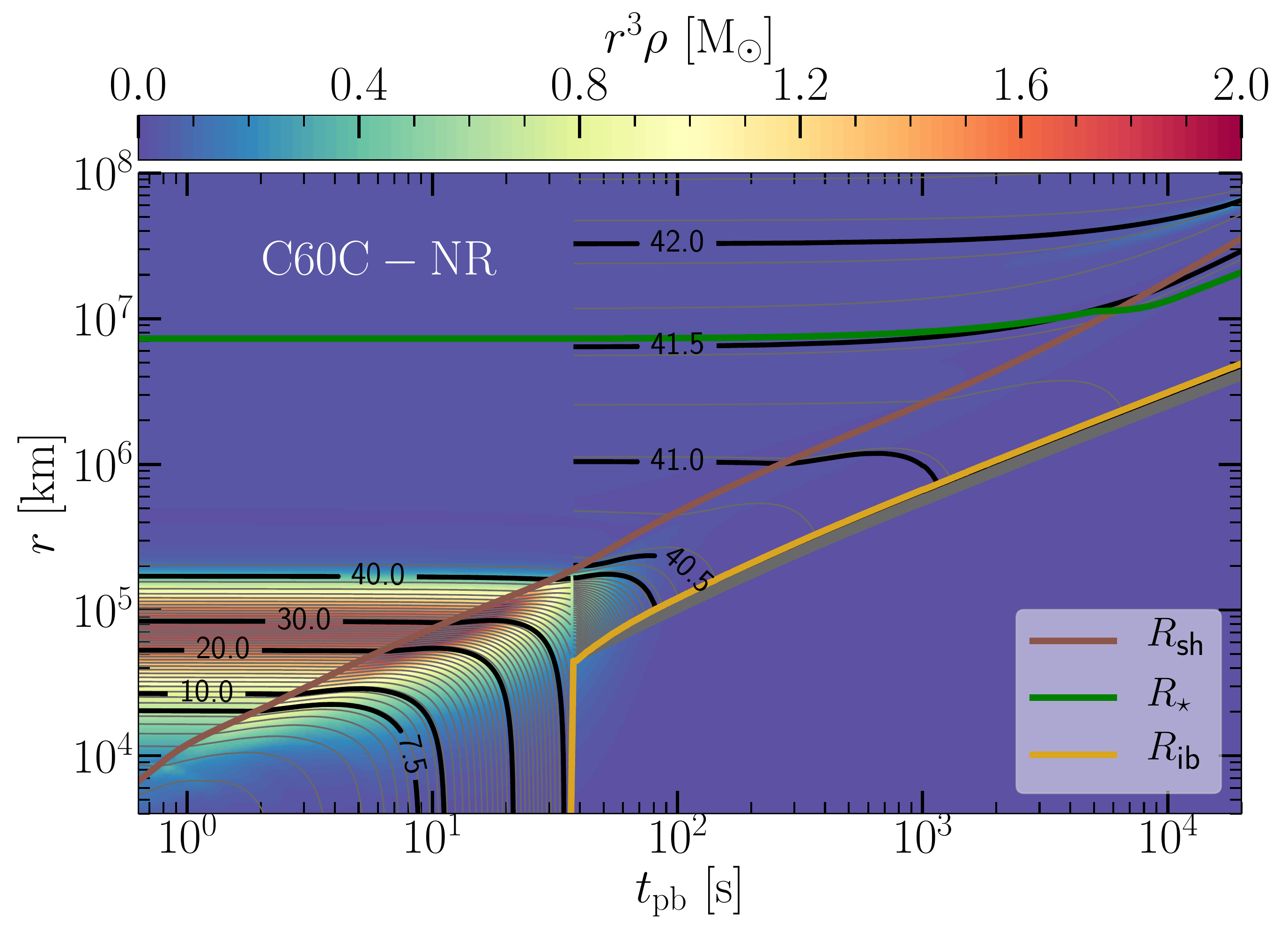}
    \vspace*{-7mm}
    \caption{Mass-shell plots for Model C60C-NR. The mass averaged entropy per baryon is color-coded in the top panel and the quantity $r^3\rho$, where $\rho$ is the baryonic mass density, is color-coded in the bottom panel. The top panel shows the angle-averaged locations of the PNS radius, $R_\mathrm{ns}$ (yellow line), the gain radius, $R_\mathrm{gain}$ (violet line), the BH radius, $R_\mathrm{bh}$ (thick black line), the radius interior to which the radial infall velocity is greater than 1000\,$\mathrm{km/s}$, $R_\mathrm{coll}$ (blue line), and sonic radius where the infall velocity exceeds the local sound speed, $R_\mathrm{cs}$ (green line), and the shock radius, $R_\mathrm{sh}$ (brown line). Note that $R_\mathrm{coll}$ and $R_\mathrm{cs}$ are only indicated in the postshock region after BH formation. The volume inside the BH is shaded in grey. In the bottom panel, in addition to the shock radius, the radii of the star, $R_\star$ (green line; initially $R_\star = R_\mathrm{prog}$; Table~\ref{tab:progenitor_property}), and of the inner boundary of the simulation domain, $R_\mathrm{ib}$ (yellow line), are marked. Several of the mass shells are highlighted by black lines with mass labels. The iron core mass is 2.37\,$\mathrm{M}_\odot$, and the interface between the Si-O layer and the O layer is at a mass coordinate of 7.54\,$\mathrm{M}_\odot$. The Fe/Si interface falls through the shock just before the stalled shock starts to expand at about 220\,ms after bounce. The outgoing shock reaches 400\,km at about $t_\mathrm{pb}$ $\approx$ 250\,ms and experiences deceleration (acceleration) where the quantity $r^3 \rho$ has a positive (negative) radial derivative.}
    \label{fig:mass_shell_C60C_NR}
\end{figure}

\begin{figure}
    \includegraphics[width=0.48\textwidth]{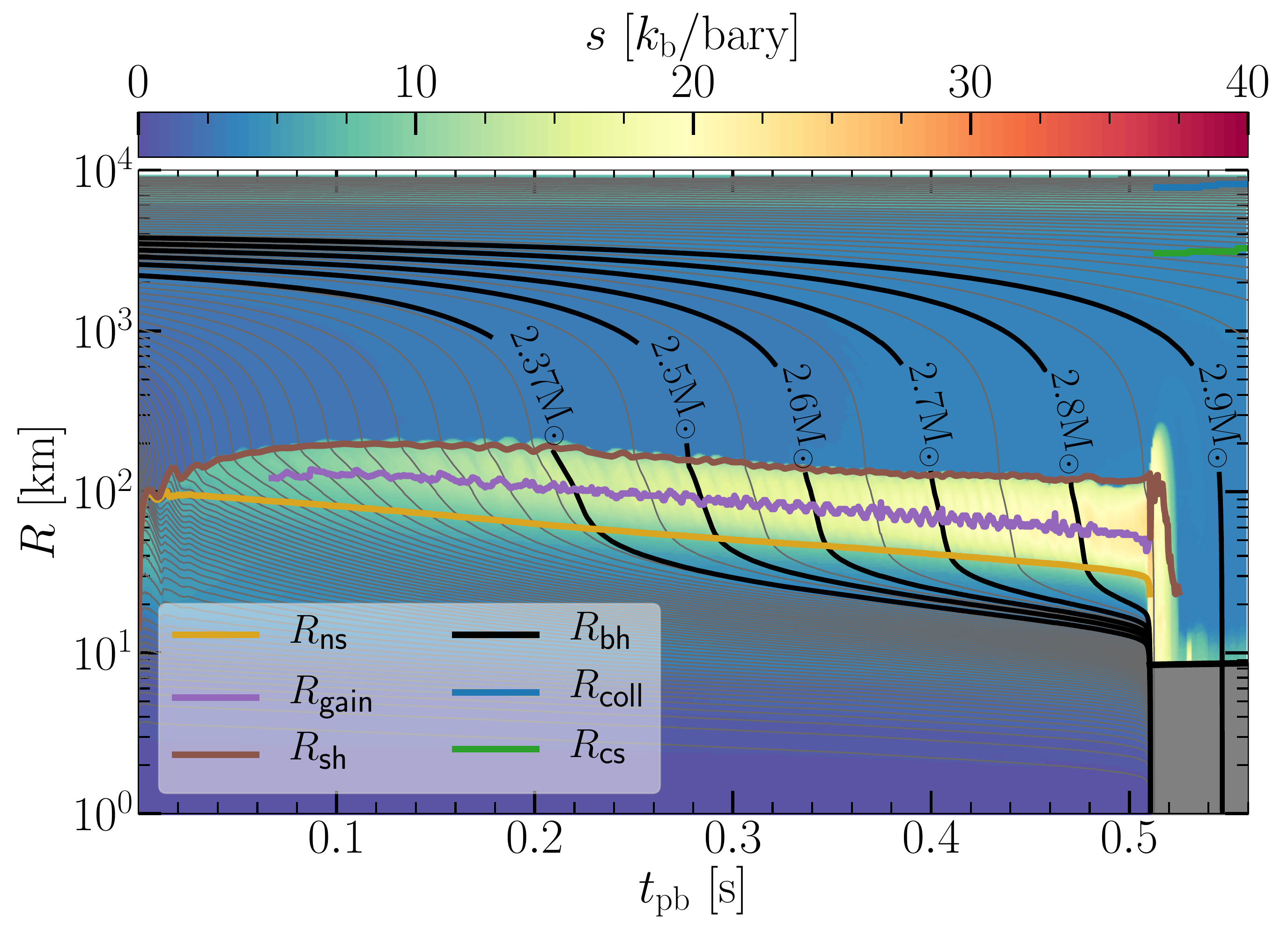}
    \vspace*{-7mm}
    \caption{Mass-shell plot for Model C60C. Quantities shown here are the same as in the top panel of Fig.\,\ref{fig:mass_shell_C60C_NR}. The iron core mass is 2.37\,$\mathrm{M}_\odot$.}
    \label{fig:mass_shell_C60C}
\end{figure}

\begin{figure}
    \includegraphics[width=0.48\textwidth]{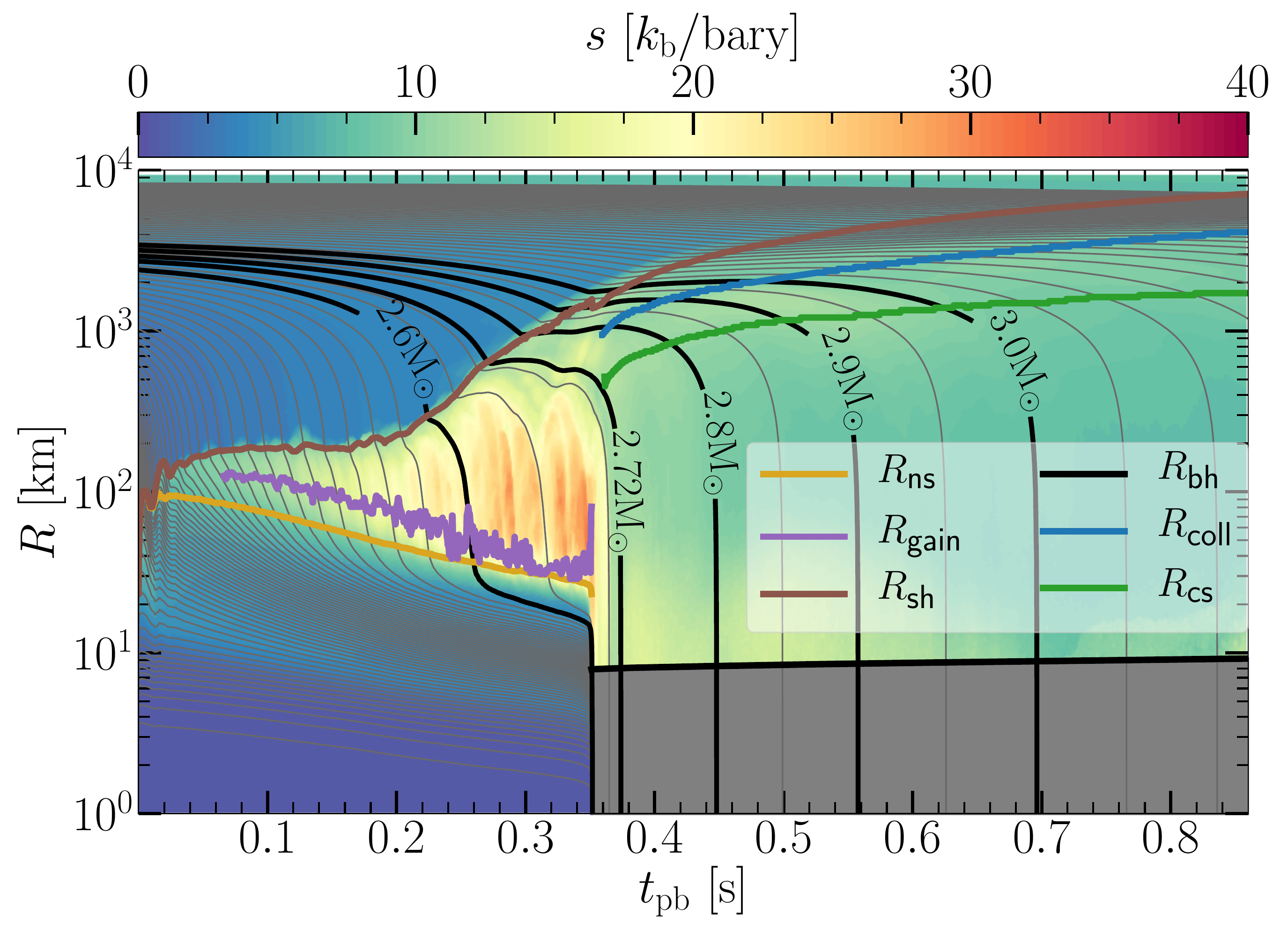}
    \includegraphics[width=0.48\textwidth]{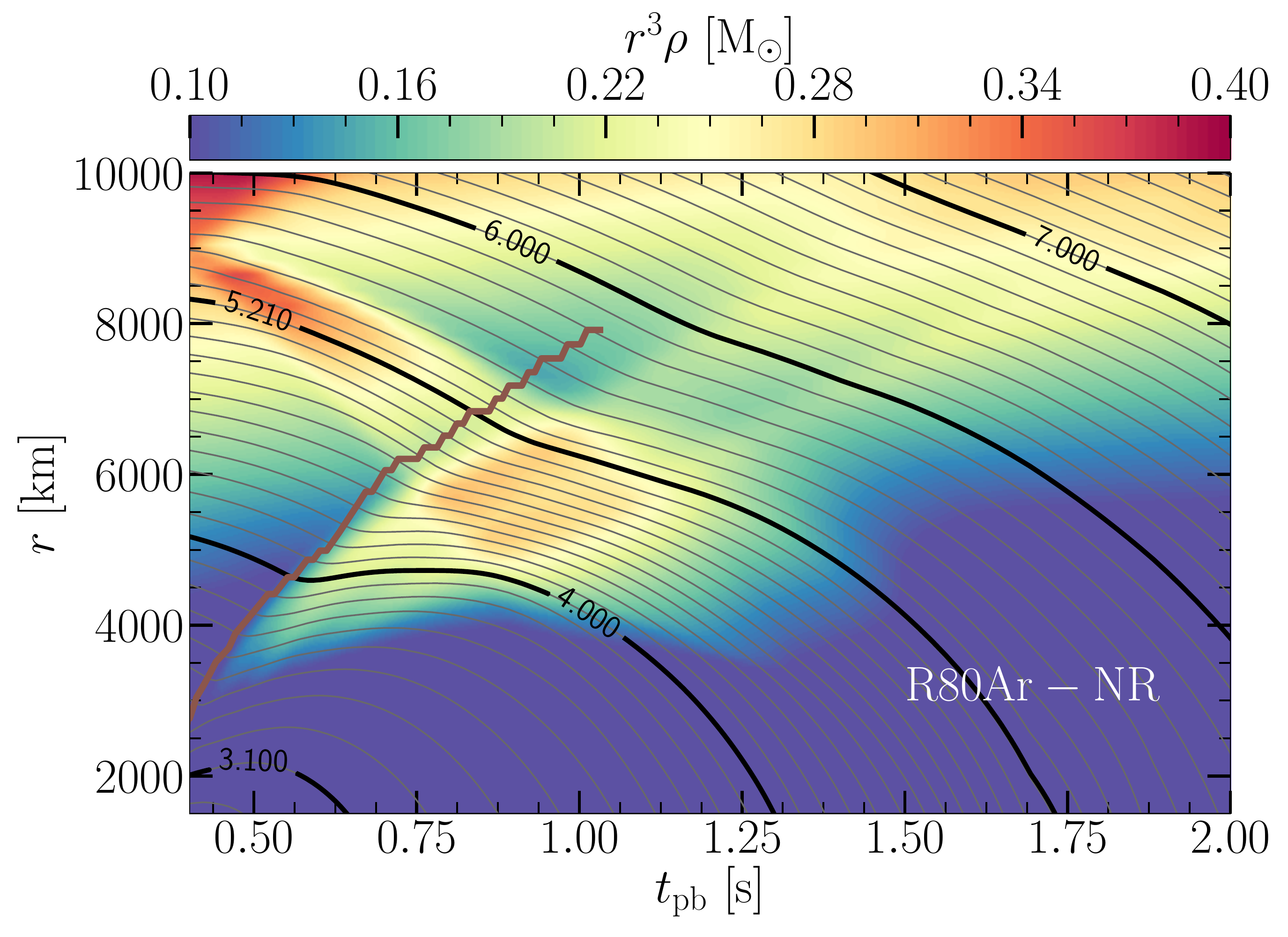}
    \vspace*{-7mm}
    \caption{Mass-shell plots for Model R80Ar-NR. Quantities shown here are the same as in Fig.~\ref{fig:mass_shell_C60C_NR}. The iron core mass is 2.72\,$\mathrm{M}_\odot$, and the interface between the Si-O layer and the O layer is at a mass coordinate near 5\,$\mathrm{M}_\odot$. The stalled shock starts to expand at about 200\,ms after bounce, and the shock reaches 400\,km at $t_\mathrm{pb}$ $\approx$ 246\,ms. The outgoing shock converts to a sonic pulse about 1\,s after core bounce.}
    \label{fig:mass_shell_R80Ar_NR}
\end{figure}

\begin{figure}
    \includegraphics[width=0.48\textwidth]{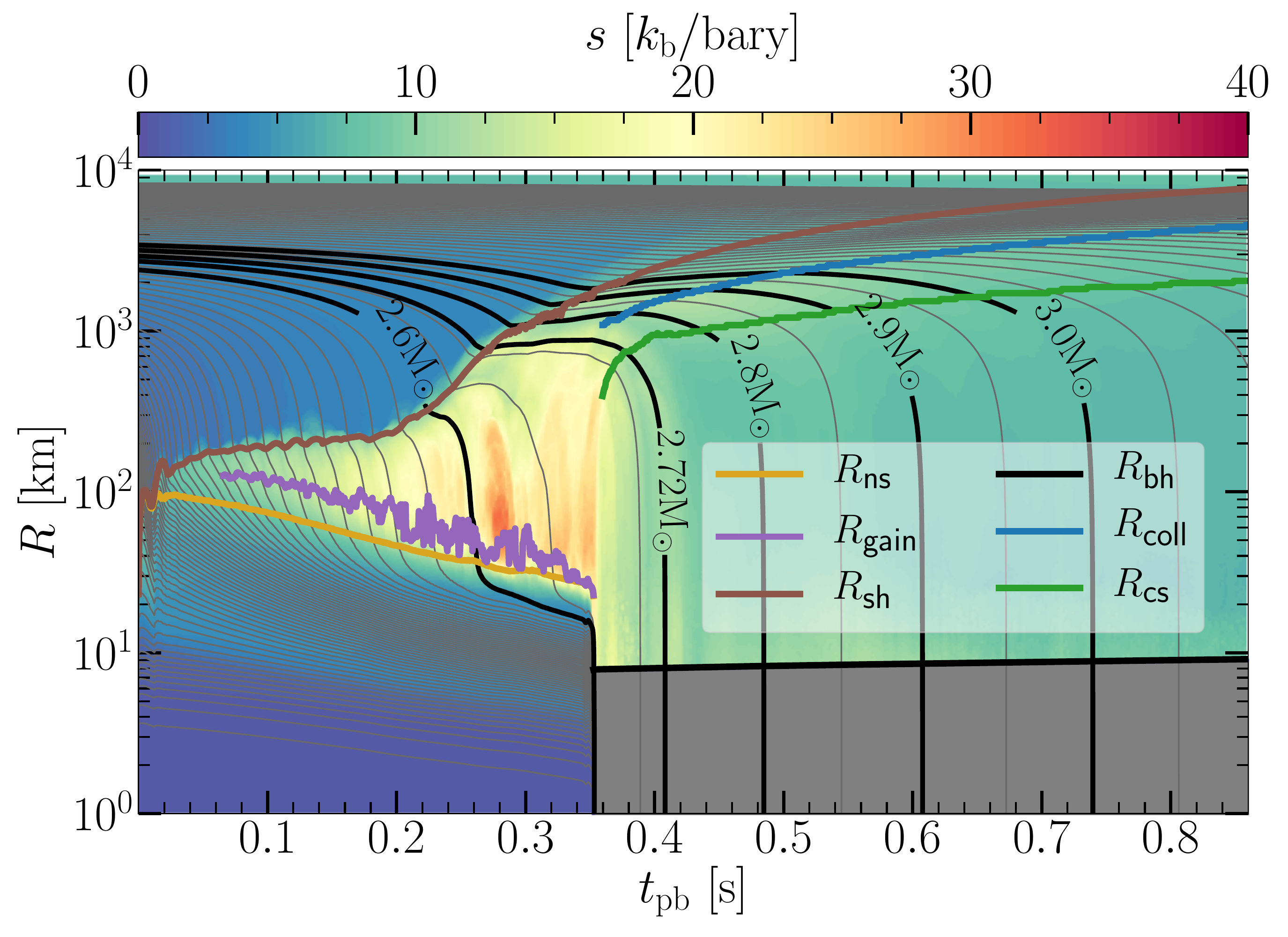}
    \includegraphics[width=0.48\textwidth]{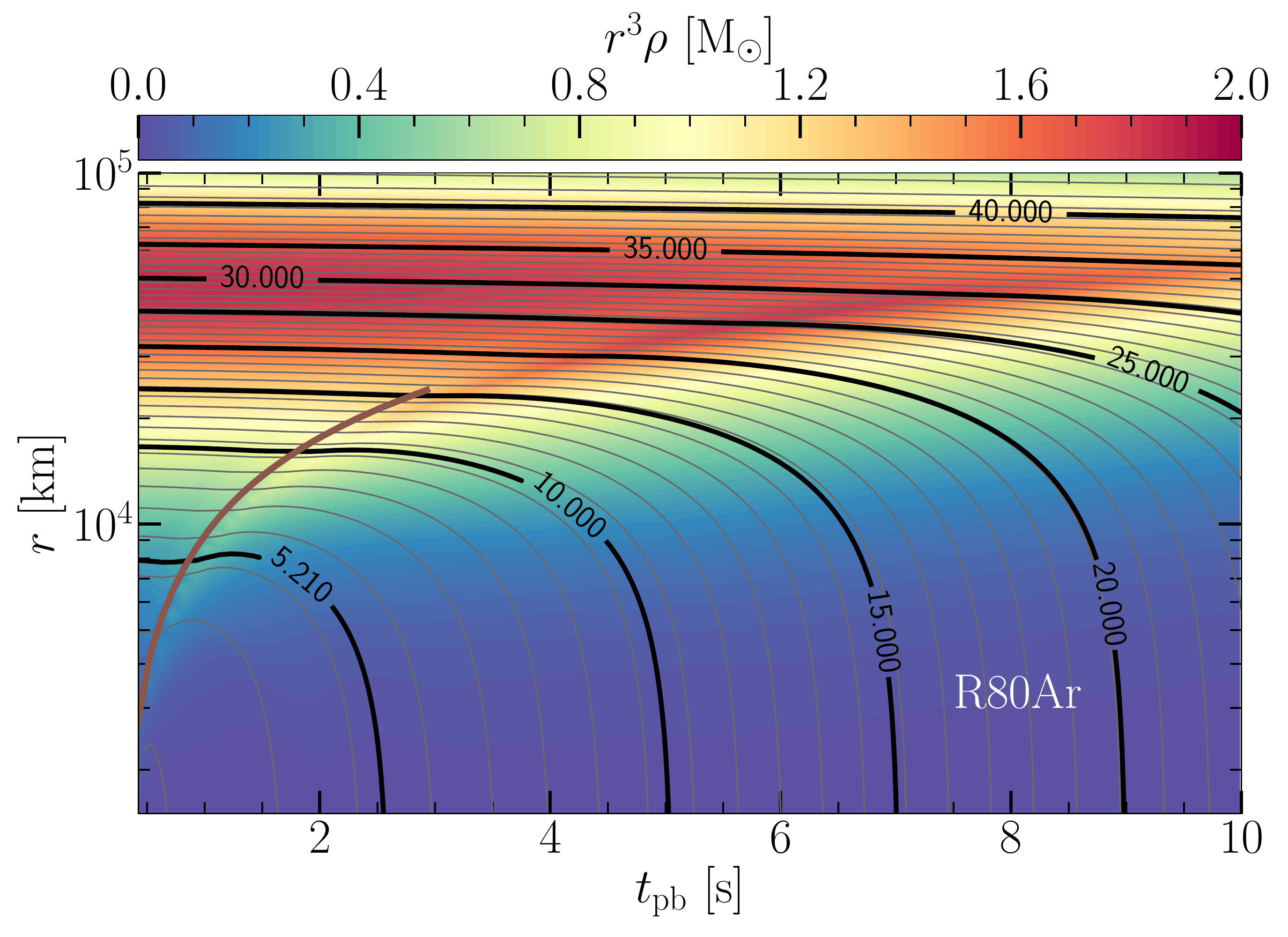}
    \vspace*{-7mm}
    \caption{Mass-shell plots for Model R80Ar. Quantities shown here are the same as in Fig.\,\ref{fig:mass_shell_C60C_NR}. The iron core mass is 2.72\,$\mathrm{M}_\odot$, and the interface between the Si-O layer and the O layer is close to a mass coordinate of 5\,$\mathrm{M}_\odot$. The stalled shock starts to expand at about 200\,ms after bounce, and the shock reaches 400\,km at $t_\mathrm{pb}$ $\approx$ 237\,ms. The outgoing shock converts to a sonic pulse several seconds after core bounce.}
    \label{fig:mass_shell_R80Ar}
\end{figure}

\begin{figure}
    \includegraphics[width=0.48\textwidth]{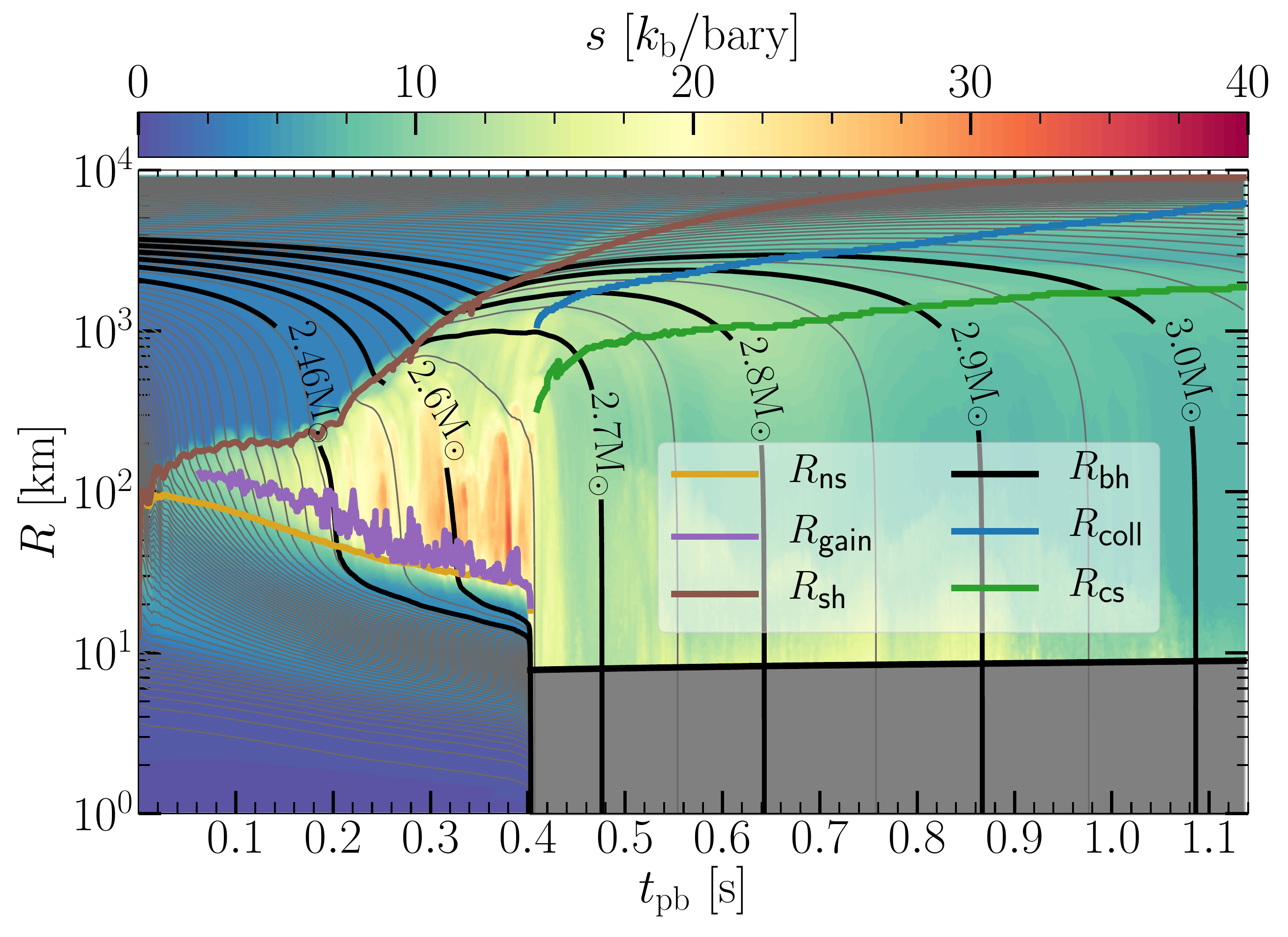}
    \includegraphics[width=0.48\textwidth]{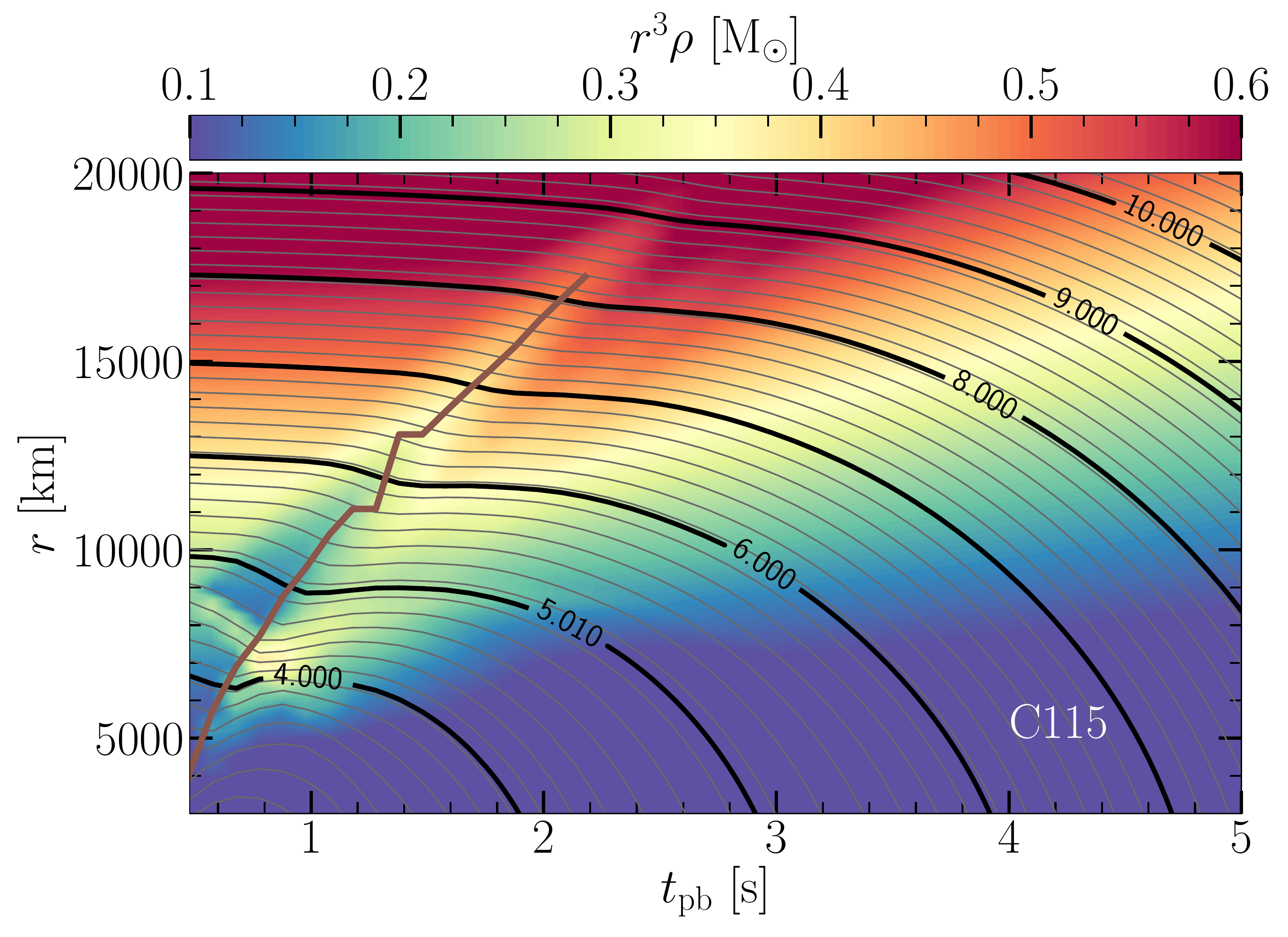}
    \vspace*{-7mm}
    \caption{Mass-shell plots for Model C115. Quantities shown here are the same as in Fig.\,\ref{fig:mass_shell_C60C_NR}. The iron core mass is 2.46\,$\mathrm{M}_\odot$, and the interface between the Si-O layer and the O layer is near a mass coordinate of 5\,$\mathrm{M}_\odot$. The Fe/Si interface falls through the shock at about 200\,ms after bounce, just before the re-expansion of the stalled shock sets in. The revived shock reaches 400\,km at $t_\mathrm{pb}$ $\approx$ 222\,ms. The outgoing shock converts to a sonic pulse about 2.5\,s after core bounce.}
    \label{fig:mass_shell_C115}
\end{figure}

In Figs.\,\ref{fig:mass_shell_C60C_NR}--\ref{fig:mass_shell_C115} we present mass-shell plots for Models C60C-NR, C60C, R80Ar-NR, R80Ar, and C115. These show angle-averaged locations of the PNS radius, $R_\mathrm{ns}$ (yellow line), the gain radius, $R_\mathrm{gain}$ (violet line), the shock radius, $R_\mathrm{sh}$ (brown line), the BH radius, $R_\mathrm{bh}$ (black line), and the average radius where the radial infall velocity exceeds 1000\,km/s, $R_\mathrm{coll}$ (blue line), and where the mean infall velocity becomes larger than the local sound speed, $R_\mathrm{cs}$ (green line). Moreover, in the lower panel of Fig.\,\ref{fig:mass_shell_C60C_NR} the radius of the star, $R_*$ (green line; initially $R_* = R_\mathrm{prog}$; Table\,\ref{tab:progenitor_property}), and of the inner boundary of the simulation domain, $R_\mathrm{ib}$ (yellow line), are marked. We stress that the concept of mass shells is introduced merely for visualization. The flow in the postshock region is highly nonspherical (see Figs.\,\ref{fig:contour_all_model_BH} and \ref{fig:contour_c60_norot}). Mass shells in the multi-dimensional case do not correspond to Lagrangian matter elements. Instead, they are defined as the (time-dependent) radii of spheres that enclose selected values of mass. But because of aspherical gas motions fluid elements can be exchanged between the thus defined mass shells.

At the times of BH formation, the shocks have reached radii of 4552\,km, 1560\,km, 1721\,km and 2239\,km in Models C60C-NR, R80Ar-NR, R80Ar, and C115, respectively (see Fig.~\ref{fig:shock_radius} and Table~\ref{tab:model_property}). After the appearance of the BHs, the shocks continue to expand initially with steady velocities of $\sim$15000\,$\mathrm{km/s}$.

The information about the BH formation propagates outward via a rarefaction wave that travels with the sonic speed of the medium. As the rarefaction reaches a certain radius in the postshock domain, the negative pressure gradient, which works against gravity, at that particular radius becomes flatter. As a result, the expansion of the postshock matter at that radius is slowed down and the matter begins to fall towards the BH with accelerated velocity. Hence, the radius $R_\mathrm{coll}$ behind the shock very approximately traces how far the rarefaction wave has travelled from the BH. In Model C60C, where the stalled shock is not revived, the rarefaction wave reaches the stagnant shock shortly after BH formation and the entire postshock layer including the shock falls into the BH within only\,$\sim$20\,ms (Fig.~\ref{fig:mass_shell_C60C}). In all other models, where re-expansion of the bounce shock is facilitated by neutrino heating, the rarefaction wave does not catch up with the outgoing shock, but the shock is still expanding until the end of the \textsc{NADA-FLD} simulation. Therefore, long-time simulations are needed to clarify the final fate of the shock, i.e., whether it continues to expand or falls back to the BH.


After the collapse of the hot PNS to a BH, we notice that the neutrino luminosities drop by 1$-$2 orders of magnitude, because the neutrino-emitting matter in and around the PNS is swallowed by the BH quickly and because most of the still infalling stellar material disappears in the BH effectively on radial paths and thus has no time to radiate neutrinos efficiently (see Section~\ref{sec:neutrino_properties}). Thus, the net neutrino heating rates diminish, too. Immediately after the BH formation, also the regions with strong neutrino heating start falling towards the BH with high velocities and are also swallowed by the BH on a short timescale. Consequently, this matter has no time to absorb neutrinos with significant efficiency and the supply of freshly neutrino-heated mass outflow comes to an end. Therefore neutrino interactions in the surroundings of the newly formed BH do not play any important role for the dynamics of the collapsing star. Eventually, on a timescale of at most a few seconds, even all of the previously neutrino-heated high-entropy matter is sucked inward and disappears in the BH. We will discuss the properties of the still emitted neutrinos in detail later in Section\,\ref{sec:neutrino_properties}.

In the rapidly rotating Model C60C, where shock revival does not occur and the shock disappears in the BH within only 20\,ms after BH formation (Fig.~\ref{fig:mass_shell_C60C}), also most of the infalling stellar matter approaches the BH effectively on free-fall trajectories. Yet, near the equatorial plane we witness indications that some of the high-$j_{z,\mathrm{eq}}$ matter tries to assemble into a thin, low-mass AD around the BH.
However, the disk mass is too small to have a considerable impact on the neutrino emission (Section~\ref{sec:neutrino_properties}) and on the GW production (Section~\ref{sec:gravitational_waves}). But there is an extended region between enclosed masses of 2.84\,M$_\odot$ (the baryonic mass initially collapsing to the BH in Model~C60C; Table~\ref{tab:model_property}) and $\sim$3.5\,M$_\odot$, where at least the progenitor's matter around the midplane has sufficient angular momentum to remain centrifugally stabilized on orbits around the BH and to increase the mass of the AD as time progresses (see Section~\ref{subsec:progenitor_property}). This will lead to enhanced neutrino production and potentially also GW emission at later epochs. The infall of the mass layers as far out as 3.5\,M$_\odot$, however, can take several seconds, which is longer than we can follow the 2D evolution in the \textsc{NADA-FLD} radiation hydrodynamics simulation. We therefore terminated the calculation for Model~C60C about 50\,ms after BH formation.

As infalling matter of the collapsing stars crosses the expanding shock in our shock-reviving models, the shock transfers energy to the gas and the infall of this shocked material is decelerated initially. However, moving inward and not receiving any strong push from below, the matter is  accelerated and eventually falls into the central BH, as we can conclude from the mass-shell plots of Figs. \ref{fig:mass_shell_C60C_NR}, \ref{fig:mass_shell_R80Ar_NR}, \ref{fig:mass_shell_R80Ar}, and \ref{fig:mass_shell_C115}. 

In the top panel of Fig.~\ref{fig:shocked_layer_properties}, we notice that the diagnostic energies of the postshock material decline after BH formation in the shock-reviving Models C60C-NR, R80Ar-NR, R80Ar, and C115, because an increasing fraction of the neutrino-heated matter falls into the BH. After the BH formation, the outflow of freshly neutrino-heated matter from the vicinity of the central compact object is stopped, but the outer parts of the high-entropy bubbles (i.e., of the high-entropy plumes produced by the neutrino heating before the BH formation) continue to expand for a transient time. The density of these bubbles is lower compared to their surroundings. As a result, the bubbles experience buoyancy forces and rise radially outward, at the same time expanding sideways. We see strong accretion flows surrounding these rising bubbles (see Figs.~\ref{fig:contour_all_model_BH} and \ref{fig:contour_c60_norot}). The expanding bubbles have high pressure compared to their surrounding material and transfer energy and momentum to this surrounding gas through mechanical work. \citet{2018ApJ...852L..19C, 2020MNRAS.495.3751C}, based on their core-collapse simulations of a 40 $\mathrm{M}_\odot$ zero-metallicity progenitor with BH formation and shock revival, concluded that the material around the high-entropy plumes can develop outward expansion and can become gravitationally unbound, provided the neutrino-energy deposition is powerful enough and the neutrino-heated matter is able to transfer sufficient energy and momentum to the overlying stellar layers. Eventually, the bubbles containing the originally neutrino-heated gas fall back to the BH, but the shock continues to expand in the models of \citet{2018ApJ...852L..19C, 2020MNRAS.495.3751C}. We see a similar behavior in our \textsc{NADA-FLD} simulations of the models with shock revival: the expanding plumes of neutrino-heated gas push the shock radially outward and thus transfer energy and momentum to the overlying stellar shells. However, this effect is by far not as strong in our simulations as in the models of \citet{2018ApJ...852L..19C, 2020MNRAS.495.3751C}, although the buoyant plumes have not yet fallen into the BH by the end of the \textsc{NADA-FLD} simulations. Hence, further long-time simulations are needed and will be discussed in Section~\ref{subsec:result_after_BH_formation_Prometheus} to clarify whether the bubbles can continue to rise outward as the density in their surrounding material drops, and whether they can transfer enough energy to unbind some of the outer or outermost stellar layers.


In the shock reviving models, matter at the base of the buoyant high-entropy plumes begins to fall into the BH right after BH formation. Yet, at the same time the outer parts of the plumes continue to expand in all directions and to engulf more and more volume in the shock-heated region. The effects of these counterworking processes can be recognized from the time evolution of the volume-filling parameter $\alpha_\mathrm{diag}$ (equation~\eqref{eq:alpha_parameter}), which is displayed in the bottom panel of Fig.\,\ref{fig:shocked_layer_properties}. The value of $\alpha_\mathrm{diag}$ is a ratio that measures the volume of the postshock matter with positive total energy and positive radial velocity relative to the total volume of the layer between the shock and either the gain radius or BH radius. Similar to the diagnostic energies, the values of $\alpha_\mathrm{diag}$ begin to drop after BH formation. This indicates that as time progresses, the accretion of matter from the high-entropy plumes into the BH becomes dominant over the expansion of the bubbles. Superimposed on the general trend of the decline are time intervals with local maxima, which correspond to short, transient periods of plume expansion.

We also notice that the beginning and the duration of the decline of both the diagnostic energy and the $\alpha_\mathrm{diag}$ parameter after BH formation vary between the models. The detailed evolution of the decline depends on the magnitude of the diagnostic energy and on the radial extension of the high-entropy plumes at the time of the BH formation. The shock radius traces the outer radii of the bubbles at least for the initial $\sim$100\,ms after the formation of the BH. Later, the shock detaches itself from the bubbles and evolves independently. 

Model C60C-NR develops the highest value of the diagnostic energy at the time of BH formation, because it exhibits the longest time interval between the onset of the shock expansion and the collapse of the PNS to a BH. Consequently, more energy can be deposited in the gain layer by neutrino heating before BH formation, and the shock and high-entropy bubbles in this model manage to expand to a larger radial distance than in all of our other models with shock revival. Because of the higher diagnostic energy and faster shock expansion (see also Fig.~\ref{fig:shock_radius} and note the log scale in the bottom panel there) the layer between $R_\mathrm{coll}$ and $R_\mathrm{sh}$ in Model C60C-NR is more extended than in the other cases (see top panel of Fig.\,\ref{fig:mass_shell_C60C_NR} and compare it with the top panels of Figs.\,\ref{fig:mass_shell_R80Ar_NR}, \ref{fig:mass_shell_R80Ar}, and \ref{fig:mass_shell_C115}). Moreover, for an initial $\sim$100\,ms after BH formation, the expansion of the bubbles in Model C60C-NR is influenced to a lesser extent by the infall of matter into the BH. Correspondingly, we observe the longest delay between the decline of the value of $\alpha_\mathrm{diag}$ and that of the diagnostic energy in Model C60C-NR compared to all other models with shock expansion (see Fig.~\ref{fig:shocked_layer_properties}). Because of the late onset and long duration of the decrease of $E_\mathrm{diag}$, the decline rate of this quantity in C60C-NR is clearly the smallest of all models.

In order to judge the results of successful or failed shock revival in our study in relation to those of \citet{2018ApJ...852L..19C, 2020MNRAS.495.3751C} and \citet{2021MNRAS.503.2108P}, a number of facts need to be taken into account. The models discussed in our paper are 2D (axisymmetric) in contrast to the 3D simulations presented by \citet{2018ApJ...852L..19C, 2020MNRAS.495.3751C} and \citet{2021MNRAS.503.2108P}. Previous studies revealed that shock revival can be facilitated by the artificial constraint of axisymmetry (see e.g., \citealt{2012ApJ...755..138H, 2016ApJ...825....6S}). One reason for this finding is the presence of the polar symmetry axis with its reflecting boundary condition, which can enable polar outflows that aid shock expansion. A second reason are the morphological differences between the toroidal geometry of structures near the equator in contrast to finger-like or cone-shaped structures near the poles, which again make polar expansion easier (e.g., \citealt{2013ApJ...775...35C}). Moreover, the cascading of turbulent kinetic energy in 2D and 3D goes in opposite directions, fostering the growth of large-scale plumes in 2D in contrast to small-scale vortex motions in 3D (see \citealt{2012ApJ...755..138H}). These effects influence the possibility of neutrino-driven shock expansion in multiple ways and often enable explosions in 2D when 3D simulations yield failures (see, e.g., \citealt{2015ApJ...808L..42M, 2018ApJ...852...28S}). Another generically 3D effect concerns rapidly rotating models, which can develop equatorial explosions in 3D because of the support by SASI spiral modes  \citep{2018ApJ...852...28S} or other triaxial spiral waves (e.g., \citealt{2020MNRAS.493L.138S}). Such phenomena do not exist in 2D. It was also found that after shock revival buoyant plumes of neutrino-heated high-entropy matter expand faster in 3D \citep{2010ApJ...714.1371H} and thus accelerate the shock expansion and enhance the diagnostic energy deposited by neutrino energy transfer \citep{2015MNRAS.453..287M, 2015ApJ...801L..24M}. Because of all of these effects, which partly work against each other, it is not straightforward to extrapolate from our 2D results to the more realistic 3D conditions and to directly compare our 2D models to previous 3D results in the literature. 

Nevertheless, a re-expansion of the stagnant shock due to neutrino energy deposition in very massive progenitors that collapse to BHs was witnessed in a larger number of previous works (see our discussion in Section~\ref{subsec:comparison_with_previous_works}) including those studying the core collapse in PPISNe \citep{2021MNRAS.503.2108P}. These results provide mutual support to each other on a qualitative level. The fact that subsequent ejection of considerable amounts of mass was possible in cases with sufficiently strong neutrino energy transfer \citep{2018ApJ...852L..19C, 2020MNRAS.495.3751C} is interesting. However, the quantitative question whether neutrino heating is powerful enough to achieve such ample mass ejection in a wider spectrum of BH-forming VMSs is still unanswered and is likely to depend on conditions that vary from case to case. ``Fallback supernovae'', i.e., explosion events in which a major fraction of the BH-forming star is expelled with high kinetic energy, may require quite fine-tuned and uncommon conditions. They were obtained by artificial enhancement of the neutrino energy deposition in the 3D simulations of \citet{2018ApJ...852L..19C, 2020MNRAS.495.3751C}. Numerically, the necessary amount of extra heating might depend on the hydrodynamical differences of simulations performed in 2D or 3D. In reality, the results are likely to depend on the details of the progenitor's core structure and may also be sensitive to still unclear properties of the nuclear EOS of hot NS matter, which can delay BH formation \citep{2021MNRAS.503.2108P,2018ApJ...857...13P} or produce a second SN shock due to a hadron-quark phase transition in high-mass PNSs~\citep{Fischer+2018}. Moreover, magnetic fields might play a non-negligible role, in particular magnetorotational effects during the collapse of rotating stars, where magnetic fields can be amplified efficiently not only by compression and turbulence but also by the magnetorotational instability, field winding, and dynamo effects.

\subsection{Long-time simulations after BH formation}
\label{subsec:result_after_BH_formation_Prometheus}

\begin{figure*}
	\includegraphics[width=\textwidth]{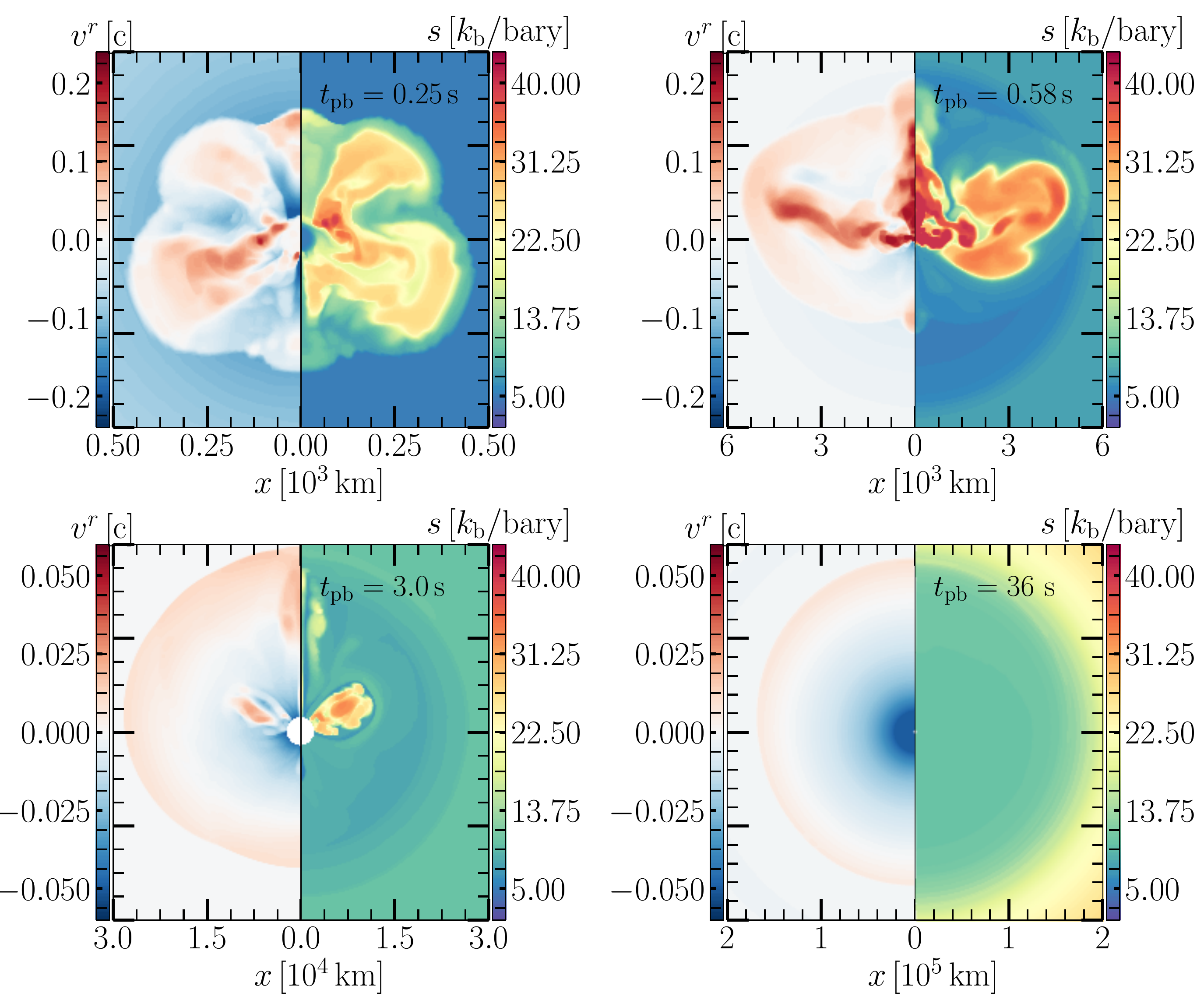}
	\vspace*{-7mm}
	\caption{Snapshots of the evolution of Model C60C-NR, depicting radial velocity $v^{r}$ (left halves of the panels) and entropy per baryon of the gas $s$ (right halves of the panels) at 0.25\,s (time of shock revival), 0.58\,s (time of BH formation), 3\,s, and 36\,s (time of shock breakout from the stellar surface) after bounce (from top left to bottom right). Although the initial high-entropy plumes of neutrino-heated matter fall back to the BH entirely, the shock continues to propagate radially outwards, which is visible by the discontinuities in the color distributions of radial velocity and entropy.}
	\label{fig:contour_c60_norot}
\end{figure*}

Let us now discuss the long-time evolution of the shock front in those of our models where shock revival happens, namely Models C60C-NR, R80Ar-NR, R80Ar, and C115. For this purpose the corresponding \textsc{NADA-FLD} models are mapped to the \textsc{Prometheus} code when the mean infall velocities of the postshock flow towards the BH become supersonic at a radius of 800\,km. The procedure of the mapping was described in Section\,\ref{subsec:numerical_setup}. Table~\ref{tab:model_property_2} lists the times of mapping, the baryonic BH masses, the mean shock radii, the diagnostic energies, and the overburden energies at the times of the mapping, the total energies radiated in neutrinos until the times of BH formation, the post-bounce times when the \textsc{NADA-FLD} simulations are stopped, and the total energies lost by neutrino radiation at the end of the \textsc{NADA-FLD} simulations. At the time of mapping, Model C60C-NR exhibits again the highest diagnostic energy of all of our shock-reviving models, as it did at the instant of BH formation (see Table~\ref{tab:model_property}). This correlates with the largest shock radius and implies the lowest (absolute) value of the overburden energy at both times. Since high-entropy plumes of neutrino-heated matter with positive radial velocities can survive for several seconds (see the case of Model C60C-NR in Figure~\ref{fig:contour_c60_norot}), the follow-up simulations with \textsc{Prometheus} are needed to determine the energy transfer from the plumes to the surrounding postshock gas. These simulations do not require any continued treatment of the neutrino effects, because the neutrino luminosities and neutrino heating have dropped by several orders of magnitude after the emergence of the BH (see Section~\ref{sec:neutrino_properties} for details).

The bottom panels of Figs.~\ref{fig:mass_shell_C60C_NR}, \ref{fig:mass_shell_R80Ar_NR}, \ref{fig:mass_shell_R80Ar}, and \ref{fig:mass_shell_C115}, provide mass-shell plots showing the shock evolution (brown lines) during the long-time simulations with the \textsc{Prometheus} code for Models C60C-NR, R80Ar-NR, R80Ar, and C115, respectively. The color coding of the background represents the quantity $r^3\rho(r)$ (see also Fig.~\ref{fig:progenitor}). In regions with positive (negative) radial derivative of this quantity, the outgoing shock, at constant energy, is expected to decelerate (accelerate). When moving through the extended Si- and O-shells, the outward propagation of the shocks in all models therefore slows down as they have to climb up the steep slope to the maximum of $r^3\rho(r)$ (Fig.~\ref{fig:progenitor}).

In Models R80Ar-NR, R80Ar, and C115, the diagnostic energies
at the time of mapping from NADA-FLD to \textsc{Prometheus} are
less than $10^{51}$\,erg already (Table~\ref{tab:model_property_2}). This is a consequence
of neutrino-heated matter falling back and being accreted by the
newly formed BH, which is an effect that is more extreme in the
models with higher overburden energies of the stellar shells swept
up by the outward moving shock. Models R80Ar-NR, R80AR, and C115
(in decreasing order) all have considerably higher overburden energies
than C60C-NR (Tables~\ref{tab:model_property} and \ref{tab:model_property_2}). The
decline of the diagnostic energies between Table~\ref{tab:model_property} (for the time
of BH formation, $t_\mathrm{BH}$) and Table~\ref{tab:model_property_2} 
(for the time of mapping, $t_\mathrm{map}$) 
can be larger or smaller than the difference of the overburden energies
in both tables, despite the fact that neutrino heating effectively
ceases at $t_\mathrm{BH}$ in all cases. The reason for these differences 
is the complex dynamics of inflows and outflows in the postshock layer. 
In particular in Model R80Ar-NR the high-entropy plumes of neutrino-heated 
gas, which carry the positive diagnostic energy, are quite fragile
and are quickly sucked inward by the gravitational attraction of the BH.
Therefore a rapidly growing fraction of their volume develops negative 
radial velocities, $v_r < 0$, and does not contribute to the integral
for $E_\mathrm{diag}$ in equation~\eqref{eq:gain_diagnostic_energy}. 

The shock in all of these models is correspondingly weak,
and it is weaker the higher the overburden energy is. Initially, 
the mass shells overrun by the shock still follow the shock in a 
transient period of expansion before they return and begin to fall
inward to the BH. But latest at about two seconds after bounce (in 
Model R80Ar-NR with the lowest diagnostic energy even already at 
$\sim$0.8\,s post bounce) the gas crossing the shock does not obtain
positive radial velocity any longer. Instead, it continues to 
collapse downstream of the shock, which has further lost
strength. Finally, the shock converts to a sonic pulse, in which 
the entropy discontinuity characteristic of shocks is absent. 
The sonic pulses in Models R80Ar-NR, R80Ar, and C115 move on
outward through the stars, carry energy, and can potentially
trigger mass loss when reaching the loosely
bound near-surface layers of the PPISN progenitors. Acoustic
pulses and waves and the implications of associated weak energy 
release (much smaller than the binding energy of the entire
star) for mass stripping from massive stars have recently been
discussed by \citet{2018MNRAS.477.1225C, 2018ApJ...863..158C, 
Coughlin+2019, 2021MNRAS.501.4266L}, and \citet{2021ApJ...908...23M}.
This possibility of shock/pulse triggered mass ejection will 
be analysed for our models later in Section~\ref{sec:massejection}.

Mass loss from the outermost layers will reduce the mass that 
ultimately ends up in the new-born BHs assembling from the 
collapsing stars. Mass shedding from the surface of the progenitor 
of Model R80Ar will also curtail the possibility of AD formation
in the collapse of this rotating model, because all shells except
the outermost layers do not carry sufficient rotational angular 
momentum to stay on orbits around the Kerr BH containing the mass
enclosed by those shells (see Section~\ref{subsec:progenitor_property} 
and Fig.~\ref{fig:progenitor}). It should be noted, however, that mass 
and angular momentum loss from the outer layers is quite uncertain 
already during the progenitor evolution because of the role of 
magnetic torques during the interpulse periods leading up to core 
collapse \citep{Woosley+2021}.

Again, the modest core rotation of model R80Ar does not cause
any fundamental differences in the shock evolution between 
Model R80Ar and its non-rotating counterpart, Model R80Ar-NR.
Both of them display similar overall dynamical behavior also 
during their long-time evolution, see Fig.~\ref{fig:shock_radius} 
for the shock radius and Fig.~\ref{fig:shocked_layer_properties} for 
the diagnostic energy. A noticeable difference, however, is 
connected to the fact that Model R80Ar possesses strong polar
high-entropy plumes and prolate shock deformation, whereas Model 
R80Ar-NR has more pronounced equatorial plumes, more massive
downflows close to the poles, and a more oblate
shape of the shock (see Fig.~\ref{fig:contour_all_model_BH}). Toroidal structures
near the equator have a larger surface to volume ratio, which 
increases the $p\mathrm{d}V$ work to be done for driving their
expansion against the infalling postshock matter. Therefore 
the polar plumes in Model R80Ar are able to expand to somewhat
larger radii, push the shock farther out, and take longer 
to be accreted into the newly formed BH. This can be concluded
from a considerably slower decline of the diagnostic energy
$E_\mathrm{diag}$ and of the volume-filling parameter of the
plumes, $\alpha_\mathrm{diag}$, in Model R80Ar compared to 
R80Ar-NR (Figure~\ref{fig:shocked_layer_properties}).

Model C60C-NR possesses the highest diagnostic energy (considerably
more than $10^{51}$\,erg) at the instant of BH formation as well 
as at the time of mapping, and its overburden energy is much lower
than in all other models (see Tables~\ref{tab:model_property} and \ref{tab:model_property_2}). Correspondingly,
the shock expands faster and remains stronger. Moreover,
the high-entropy plumes of neutrino-heated matter are inflated
behind the shock to much larger radii (Fig.~\ref{fig:contour_all_model_BH}). This is
facilitated by a longer-lasting supply of freshly heated, buoyant gas due to the significantly longer survival time of the PNS after the shock revival and before BH formation
(Table~\ref{tab:model_property}).

\begin{figure}
	\includegraphics[width=0.48\textwidth]{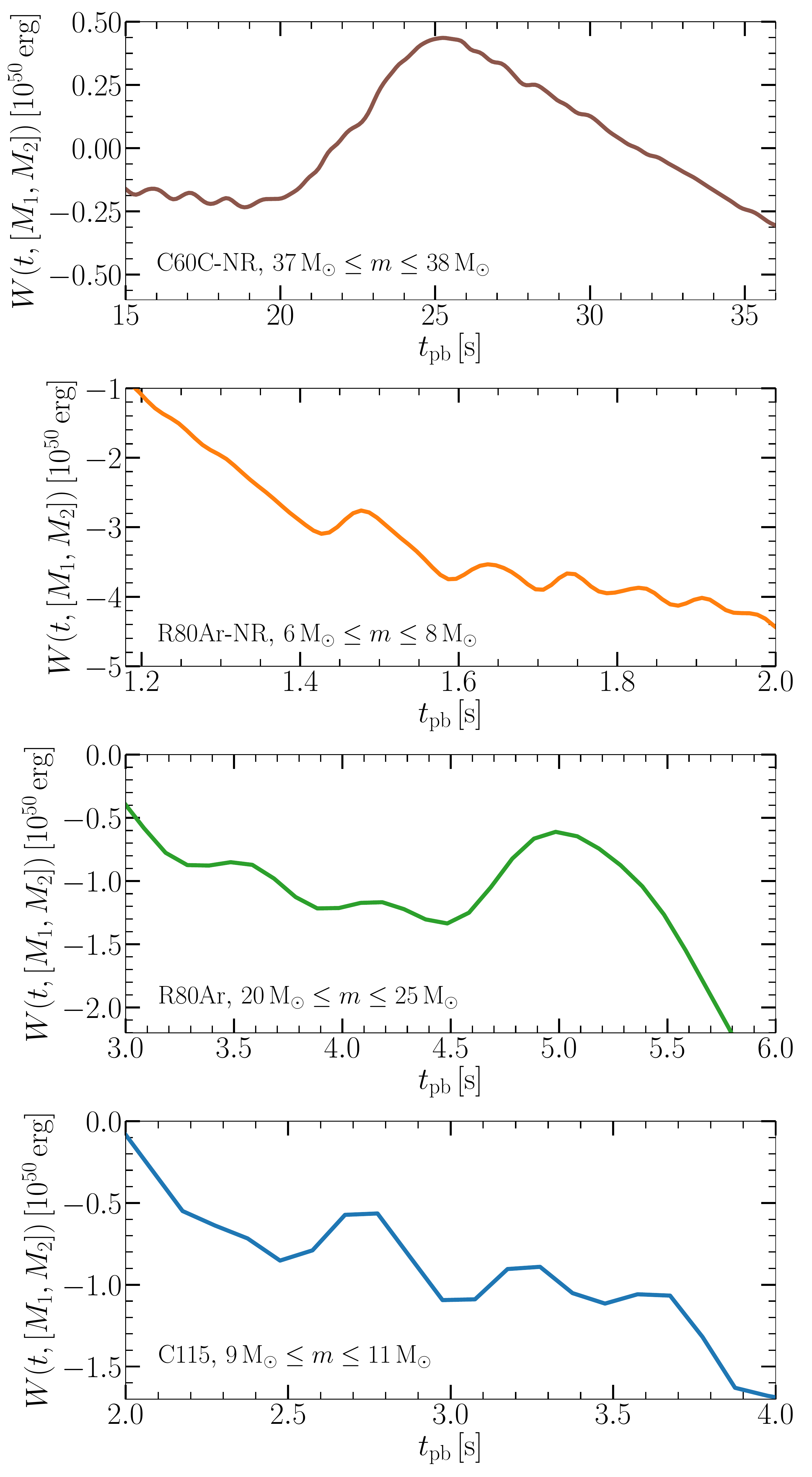}
	\vspace*{-7mm}
	\caption{Time evolution of the work done by the outgoing shock or sonic pulse on specified Lagrangian mass shells according to equation~\eqref{eq:energy_sonic_pulse_F}. The inner and outer boundaries of the mass intervals considered for the different models are indicated by the labels in the panels. The pronounced increase of the total energy starting at about 20\,s, 1.43\,s, 4.5\,s, and 2.5\,s after bounce and the subsequent decrease beginning at about 25\,s, 1.48\,s, 5.0\,s and 2.8\,s in Models C60C-NR, R80Ar-NR, R80Ar, and C115 (from top to bottom), respectively, signal the passage of the outgoing shock or sonic pulse. There is a superimposed long-time trend of the shell energy connected to the infall of the mass shells and the associated compression work.}
	\label{fig:energy_sonic_pulse}
\end{figure}

Figure~\ref{fig:contour_c60_norot} shows snapshots of the structure of Model C60C-NR
at the time of shock revival, the instant of BH formation,
3\,s after core bounce, and 36\,s after bounce.
When the PNS collapses to a BH, extended plumes exist near
the equatorial plane and along to north-polar direction,
but not near the south-polar direction. This
causes a strongly dipolar asymmetry of the expanding shock
(the dipole amplitude reaches 20\% of the angle-averaged
shock radius). But on its way out the shock detaches
from the high-entropy plumes and becomes
progressively more spherical (compare the upper right and 
lower left plots in Fig.~\ref{fig:contour_c60_norot}). After the BH formation at
0.58\,s post bounce, the high ends of the plumes continue to push
outward, driven by buoyancy forces, although fresh 
neutrino-heated matter is no longer added at their base.
They need to perform $p\mathrm{d}V$ work against the 
infalling matter that is swept up by the outgoing shock and
channeled into dense, low-entropy downdrafts between the plumes.
In the long run, however, the buoyancy-driven rise of the plumes 
is overwhelmed by the persistent ram pressure exerted by the infalling matter 
in addition to the loss of support at the base when more and 
more gas gets sucked into the BH. Eventually, the plumes break
down and are entirely swallowed by the BH (see bottom right panel
in Fig.~\ref{fig:contour_c60_norot}).

Subsequently, the shock continues to propagate radially outward,
irrespective of the disappearance of the plumes, which have 
transferred enough expansion work to the overlying shocked
matter such that the shock is able to reach the outermost
layers of the star (see lower panels of Figs.~\ref{fig:shock_radius} 
and \ref{fig:mass_shell_C60C_NR}). During all of its outward motion 
the shock is sufficiently strong to enforce a temporary co-expansion 
of the shocked stellar shells before these shells decelerate again 
and ultimately fall back to be absorbed by the inner grid 
boundary.\footnote{As described in Section~\ref{subsec:numerical_setup}, the inner
boundary of the computational grid is progressively moved 
radially outward during the \textsc{Prometheus} run, but it
is always placed at a location where the velocity of the
infalling matter has become supersonic.}
After the deceleration of the shock in the region of a positive 
gradient of $r^3\rho$ until about 28\,M$_\odot$ (see Fig.~\ref{fig:progenitor}),
the shock accelerates in a region of negative
$r^3\rho$-derivative and experiences no deceleration afterwards,
because the $r^3\rho$-profile flattens towards the
stellar surface (Fig.~\ref{fig:progenitor}). Therefore a reverse shock does not
form before the outgoing shock leaves the star. We mention in
passing that the near-surface layers of the C60C progenitor 
are slowly expanding before they are hit by the outgoing shock.
This is an effect that is connected to the mass-loss episodes
in course of the PPI phases that the
star experiences prior to its collapse. Moreover, the shock 
in none of our simulations including Model C60C-NR is 
strong enough to heat oxygen and silicon to sufficiently high
temperatures for explosive nuclear burning, which we include 
in the long-time runs with the \textsc{Prometheus} code
via a small $\alpha$-network (see Section~\ref{subsec:numerical_setup}).

\subsection{Mass ejection estimates for expanding shocks}
\label{sec:massejection}

In the following we shall attempt to estimate the mass that can 
become unbound when the shock in Model C60C-NR and the acoustic
pulses in Models R80Ar-NR, R80Ar, and C115 reach the outer stellar
layers. Our effort, however, is hampered by several aspects.
First, it is difficult to track the evolution of the outgoing 
sonic pulse with high accuracy, because the wave gets smeared 
and thus its decreasing amplitude enhances the possible influence 
of dissipative numerical effects. Second, the KEPLER progenitor 
models exhibit fairly low resolution in radial space near the 
stellar surfaces because of their use of a Lagrangian grid in 
the mass coordinate. Therefore the density profile is not 
reliably represented in the low-density outer layers of the stars,
in particular because PPISN mass loss stretches expanding mass
zones in radius. Third, due to the mass loss in such PPISN episodes,
the near-surface layers are not strictly in hydrostatic equilibrium.
All these facts limit the possibility to analyse our models in close
connection to the analytical considerations by 
\citet{2018MNRAS.477.1225C, 2018ApJ...863..158C, 2021MNRAS.501.4266L},
and \citet{2021ApJ...908...23M}, where power-law and polytropic hydrostatic
background structures were considered. 

For these reasons we will
refer to a simple criterion by comparing the energy of the 
outgoing shock or sonic pulse with the binding energy of the 
outermost stellar layers. We will consider the mass that possesses
a binding energy equal to the energy of the weak explosion
wave as an upper limit to the amount of matter that can become 
unbound, ignoring possible radiative losses as well as excess
kinetic energy of the outflow at infinity. Moreover, we also
ignore the possibility of additional or enhanced mass loss caused
by the hydrodynamic response of the star to the gravitational-potential 
reduction associated with the neutrino emission from the transiently 
stable PNS (\citealt{1980Ap&SS..69..115N, 2013ApJ...769..109L, 2018MNRAS.477.1225C, 
2018MNRAS.476.2366F}). Because of the short life time of the PNSs 
(they collapse to BHs within less than $\sim$0.6\,s after bounce at the latest; 
Table~\ref{tab:model_property}), the mass equivalent of the total energy 
radiated in neutrinos is less than $\sim$0.1\,M$_\odot c^2$ in all models 
(see Table~\ref{tab:model_property_2}). The maximum kinetic energy
of the acoustic pulse triggered by this decrement of the gravitational 
mass was found to be at most a few $10^{46}$\,erg in the models of 
\citet{2021ApJ...911....6I}. This is orders of magnitude lower than the 
energy of the revived bounce shock or its relic sonic pulse, as discussed
in the following section.

In order to estimate the energy carried by the outgoing shock or 
sonic pulse at a late stage of the \textsc{Prometheus} simulations,
we evaluate the work done by the shock/sonic pulse on a
Langrangian mass shell between enclosed masses $M_1$ and $M_2 > M_1$,
which are passed by the shock/pulse and that we specify individually
for each of our shock-reviving models as labelled in Fig.~\ref{fig:energy_sonic_pulse}.
This work can be written as:
\begin{eqnarray}
W(t;[M_1,M_2]) &=& E_\mathrm{shell}(t;[M_1,M_2]) - E_\mathrm{shell}(0;[M_1,M_2]) \nonumber \\
&=& -\, 4\pi \int_0^t\mathrm{d}t' R_2^2(t')p_2(t')\dot R_2(t') \nonumber \\
&& +\, 4\pi \int_0^t\mathrm{d}t' R_1^2(t')p_1(t')\dot R_1(t') \,,
\label{eq:energy_sonic_pulse_F}
\end{eqnarray}
where $R_1$ and $R_2$ are the radii corresponding to the enclosed 
masses $M_1$ and $M_2$ and $p_1$ and $p_2$ are the corresponding
values of the gas pressure at these radii. Since the shock/pulse running out through the star spherisizes quickly (see discussion above), integrals over spherical volumes for given
enclosed mass values provide Lagrangian information as in the spherically symmetric case.

Figure~\ref{fig:energy_sonic_pulse} displays the time evolution of $W(t;[M_1,M_2]) = E_\mathrm{shell}(t;[M_1,M_2]) - E_\mathrm{shell}(0;[M_1,M_2])$ for our set of relevant models. 
When the outgoing shock/pulse reaches the shell at $R_1$, it compresses the shell and exerts $p\mathrm{d}V$ work, leading to an increase of the total energy contained by the shell.
Inversely, when the shock/pulse leaves the shell at $R_2$, expansion 
work causes a decline of the shell energy again. This effect can be
seen in Fig.~\ref{fig:energy_sonic_pulse} as a transient increase of the energy in the shell. We can thus estimate the work done by the shock in Model C60C-NR
as roughly 7$\times10^{49}$\,erg. The sonic pulses in Models R80Ar-NR, R80Ar, 
and C115 possess, approximately, energies of 4$\times10^{49}$\,erg, 7$\times10^{49}$\,erg, and 4$\times10^{49}$\,erg, respectively. The corresponding time intervals when 
these energies are measured in Models C60C-NR, R80Ar-NR, R80Ar, and
C115 are 20.0--25.0\,s, 1.43--1.48\,s, 4.5--5.0\,s, and 2.5--2.8\,s, 
respectively. There is a general, long-time trend of a decrease of the 
shell energy superimposed on this transient increase of the energy in 
the considered mass shells. This trend is connected to the infall of the shell in
the course of the stellar collapse. Since the inner shell radius $R_1$
falls faster and is associated with a higher pressure, the shell exerts 
$p\mathrm{d}V$ work on the volume at $r < R_1$. This overall trend
is less pronounced in Model C60C-NR than in the other models, because
the outgoing shock in this case heats the shell, i.e., it deposits 
thermal energy, and the $p\mathrm{d}V$ work when the shock exits the
shell is smaller than at its entry.

For coming up with a rough estimate of the mass that can be made
unbound by the outgoing shock or sonic pulse, we assume that our
measured shock/pulse energy is conserved and carried outward to
the near-surface layers. Since the shock in Model C60C-NR dissipates
energy on the way out and heats the swept-up shells, we use the 
energy estimated from the shock's exit of the considered mass shell
in Fig.~\ref{fig:energy_sonic_pulse} as a rough proxy of the available energy for unbinding
outer stellar layers. Again, this implies that our estimated ejecta
mass is an upper limit, because the outgoing shock loses
further energy by dissipative heating of the stellar matter swept 
up on its way to the surface, but most of this heated gas will
ultimately fall back to the BH.

In Models R80Ar-NR and C115 the shock and then the
sonic pulse cannot be well identified after only a short
period of time, and we therefore stopped our simulations quite
early. Also in these models the pulse energies estimated at these 
early times may just be optimistic upper limits of the pulse energy
carried outward, because the rear parts of the acoustic pulses
may be pulled inward in the supersonic infall of stellar matter
whose collapse is triggered by the rarefaction wave from the BH 
formation. Our estimated ejecta masses in all cases should therefore
be considered as generous upper limits. 

Using the shock/pulse energies
mentioned above and equating them with the 
binding energies of the outer stellar layers in our models, we
thus obtain the following upper bounds for the ejecta masses in
Models C60C-NR, R80Ar-NR, R80Ar, and C115:
0.14\,M$_\odot$, 1.1\,M$_\odot$, 3.5\,M$_\odot$, and 0.07\,M$_\odot$,
respectively.\footnote{In Model C60C-NR in particular, but to some
extent also in the other models, the radial resolution of the 
near-surface layers of the progenitor is rather coarse in the 
KEPLER calculations. This is a downside of the expanding radii of
the cells of the Lagrangian mass grid used during the phases of PPISN 
outbursts. For this reason our estimates of the binding energies of the
outermost stellar layers are only crude. In Model C60C-NR these layers 
also expand continuously before the shock reaches the surface (see
Figs.~\ref{fig:shock_radius} and \ref{fig:mass_shell_C60C_NR}). 
This effect is caused by non-vanishing velocities in the aftermath 
of a PPISN phase, but it may partly also be connected to an imperfect 
representation of the poorly resolved pressure
gradient after mapping from KEPLER to \textsc{Prometheus}. 
Because of these shortcomings we refrain from determining
the ejecta mass and energy directly by the hydrodynamical results
of our simulation of the shock breakout in Model C60C-NR.}

The masses that might be stripped off the stars when the shock or 
sonic pulse reaches the surface of the stars are quite considerable,
although they are only small fractions of the total stellar masses at 
the onset of collapse and will reduce the masses of the newly forming
BHs only moderately. Our simple criterion for the expelled masses does
not allow us to predict the kinetic energies associated with
the ejecta. But even if most of the energy of the shock or sonic pulse
(several $10^{49}$\,erg, see above)
is consumed for unbinding the outermost stellar layers and only a minor
part ends up as kinetic energy, luminous outbursts of radiation 
could be the consequence when this kinetic energy is dissipated in
the collision of the ejecta with the circumstellar shells generated
during the PPISN phases. The duration of such displays will be 
affected, however, by the fact that no radioactive nuclei such as 
$^{56}$Ni are thrown out in the final explosions, because all of the
neutrino-heated matter as well as the shock-heated material in the
core of the star fall back and are swallowed by the BH. Mass ejection 
and optical light curves for very low energy Type-II SNe have been 
calculated by \citet{Lovegrove+2017}, and these might have similar 
characteristics.

\begin{figure}
	\includegraphics[width=0.48\textwidth]{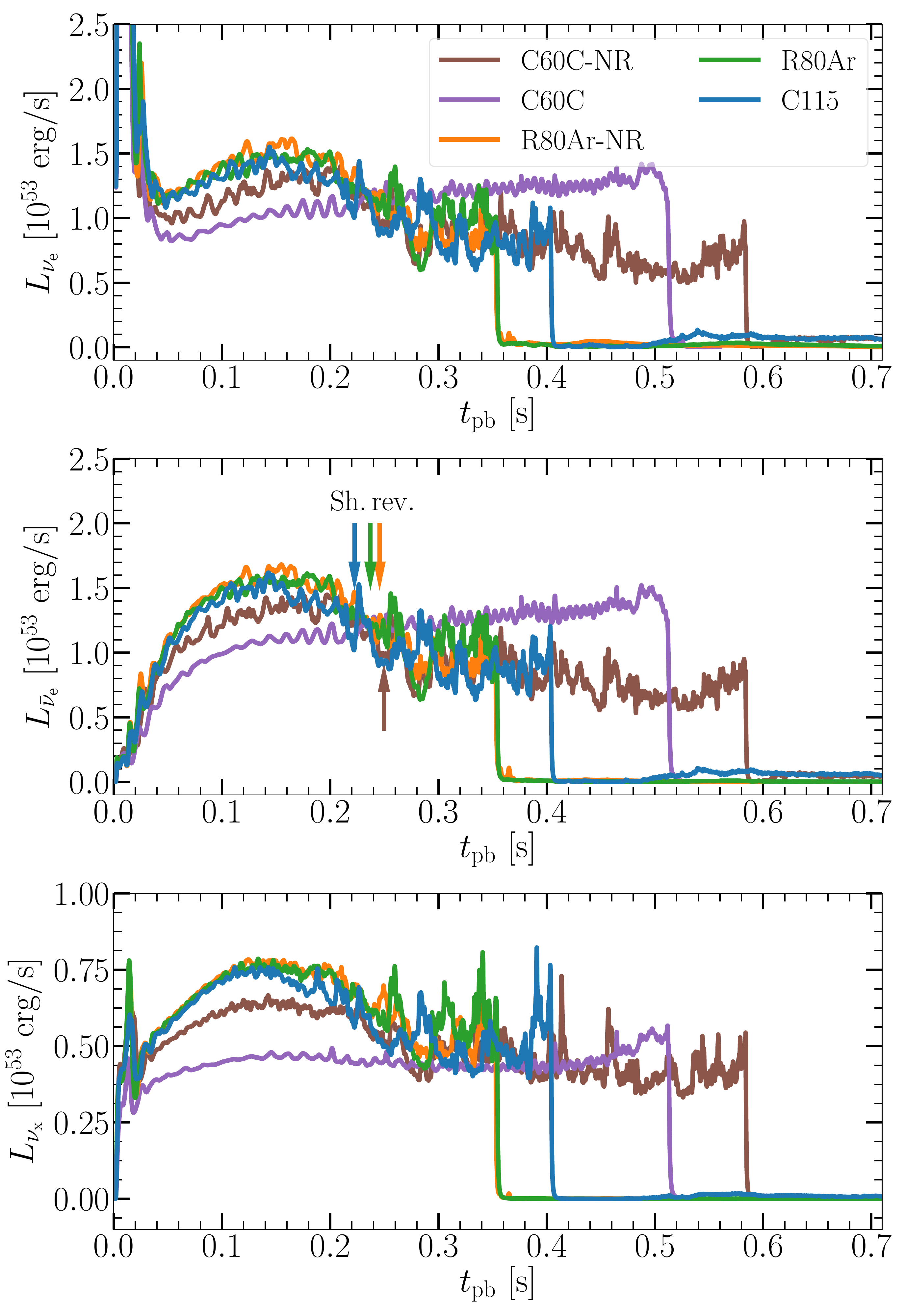}
	\vspace*{-7mm}
	\caption{Time evolution of the luminosities (spherical integrals as defined by equation \eqref{eq:luminosity}) of electron neutrinos (top), electron antineutrinos (middle), and a single species of heavy-lepton neutrinos (bottom) for an observer at rest at infinity, evaluated at a radius of 500 km. The color scheme for the different models is the same as in Fig.~\ref{fig:shock_radius}. Before the onset of neutrino-driven shock expansion and the corresponding decline of the accretion luminosities, the electron neutrino and antineutrino luminosities are higher in the shock-reviving Models C115, R80Ar-NR, R80Ar, and C60C-NR than in the rapidly rotating Model C60C, where neutrino heating does not trigger shock expansion. The arrows in the middle panel indicate the times of shock revival and the steep drop of the neutrino luminosities marks the instant of BH formation. The corresponding mean neutrino energies and RMS energies are shown in Figs.~\ref{fig:emean} and \ref{fig:erms}, respectively.}
	\label{fig:luminosity}
\end{figure}
\begin{figure}
	\includegraphics[width=0.48\textwidth]{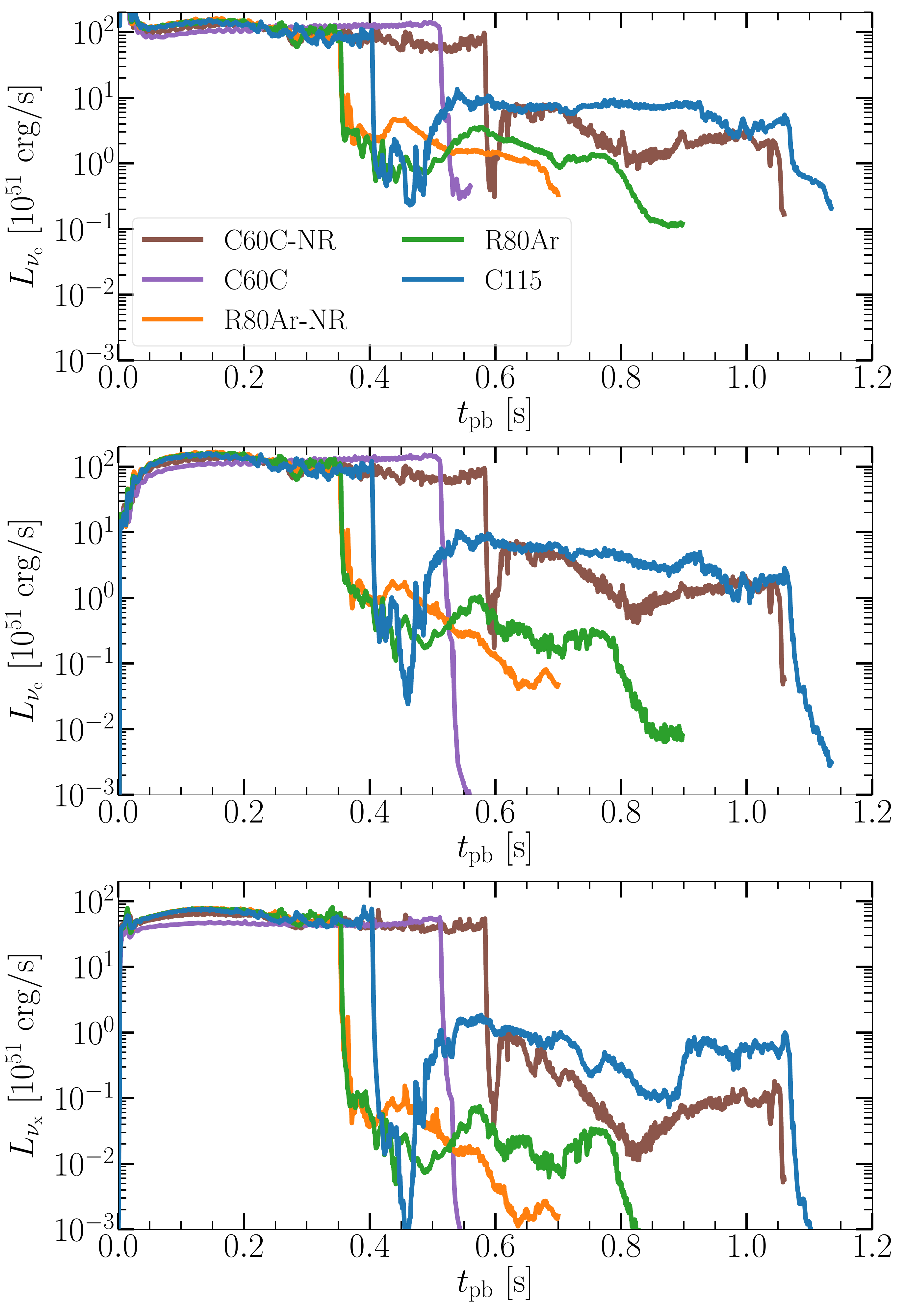}
	\vspace*{-7mm}
	\caption{Same as Fig.~\ref{fig:luminosity}, but with a logarithmic scale for the luminosities and the abscissa extended to later times after BH formation. The luminosities drop strongly after BH formation but continue on a one to two orders of magnitude lower level because of aspherical accretion of neutrino-heated and shock-heated gas by the BH. The final decline of the luminosities is connected to the effectively spherical collapse of the stellar layers that fall inward from increasingly larger radial distances.}
	\label{fig:luminosity_abhf}
\end{figure}
\begin{figure}
	\includegraphics[width=0.48\textwidth]{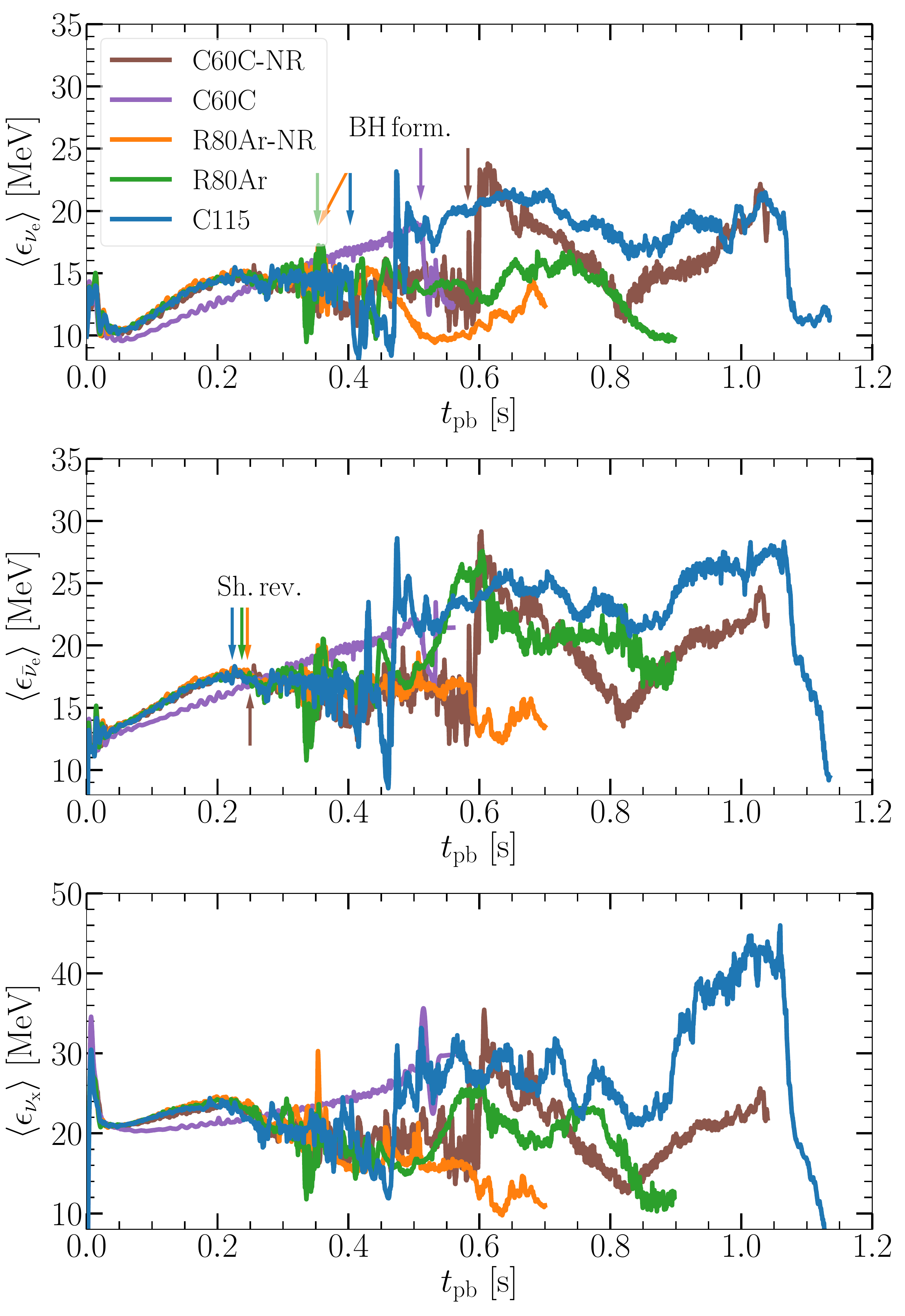}
	\vspace*{-7mm}
	\caption{Time evolution of the radiated mean neutrino energies (spherical averages as defined by equation~\eqref{eq:emean}) of electron neutrinos (top), electron antineutrinos (middle), and heavy-lepton neutrinos (bottom) for an observer at rest at infinity, evaluated at a radius of 500\,km. The color scheme for the different models is the same as in Fig.~\ref{fig:shock_radius}. The arrows mark the times of BH formation (top panel) and shock revival (middle panel).}
	\label{fig:emean}
\end{figure}
\begin{figure}
	\includegraphics[width=0.48\textwidth]{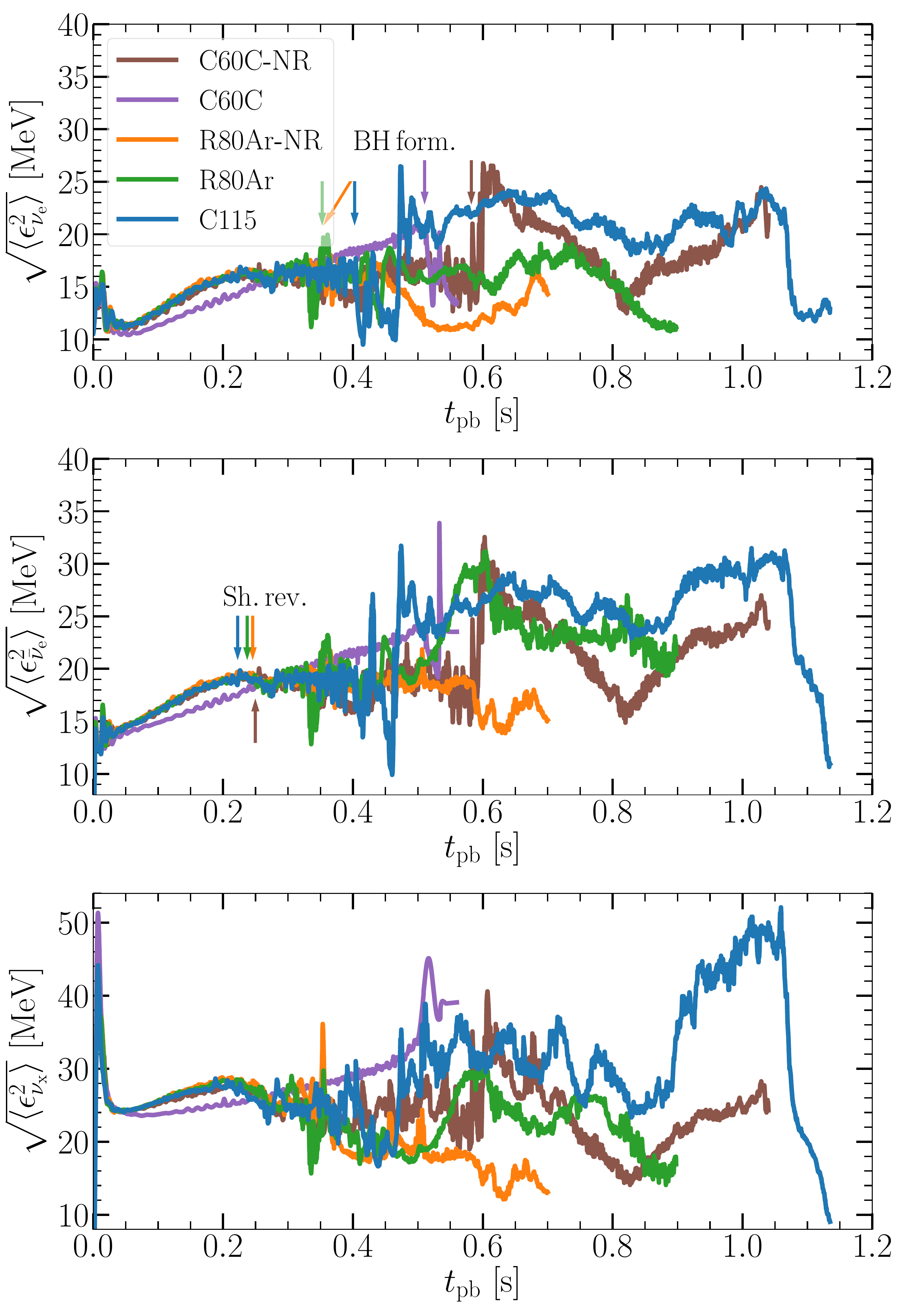}
	\vspace*{-7mm}
	\caption{Same as Fig.~\ref{fig:emean} but for the RMS energies of the neutrino number flux (spherical averages as defined in equation~\eqref{eq:erms}) for an observer at rest at infinity.}
	\label{fig:erms}
\end{figure}
\begin{figure*}
	\includegraphics[width=\textwidth]{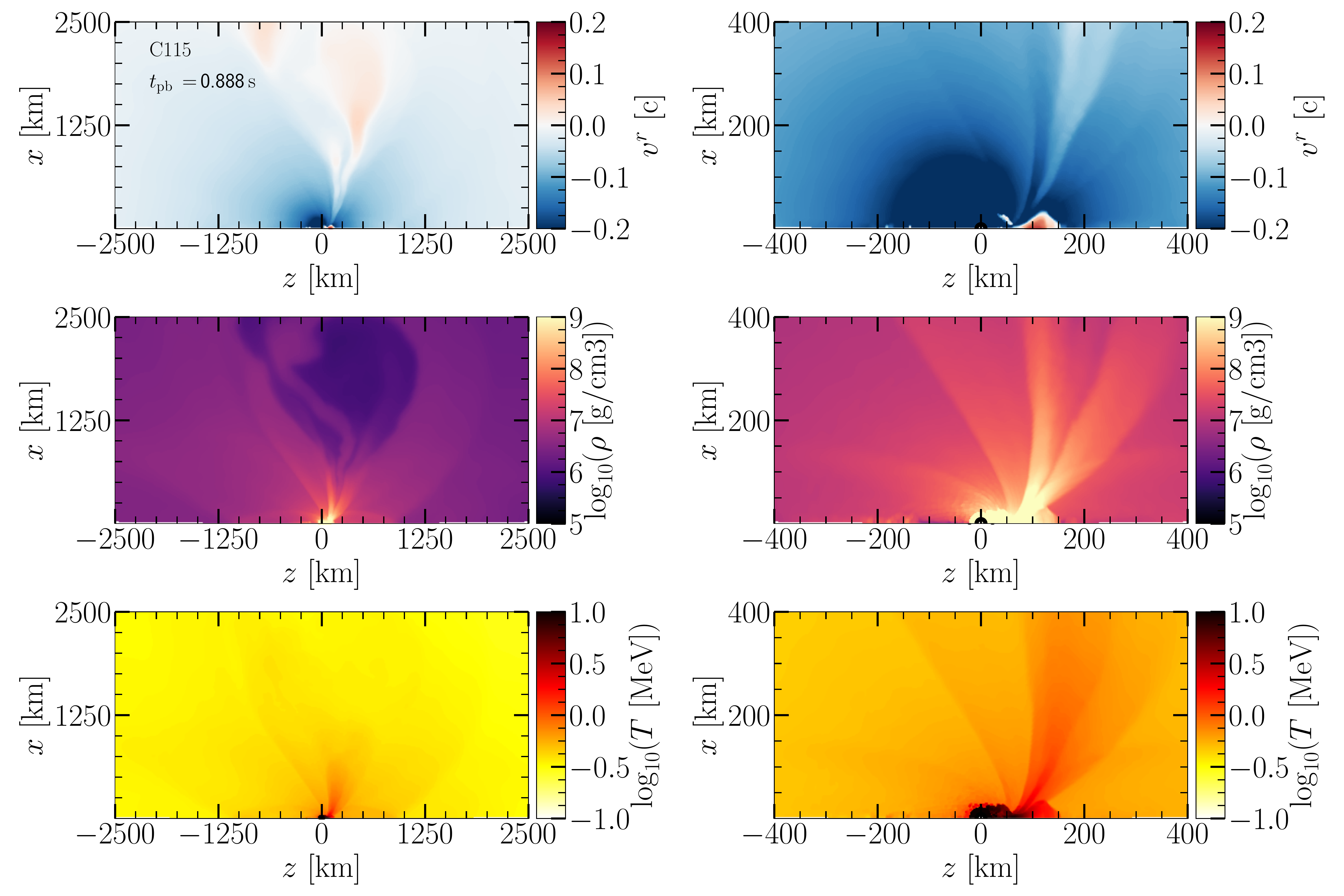}
	\vspace*{-7mm}
	\caption{Cross-sectional cuts of the radial velocity (top), density (middle), and temperature (bottom) for Model C115 at an exemplary time of 0.888\,s after bounce, which is nearly 0.5\,s after the BH formation in this model. The right panels are close-ups of the left panels, showing the immediate vicinity of the BH. The symmetry axis ($z$-axis) of the 2D simulation is plotted as abscissa. In right panels, the BH radius is marked by black lines. A buoyant, low-density, high-temperature and high-entropy plume near the equatorial plane (see also Figure~\ref{fig:contour_all_model_BH}, bottom panel) funnels the collapsing matter mainly towards one pole, where the converging flow is compressed into a high-temperature and high-density hot-spot region that obstructs the accretion. Because of the outward push of this clump of matter, shocks occur at the interfaces between the high-entropy plume and the accretion flows falling towards the center from larger distances.}
	\label{fig:c115_contour_abhf_hydro}
\end{figure*}
\begin{figure}
	\includegraphics[width=0.48\textwidth]{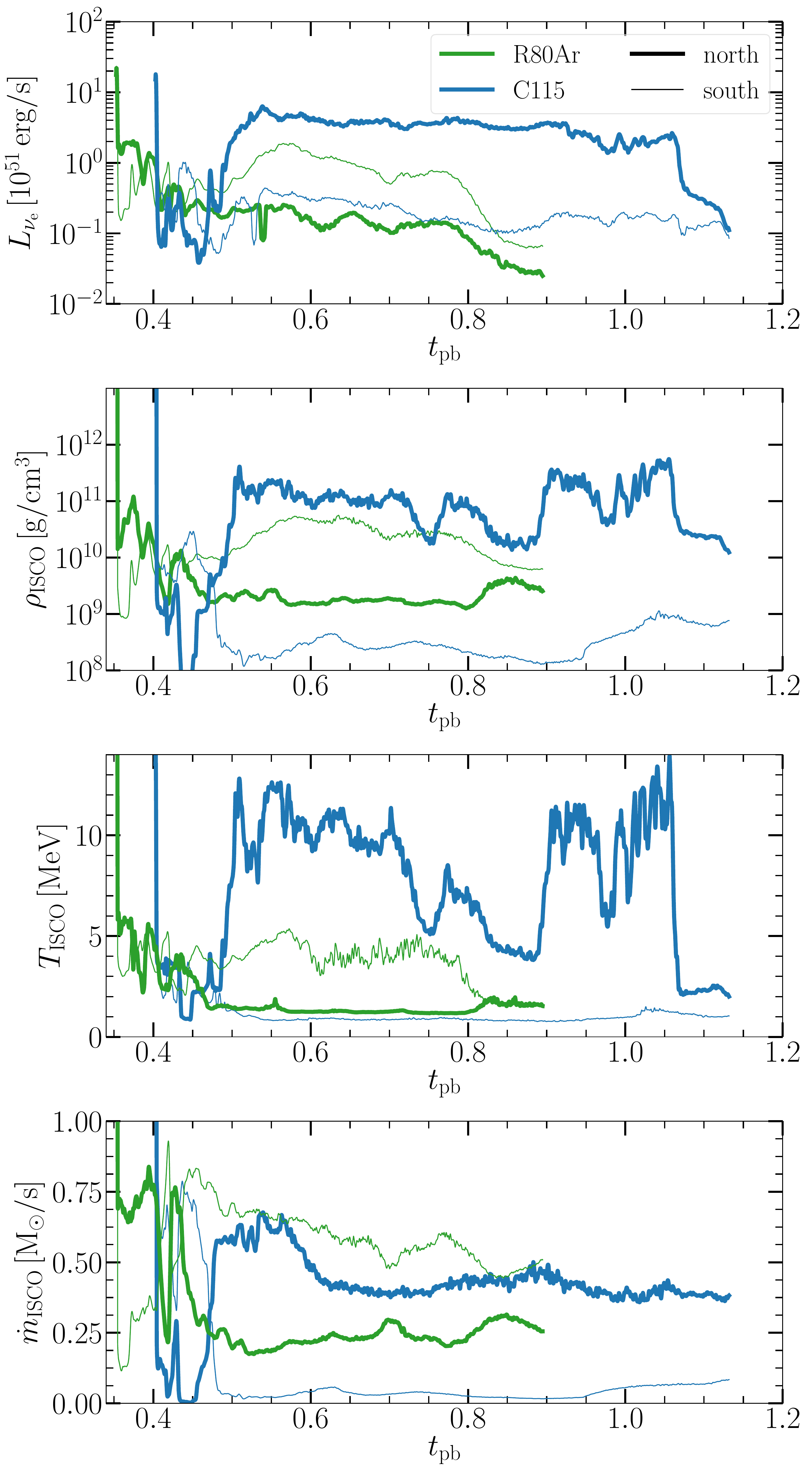}
	\vspace*{-7mm}
	\caption{Time evolution of the hemispherically averaged values of quantities characterizing the asymmetric accretion and neutrino emission in Models R80Ar (green lines) and C115 (blue lines) after BH formation. Top: Lab-frame electron neutrino luminosity evaluated at $r=100$\,km according to equation (32); second row: density at the ISCO; third row: temperature at the ISCO; bottom: hemispheric mass-accretion rate at the ISCO. Quantities of the northern hemisphere are shown by thick lines, those of the southern hemisphere by thin lines.}
	\label{fig:neutrino_hemidiff}
\end{figure}
\begin{figure}
	\includegraphics[width=0.48\textwidth]{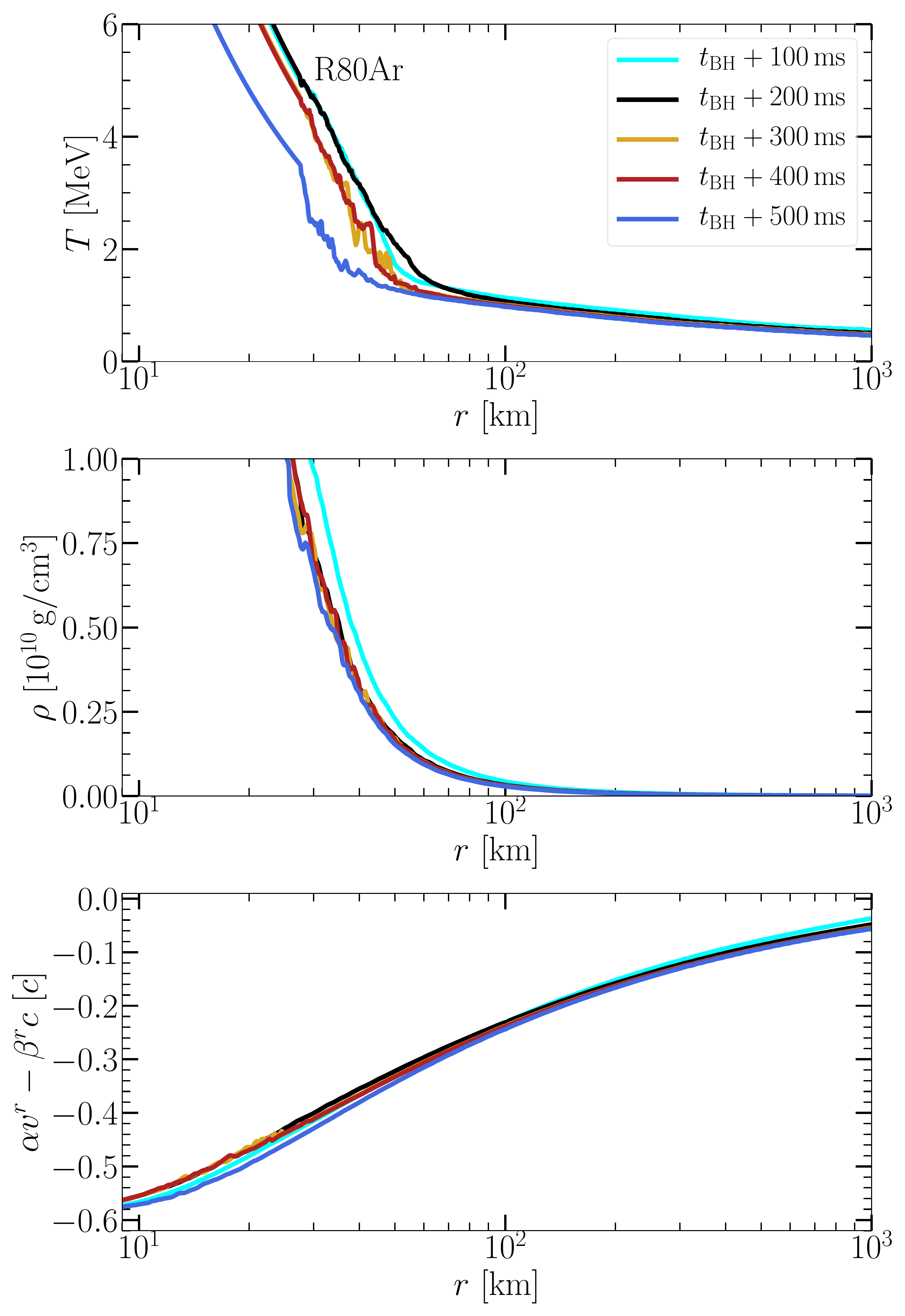}
	\vspace*{-7mm}
	\caption{Radial profiles of the angle-averaged temperature (top panel), density (middle panel), the radial fluid velocity in the local rest frame, $\alpha v^{r}/c-\beta^{r}$, for Model R80Ar at 100 ms (cyan), 200 ms (black), 300 ms (yellow), 400 ms (red), and 500 ms (blue) after BH formation. The plots show the immediate vicinity around the BH, which has a radius of less than 10\,km. Here, $v^{r}$, $\beta^{r}$, and $\alpha$ are the angle-averaged radial fluid velocity, the radial shift vector, and the lapse function, respectively.}
	\label{fig:r80_norot_radial_profile_hydro}
\end{figure}
\begin{figure}
	\includegraphics[width=0.48\textwidth]{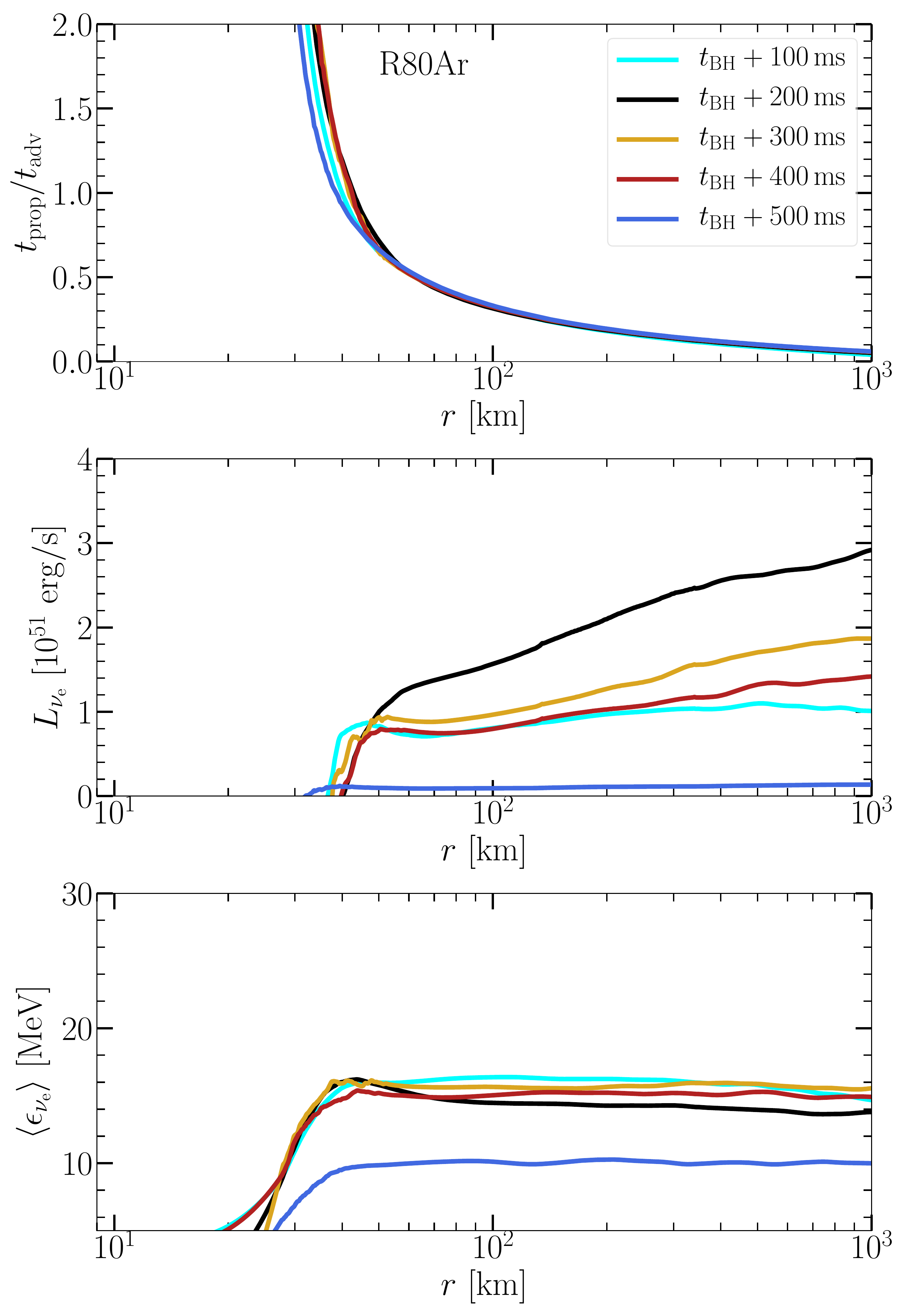}
	\vspace*{-7mm}
	\caption{Radial profiles of angle-averaged quantities for electron neutrinos in Model R80Ar at 100 ms (cyan), 200 ms (black), 300 ms (yellow), 400 ms (red), and 500 ms (blue) after BH formation. Top panel: ratio of neutrino propagation timescale, $t_\mathrm{prop}$ (see equation \eqref{eq:neutrino_propagation_timescale}), and the advection timescale, $t_\mathrm{adv}$ (see equation \eqref{eq:neutrino_advection_timescale}). Middle panel: lab-frame neutrino luminosity. Bottom panel: lab-frame mean neutrino energy of the neutrino number flux. The plots show the immediate vicinity around the BH, which has a radius of less than 10\,km.}
	\label{fig:r80_norot_radial_profile_neutrino}
\end{figure}

\section{Neutrino emission and BH kicks}
\label{sec:neutrino_properties}

In this section we will discuss the neutrino signals of our models,
including ---for the first time--- the emission properties beyond the 
instant of BH formation as well as the implications of anisotropic
neutrino emission for BH kicks. To this end we continue the \textsc{NADA-FLD}
simulations for several 100\,ms beyond the time when the PNS collapses
to a BH and also beyond the mapping times for the \textsc{Prometheus} 
simulations ($t_\mathrm{map}$, as listed in Table~\ref{tab:model_property_2}), in order to
determine the (lower-level) neutrino emission connected to aspherical
accretion by the BH.

\subsection{Neutrino signal before BH formation}

Figures~\ref{fig:luminosity} and \ref{fig:luminosity_abhf} provide the ($4\pi$-integrated) luminosities in the lab frame for all neutrino species (according to equation~\eqref{eq:luminosity}) and Figs.~\ref{fig:emean} and \ref{fig:erms} display the corresponding mean energies and the RMS energies, respectively, of the neutrino number emission (according to equations~\eqref{eq:emean} and \eqref{eq:erms}). 
As usual, the neutrinospheric emission of the PNS can be imagined to be composed of a core component, which is fed by neutrinos diffusing out from the neutrino-opaque high-density
core of the PNS, and an accretion component, which originates from
the less opaque and semi-transparent, hot PNS mantle. The accretion
luminosity $L_\mathrm{acc}$ depends on the mass accretion rate 
$\dot M$ and the gravitational potential of the PNS and can be coined
as:
\begin{eqnarray}
L_\mathrm{acc} = \xi\,\frac{GM_\mathrm{ns}\dot M}{R_\mathrm{ns}}\,,
\label{eq:lum_rel}
\end{eqnarray}
where $M_\mathrm{ns}$ and $R_\mathrm{ns}$ are the PNS mass and radius, 
respectively. The dimensionless factor $\xi$ is of order unity and is
found to be around 0.5 for all of our models before the neutrino-driven
shock expansion sets in (with $\dot M$ measured at $r = R_\mathrm{ns}$),
in agreement with previous studies by \citet{2014ApJ...788...82M}.

Because of the high mass accretion rates (Fig.~\ref{fig:mass_accretion}) and
the correspondingly rapidly growing PNS mass and shrinking PNS radius
(Fig.~\ref{fig:NS_property}), the accretion luminosities of $\nu_\mathrm{e}$ and 
$\bar\nu_\mathrm{e}$ during the first 200\,ms after bounce are very
high, namely for each of these neutrinos up to more than 
$(1.3-1.5)\times 10^{53}$\,erg/s (Figs.~\ref{fig:luminosity}, 
\ref{fig:luminosity_abhf}), before the neutrino-driven shock revival 
happens in Models C60C-NR, R80Ar-NR, R80Ar, and C115. When the 
neutrino-driven expansion of the shock sets in (Fig.~\ref{fig:shock_radius}),
the mass accretion onto the PNS is reduced and the luminosities begin to
decline gradually. During the phase of maximum accretion luminosities
of electron neutrinos and antineutrinos, the muon and tau neutrino
luminosities and those of their antineutrinos have (individually)
about half the size of the $\nu_\mathrm{e}$ and $\bar\nu_\mathrm{e}$
luminosities. The mean energies and RMS energies roughly follow the 
trend in the time evolution of the luminosities. The mean (RMS) 
energies level off after the onset of shock revival and before BH
formation, in which phase they reach up to $\sim$15\,MeV ($\sim$18\,MeV) for 
$\nu_\mathrm{e}$, $\sim$18\,MeV ($\sim$20\,MeV) for $\bar\nu_\mathrm{e}$,
and 23--25\,MeV (27--29\,MeV) for heavy-lepton neutrinos 
(Figs.~\ref{fig:emean}, \ref{fig:erms}).

The rapidly rotating and non-exploding Model C60C exhibits considerably
lower luminosities and mean energies of all neutrino species during the 
phase prior to shock expansion in the other models (Fig.~\ref{fig:luminosity}). 
This tendency is partly witnessed also for the neutrino luminosities of its non-rotating counterpart C60C-NR already, because both models possess lower mass accretion rates
and a correspondingly more slowly growing PNS mass than the rest of
our model set (see Figs.~\ref{fig:mass_accretion} and \ref{fig:NS_property}). The bigger effect reducing the neutrino luminosities, however, is connected to the centrifugal 
deformation of the PNS in Model C60C, whose average radius is much bigger
than in all other models. This centrifugal stretching also decreases the
average temperature near the neutrinosphere, mainly close to the equator,
for which reason the ($4\pi$-averaged) mean energies and RMS energies of Model C60C are
significantly lower than those in all the other models before these
latter models experience shock revival (Figs.~\ref{fig:emean}, \ref{fig:erms}; see also
Section~\ref{subsec:impact_of_rotation}). The slow rotation of Model R80Ar does not cause any
systematic or significant differences in the neutrino emission 
properties compared to the non-rotating case of R80Ar-NR. Differences
between these two models are therefore likely to be of stochastic nature.

In contrast to all other models, the steady increase of the radiated
$\nu_\mathrm{e}$ and $\bar\nu_\mathrm{e}$ luminosities and of the mean 
energies of all neutrino species continues in Model C60C from some
10\,ms after bounce until BH formation, because shock expansion is absent in this model
and the PNS mass grows monotonically by a high rate of accretion, 
along with the monotonic contraction of the PNS radius (Figs.~\ref{fig:mass_accretion}, \ref{fig:NS_property}). Since heavy-lepton neutrinos are not efficiently produced in the 
accretion mantle of the PNS but leak out mostly from the PNS core
and decouple deeper inside the PNS, their luminosities in Model C60C 
behave differently with time than those of $\nu_\mathrm{e}$ and 
$\bar\nu_\mathrm{e}$. The $\nu_\mathrm{x}$ luminosities in C60C
reach a flat, broad peak with nearly constant level between about 100\,ms and 200\,ms after bounce, then decline only slightly over the next $\sim$200\,ms, before they rise again
during the last $\sim$100\,ms before BH formation. This time evolution
is explained by the fact that the core-emission of heavy-lepton 
neutrinos obeys approximately the Stefan-Boltzmann law for blackbody
radiation, scaling with $R_\mathrm{ns}^2 T_\nu^4$ for a neutrinospheric
temperature $T_\nu$. Therefore there is a competition between shrinking 
PNS radius and rising neutrinospheric temperature (reflected by the 
steadily increasing mean neutrino energies), which leads to a nearly
constant $\nu_\mathrm{x}$ luminosity for roughly 400\,ms, and only 
shortly before BH formation the more rapid growth of the temperature
in the compressed PNS core wins and triggers a moderate rise of the
$\nu_\mathrm{x}$ luminosity. 

The neutrino luminosities and mean energies of electron neutrinos
and antineutrinos, and to a smaller extent also those of the heavy-lepton
neutrinos, exhibit quasi-periodic large-amplitude fluctuations before
shock revival with minimum-maximum variations of up to 20\% of the
$4\pi$-averaged $\nu_\mathrm{e}$ and $\bar\nu_\mathrm{e}$ luminosities.
These fluctuations are correlated with periods of shock expansion
and contraction during the post-bounce accretion phase
(see Fig.~\ref{fig:shock_radius}), and they are caused by large-scale SASI and/or
convective mass motions in the postshock layer, which modulate the 
accretion flow between shock and PNS (see, e.g., \citealt{2009A&A...496..475M} for similar results in non-exploding 2D models of lower-mass progenitors). In the non-exploding Model C60C the variations of the neutrino luminosities and mean energies are
particularly regular and continue until BH formation with a steadily
rising frequency. This is compatible with strong SASI activity in this 
model, whose dominance is fostered by the continuous contraction of
the shock and PNS radii (Figs.~\ref{fig:shock_radius}3, \ref{fig:NS_property}5), which leads to a growing frequency of the SASI sloshing motions of $f_\mathrm{SASI} \propto
R_\mathrm{sh}^{-3/2}[\ln(R_\mathrm{sh}/R_\mathrm{ns})]^{-1}$ 
\citep{2014ApJ...788...82M}. The neutrino emission peaks twice in one 
SASI cycle when matter is channelled onto the PNS during the shock
contraction phases in both hemispheres. Therefore the oscillations
in the $4\pi$-averaged neutrino luminosities and mean energies 
appear with twice the SASI frequency.

After shock revival and before BH formation in Models C60C-NR, R80Ar-NR,
R80Ar, and C115, the excursions in the luminosities and mean energies
of all kinds of neutrinos become more irregular in time and their 
amplitudes much larger (minimum-maximum variations up to 50\% of the 
$4\pi$-averaged luminosities) than before shock revival. This phenomenon
is connected to the stochastically occurring accretion downflows that 
impact on the PNS surface and that can be temporarily constricted
or quenched by the rising plumes of neutrino-heated matter. The
violence of these downflows and therefore the variation amplitude of 
the neutrino-emission properties grows with time as the shock
expands and the downflow funnels carry matter towards the PNS from 
increasingly larger distances and with increasingly higher infall 
velocities.

\subsection{Neutrino signal after BH formation}

When the PNS collapses to a BH, the luminosities of all neutrino species
plummet by at least one to two orders of magnitude to average levels of 
$10^{51}$--$10^{52}$\,erg/s for $\nu_\mathrm{e}$ and $\bar\nu_\mathrm{e}$
and below $\sim$\,$10^{51}$\,erg/s for heavy-lepton neutrinos. 
The steep drop takes place within only a few milliseconds.
In Fig.~\ref{fig:luminosity_abhf} the luminosities are displayed logarithmically
to improve the visibility of the emission after the BH formation. 
The mean and RMS energies do not show any such dramatic decline at the 
time when the BH forms (Figs.~\ref{fig:emean}, \ref{fig:erms}). 
During the subsequent evolution the mean energies of electron 
neutrinos and antineutrinos
remain on roughly constant levels in Models R80Ar and R80Ar-NR,
drop by $\sim$5\,MeV in Model C60C, or even increase by more than
$\sim$5\,MeV in Models C60C-NR and C115 (for $\bar\nu_\mathrm{e}$ this
happens even more extremely also in Model R80Ar roughly 
200\,ms after BH formation). But in all cases they continue to 
exhibit large fluctuations. For heavy-lepton neutrinos there is a 
tendency of an increase of the mean energies during the evolution
after BH formation in all models except in R80Ar-NR and C60C, but
a growing amplitude of excursions of the mean energies of
$\nu_\mathrm{x}$ after BH formation is an even more conspicuous
phenomenon.

These time-dependent features in the mean energies of the radiated
neutrinos correlate with temporal changes of the luminosities of 
$\nu_\mathrm{e}$ and $\bar\nu_\mathrm{e}$ over up to one to two
orders of magnitude and of heavy-lepton neutrinos over up to 
three orders of magnitude. Such dramatic fluctuations are connected
to large variations of the mass accretion rate by the new-born
BH, which receives matter from massive downflows that
penetrate to the BH anisotropically between the still existing,
extended high-entropy plumes of neutrino-heated matter 
(see Fig.~\ref{fig:contour_c60_norot}). The accretion is extremely 
variable because the downflows are unstable and unsteady in their 
locations due to their interaction
with the surrounding bubbles. They can dive more or less directly into
the BH, thus not efficiently emitting neutrinos because of the short
time for experiencing such energy loss. 
But they may also collide with each other to accumulate mass in
hot clumps and high-density belts around the BH or above the poles
of the BH. 

An example is shown in Fig.~\ref{fig:c115_contour_abhf_hydro} for Model~C115 at 0.888\,ms after bounce, when the $\nu_\mathrm{x}$ luminosity in this model increases by an order of
magnitude (Fig.~\ref{fig:luminosity_abhf}) and the mean energy of the emitted 
$\nu_\mathrm{x}$ displays a steep rise to a very high peak (Figs.~\ref{fig:emean}, \ref{fig:erms}). 
Figure~\ref{fig:c115_contour_abhf_hydro} visualizes how a massive, centrally converging accretion downdraft, which is confined by a rising high-entropy plume of neutrino-heated gas, feeds a high-density, high-temperature region above the north pole of the BH. This dense
clump of matter there has a length of more than 100\,km in $z$-direction 
and a diameter of more than 50\,km in the perpendicular directions.
It is surrounded by accretion shocks, and the shocked-heated, compressed
gas reaches densities of up to several $10^{11}$\,g/cm$^3$ and temperatures 
up to 14\,MeV (see Fig.~\ref{fig:neutrino_hemidiff}), efficiently radiating high-energy muon
and tau neutrinos and antineutrinos created through electron-positron
pair annihilation. These neutrinos can escape effectively unhindered
from the environment, which has densities lower than the typical
neutrinospheric densities ($10^{13}$\,g/cm$^3$ and higher) of 
heavy-lepton neutrinos.\footnote{We need to point out here that pair 
production processes are taken into account in our transport solver only 
for heavy-lepton neutrinos (see Table~\ref{tab:neutrino_opacity}). 
While usually charged-current beta-processes dominate the production of $\nu_\mathrm{e}$ and 
$\bar\nu_\mathrm{e}$, it is possible that in the low-density,
high-temperature environment of the compressed accretion flows
also the production of $\nu_\mathrm{e}$-$\bar\nu_\mathrm{e}$ pairs
by $\mathrm{e}^+\mathrm{e}^-$ annihilation might contribute 
significantly to the creation of high-energy electron neutrinos
and antineutrinos. The presence of a large number density of positrons
and the corresponding relevance of positron captures on neutrons is 
suggested by a local peak of the $\bar\nu_\mathrm{e}$ luminosity
in Model C115 around 0.9\,s after bounce (see Fig.~\ref{fig:luminosity_abhf}). It is therefore possible that our results underestimate the luminosities of 
$\nu_\mathrm{e}$ and $\bar\nu_\mathrm{e}$, specifically connected 
to an underestimation of the high-energy tails of their spectra.}

Figures~\ref{fig:r80_norot_radial_profile_hydro} and \ref{fig:r80_norot_radial_profile_neutrino} visualize in
more detail the typical physical conditions and neutrino-emission properties
in the accretion flow to the BH by angle-averaged profiles
of Model~R80Ar at different times over half a second after BH formation. 
This model was chosen as another exemplary case for plotting the angle-averaged
profiles, because the hemispheric differences are not quite as extreme in this
case as they are in Models~C60C-NR and C115.
Density and temperature exhibit considerable variations with time (Figs.~\ref{fig:neutrino_hemidiff}, \ref{fig:r80_norot_radial_profile_hydro}), which is a consequence of the accretion fluctuations described
above. In contrast, the profile of the radial velocity in local rest frames (Fig.~\ref{fig:r80_norot_radial_profile_hydro}, bottom
panel) is nearly uniform, because the angular average is dominated by the
rapidly infalling downdrafts. These move under free-fall conditions,
which change only slowly because of the modest growth of the BH mass
over the considered time span. 

Neutrinos that are freshly produced in the accretion flow can escape
from the infalling matter only as long as their outward propagation 
timescale is shorter than the inward advection timescale.
The former is defined by
\begin{eqnarray}
	t_\mathrm{prop} = \frac{W N\,\Delta V}{\alpha F^r\,\Delta A}\,,
	\label{eq:neutrino_propagation_timescale}
\end{eqnarray}
where $W$, $\alpha$, $N$, and $F^r$ are the Lorentz factor, the lapse function, the angle-averaged number density and
the angle-averaged radial number flux density of neutrinos, respectively,
in the comoving frame for a spherical shell with (outer) surface area
$\Delta A$ and volume $\Delta V$. The neutrino advection timescale for
this spherical shell located at radius $r$ and with an infall velocity
of $\alpha v^r - \beta^r c$ is estimated by
\begin{eqnarray}
	t_\mathrm{adv} = \frac{r}{\alpha v^r - \beta^r c}\,,
	\label{eq:neutrino_advection_timescale}
\end{eqnarray}
where $v^r$ and $\beta^r$ are the
radial fluid velocity and the radial shift vector, respectively. 
Here, $v^r$ and $\alpha v^r - \beta^r c$ are the radial fluid velocity in the Eulerian and local rest frame observer (coordinate observer), respectively (see, e.g., \citealt{baumgarte_shapiro_2010} for a discussion of Eulerian observer quantities).

Figure~\ref{fig:r80_norot_radial_profile_neutrino} shows that the electron neutrino
luminosity as measurable by a distant observer in the lab frame 
begins to rise steeply around the radius where
$t_\mathrm{prop}/t_\mathrm{adv}$ drops below unity. The 
corresponding mean energy peaks roughly where the luminosity rise levels
off, and it decreases towards the BH because of gravitational 
redshifting and neutrino trapping, which permits only low-energy 
neutrinos to leak out. The slower rise of the luminosities at distances 
$r \gtrsim 100$\,km is not connected to local neutrino production
at such large distances, where the density and temperature are too low
for efficient neutrino reactions. Instead, it is an effect of the
time-dependent evolution of the luminosities. 
The average density in the region where $t_\mathrm{prop}/t_\mathrm{adv} \sim 1$
is below $10^{10}\,$g/cm$^3$, which implies that the neutrinos in this
region are not trapped and thus not dragged inward despite the fast motion
of the infalling matter. The average temperature in this region is only
around 2--3\,MeV, whereas the mean energy of the escaping electron neutrinos
is (redshifted) around 14--16\,MeV at early times after the BH formation
(see also Fig.~\ref{fig:emean}). This suggests high values of the electron degeneracy
(degeneracy parameters $\eta_\mathrm{e}\gg 1$) in the compressed, 
low-entropy, unshocked accretion downdrafts, which favor the emission
of non-thermal $\nu_\mathrm{e}$. High degeneracy also suppresses the presence of 
positrons in the infalling flows and therefore quenches
the emission of electron antineutrinos (produced by positron captures) 
and heavy-lepton neutrinos, whose luminosities in Models R80Ar and R80Ar-Nr are considerably lower than the $\nu_\mathrm{e}$ luminosity, for heavy-lepton neutrinos by roughly two orders of magnitude (see Fig.~\ref{fig:luminosity_abhf}). 

Models C60C-NR and C115 constitute exceptions to such conditions 
because of the collision 
and shock-heating of converging accretion flows in the vicinity of the
BH (see Fig.~\ref{fig:c115_contour_abhf_hydro} and discussion above). Since the
thus formed clumpy regions of decelerated, partly expanding, shock-heated
gas around the BH or above its poles produce neutrinos very efficiently,
the $\nu_\mathrm{e}$ luminosities in these two models after the
BH formation are roughly five times higher (around $10^{52}$\,erg/s)
than in Models R80Ar and R80Ar-NR. Similarly, the $\bar\nu_\mathrm{e}$
luminosities are only moderately lower than those of $\nu_\mathrm{e}$,
and the heavy-lepton neutrino luminosities are only one order of magnitude
lower instead of two orders of magnitude in Models R80Ar and R80Ar-NR.
Also the mean energies of all species of emitted neutrinos, but in 
particular those of electron antineutrinos and heavy-lepton neutrinos, 
are considerably higher in Models C60C-NR and C115. This reflects the 
contributions from the hot-spot emission by the high-temperature,
high-density clumps near the BH. 

Presently it is not clear how these
effects depend on the 2D nature of our simulations with their artificial
constraint of axisymmetry, in which the accretion downflows are toroidal
sheets of matter instead of 3D funnels. The latter can move in any angular
direction and not just in the latitudinal direction as in the 2D case. 
Future 3D core-collapse calculations of BH forming stars, also of
rotating very massive progenitors, are needed to answer the question
whether our results for the neutrino emission after BH formation
are independent of the dimensionality of the modeling. In particular
the emission phases of high neutrino luminosities in combination
with very high energies of the radiated neutrinos are an interesting 
post-BH formation phenomenon witnessed for the first time in our simulations.

Overall, despite the extreme variations between different models and
despite the huge variability in time, the steep drop of the neutrino 
luminosities that happens generally at the time when the PNS collapses 
to a BH (Fig.~\ref{fig:luminosity_abhf})
is explained by the decreasing densities and the rapid infall
of the matter accreted into the BH. This also implies a dramatic decline of
the neutrino heating of the remaining high-entropy plumes, because the
reduced neutrino luminosities are not compensated by the rise of
the mean energies of the radiated neutrinos. The plumes therefore lose
support by neutrino energy input at their base, and the gravitational pull 
by the BH begins to decelerate the outward expansion of the gas in the plumes
until the gas motion is reversed to infall and all of the initially 
neutrino-heated matter in our models gets swallowed by the newly formed BHs 
(see Section~\ref{sec:result_after_BH_formation} for a detailed discussion).

Finally, after most of the plumes have fallen back, the accretion of the
overlying stellar layers becomes essentially spherical and the neutrino
luminosities experience another sharp decline with a steep negative 
derivative at around the times when we stop our NADA-FLD simulations
(Fig.~\ref{fig:luminosity_abhf}). This happens latest, namely at about 1.1\,s
after bounce, in Models C60C-NR and C115, which have the highest 
diagnostic explosion energies initially and the longest survival time
of the high-entropy plumes (Fig.~\ref{fig:shocked_layer_properties}). 
For $\nu_\mathrm{e}$ the luminosities then fall to $\sim$\,$10^{50}$\,erg/s,
and for all other species to even much lower values,
because the collapsing matter is degenerate and does not get shocked
before passing the BH radius, thus producing only $\nu_\mathrm{e}$ 
at relevant rates. This final drop in Model C60C, where shock revival
is absent, happens within only a few milliseconds after
the BH forms at 0.51\,s post bounce. Rotation in this model is not
sufficiently fast to permit AD formation from any relevant amount 
of matter at this early epoch of the evolution, i.e., the BH accretion 
is effectively spherical also in this model shortly after BH formation
(see Section~\ref{subsec:result_after_BH_formation_NADA}).

\begin{figure}
	\includegraphics[width=0.48\textwidth]{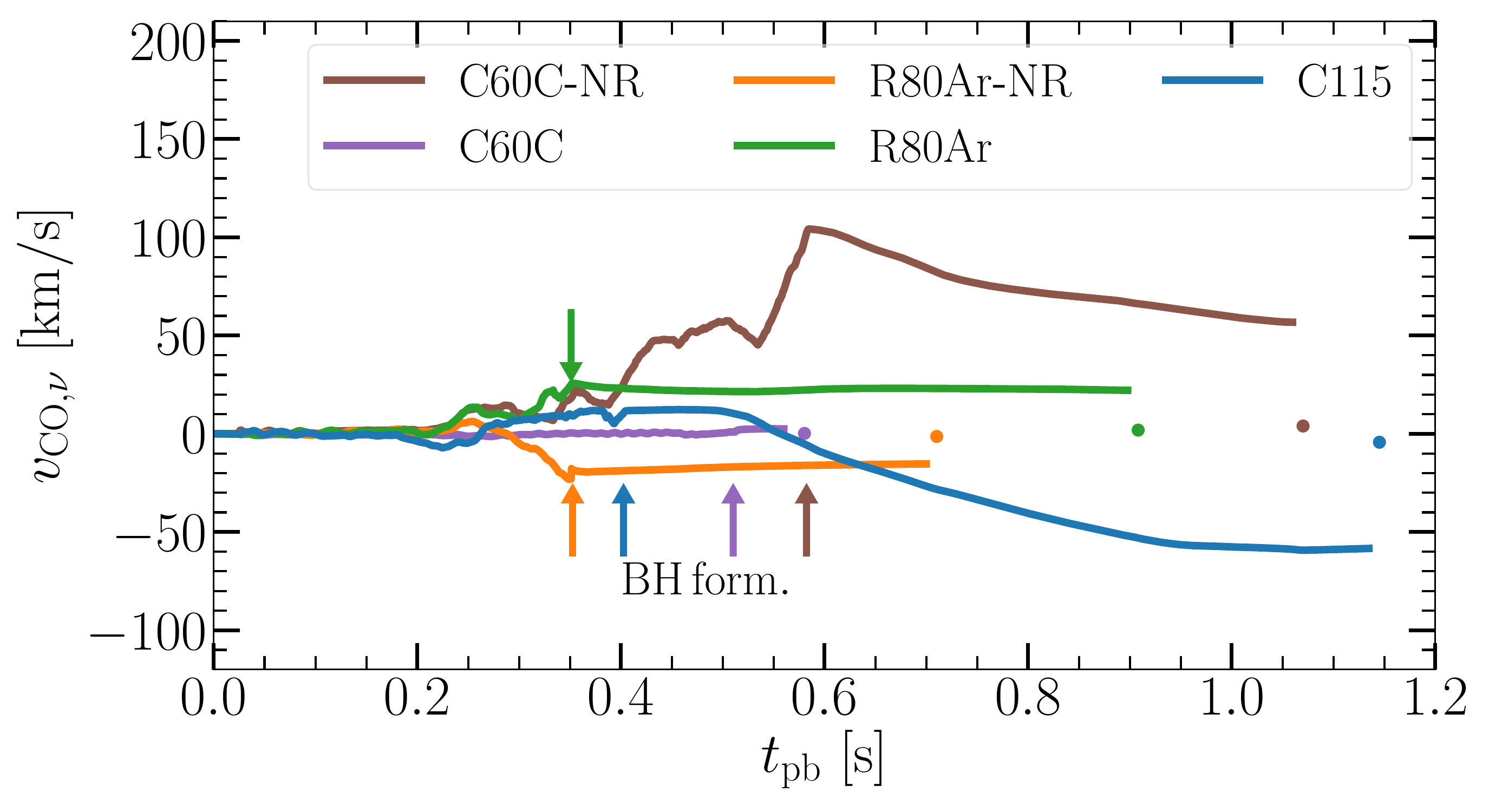}
	\vspace*{-7mm}
	\caption{Time evolution of the neutrino-induced kick velocities of the compact remnants in our set of models. Because aspherical mass ejection is not expected to be triggered by the outward propagating shocks (or sonic pulses) in the shock-reviving models, asymmetric neutrino (and GW) emission is the only mechanism to kick the relic compact object. The kick velocity  typically grows due to the high neutrino luminosities until the PNS collapses to a BH. Afterwards it declines because of momentum conservation and the growing mass of the BH accreting infalling stellar matter. In Model C115 the kick velocity continues to grow even after the BH formation because of still considerable, asymmetric neutrino luminosities and a relatively lower gain rate of mass by the BH compared to the other models (see also Figs.~\ref{fig:c115_contour_abhf_hydro}, \ref{fig:neutrino_hemidiff}). The final kick velocities (after accretion of the entire, not ejected progenitor mass) are marked by dots and are 3.95\,km/s, 0.18\,km/s, $-$1.38\,km/s, 1.86\,km/s, $-$4.31\,km/s for Models C60C-NR, C60C, R80Ar-NR, R80Ar, and C115, respectively. The arrows mark the times of BH formation.}
	\label{fig:neutrino_driven_kick}
\end{figure}

\subsection{BH kicks by anisotropic neutrino emission}

Since the initial, neutrino-heated ejecta with their hydrodynamic
asymmetries fall back to the BH entirely, there is no asymmetric 
mass ejection from our models connected to the explosion mechanism.
Therefore, the BH can receive a natal kick only by asymmetric 
emission of neutrinos and GWs. We focus here on the former, because
neutrinos carry away many orders of magnitude more energy and momentum
from collapsing stars than GWs do (see also Section~\ref{sec:gravitational_waves}). 

Figure~\ref{fig:neutrino_driven_kick} displays the kick velocities of the
compact remnants due to anisotropic neutrino emission for all of our
simulations as functions of time. They are computed by following the
analysis in Appendix~D of Stockinger et al. (2020), i.e., by time-integrating
equation~(D2) there and dividing the thus obtained time-dependent momentum
associated with asymmetrically emitted neutrinos by the instantaneous
baryonic mass of the NS or BH as in equation~(D3).

In Models R80Ar-NR, R80Ar, and C60C the kick velocities level off 
at the times of BH formation at values that can reach 20--30\,km/s and
then begin to monotonically decline on long timescales. There are two
contributing reasons for this behavior. On the one hand the 
low or modest asymmetry of the remaining neutrino emission from the
accretion flows to the BH implies that the recoil momentum obtained
until the instant of BH formation is effectively the final value. On
the other hand the spherically symmetric (or equatorially symmetric
in the rotating Models R80Ar and C60C) collapse of 
fallback matter does not change the BH momentum but just leads to 
a continuous growth of the BH mass. Both effects together are responsible
for the slow monotonic decrease of the BH's kick velocity and the very
low final kicks of only up to 2\,km/s
(marked by dots in Fig.~\ref{fig:neutrino_driven_kick}).
We note that our final kick velocities are upper limits, because they
are computed with fallback masses that are corrected for our optimistic
estimates of the mass loss triggered by the breakout of the shock wave 
or sonic pulse from the stellar surface (see Section~\ref{sec:massejection}).

Models C60C-NR and C115 exhibit a different evolution compared to the
other models (Fig.~\ref{fig:neutrino_driven_kick}). Their neutrino-induced kick velocities are initially much higher, and even after the collapse and accretion
of all gravitationally bound stellar matter, the BH kicks are still
4--5\,km/s and thus higher than in the other models. 
In Model C60C-NR a kick velocity in excess of 100\,km/s is
transiently reached. The main acceleration of the compact object sets
in at about 400\,ms after bounce and proceeds in two episodic steps until
the moment when the PNS collapses to a BH at 580\,ms post bounce. At
that time the growth of the kick velocity ends and its typical decline
for the accreting BH sets in. The episodic increase is connected to the extreme
asymmetry between northern and southern hemisphere in Model C60C-NR 
before and around the BH formation, where a strong downflow exists in 
the southern hemisphere and plumes of neutrino-heated matter expand
mostly above the equator in the northern hemisphere (see Fig.~\ref{fig:contour_all_model_BH}).
The stronger neutrino emission by the episodic accretion in the southern
hemisphere kicks the compact remnant in the northward direction. 

In contrast, Model C115 experiences
the main acceleration more than 100\,ms after the BH formation. The
kick velocity transiently reaches a value around $-$60\,km/s between 
1000\,ms and 1100\,ms after bounce. Only afterwards the neutrino emission
plummets (Figs.~\ref{fig:luminosity_abhf}, \ref{fig:neutrino_hemidiff}) and the slow decline of
the kick velocity of the spherically accreting BH sets in.
This special evolution is explained by the pronounced north-south
asymmetry of the accretion and neutrino emission in Model C115 after 
the NS has collapsed to the BH, as shown
by Figs.~\ref{fig:c115_contour_abhf_hydro} and \ref{fig:neutrino_hemidiff}. Compared to Model R80Ar, for 
example (in both cases the BHs form at similar times around or before
400\,ms p.b.; Table~\ref{tab:model_property}), C115 possesses a much more extreme north-south asymmetry of the accretion properties and much higher neutrino luminosities. This is particularly obvious when one compares the 
hemispheres where the accretion and neutrino emission are stronger.
In Model C115 the temperatures there reach up to nearly 15\,MeV
and the densities more than $5\times 10^{11}$\,$\mathrm{g/cm^3}$ (Fig.~\ref{fig:neutrino_hemidiff}). Again, this
can be understood by the shock heating of the converging and 
colliding accretion downflows above the north pole of the BH in this 
model (see Fig.~\ref{fig:c115_contour_abhf_hydro}).
Model C60C-NR also radiates neutrinos asymmetrically after BH formation,
similar to Model R80Ar but even with higher luminosities (Fig.~\ref{fig:luminosity_abhf}).
Nevertheless, a growth of the kick velocity after BH formation cannot
be witnessed in C60C-NR, because its neutrino luminosities, especially
those of the heavy-lepton neutrinos, remain considerably below the 
luminosities in Model C115 during most of the time (Fig.~\ref{fig:luminosity_abhf}),
whereas the mass accretion rates by the BHs in both models
are nearly the same (see Fig.~\ref{fig:mass_accretion}).

Despite the low final values of the BH kick velocities (Fig.~\ref{fig:neutrino_driven_kick}),
the transient kicks can have the interesting consequence that the BH is pushed
out of the center of the collapsing star, which in extreme cases might influence
the fallback and accretion of matter by the BH \citep[for a discussion, see][]{Janka+2021}.

\begin{figure}
    \includegraphics[width=0.48\textwidth]{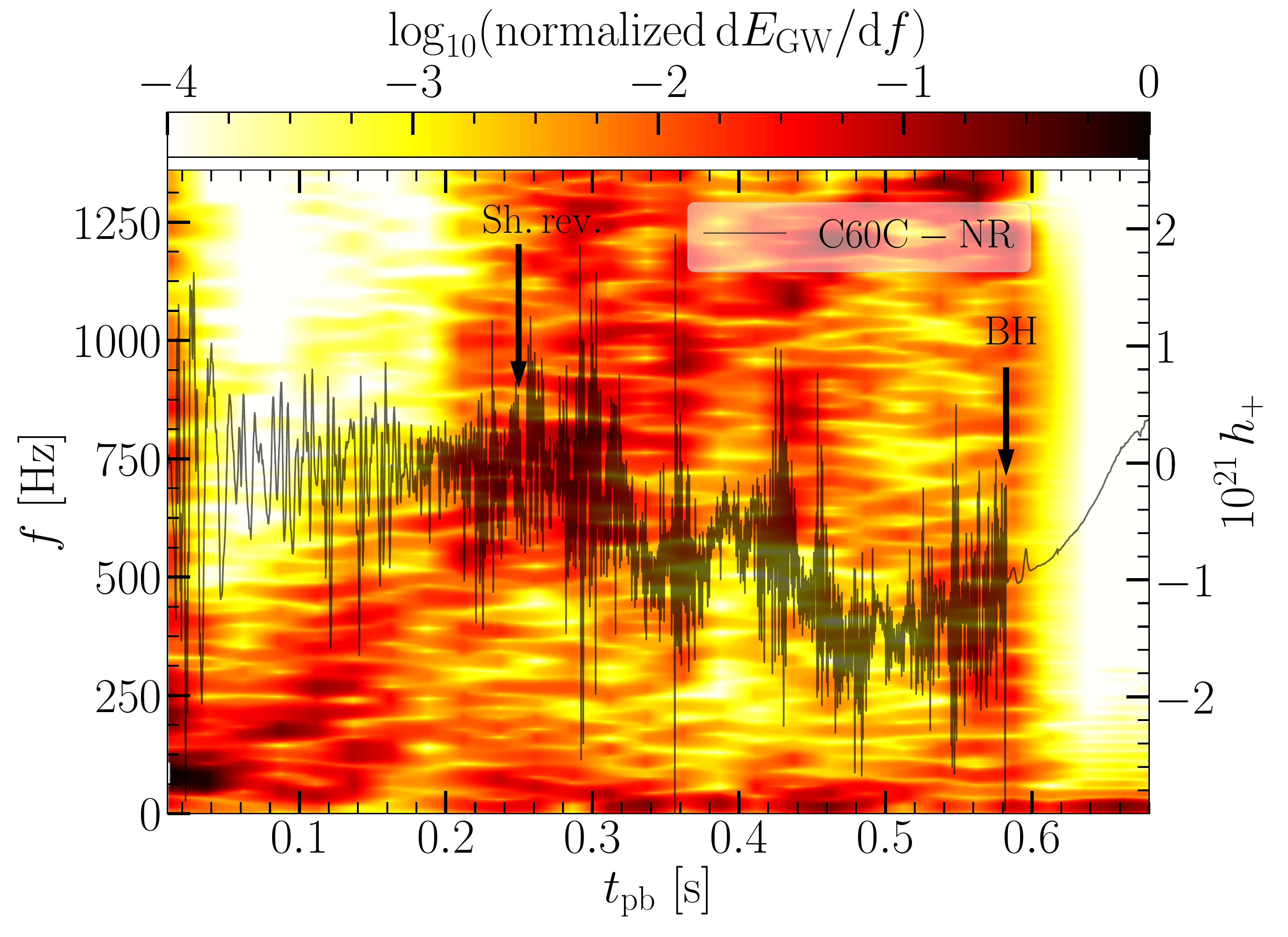}
    \vspace*{-7mm}
    \caption{Time evolution of the dimensionless GW strain (grey line) and the GW spectrogram for the non-rotating Model C60C-NR. The assumed distance between GW source and detector is 10\,kpc. The times of shock revival and BH formation are marked by arrows. The short-time Fourier transform for the spectrogram applies a sliding window of 50\,ms. The spectral energy density, $\mathrm{d}E_\mathrm{GW}/\mathrm{d}f$, is normalized by its maximum value.}
    \label{fig:GW_c60_norot}
\end{figure}
\begin{figure}
    \includegraphics[width=0.48\textwidth]{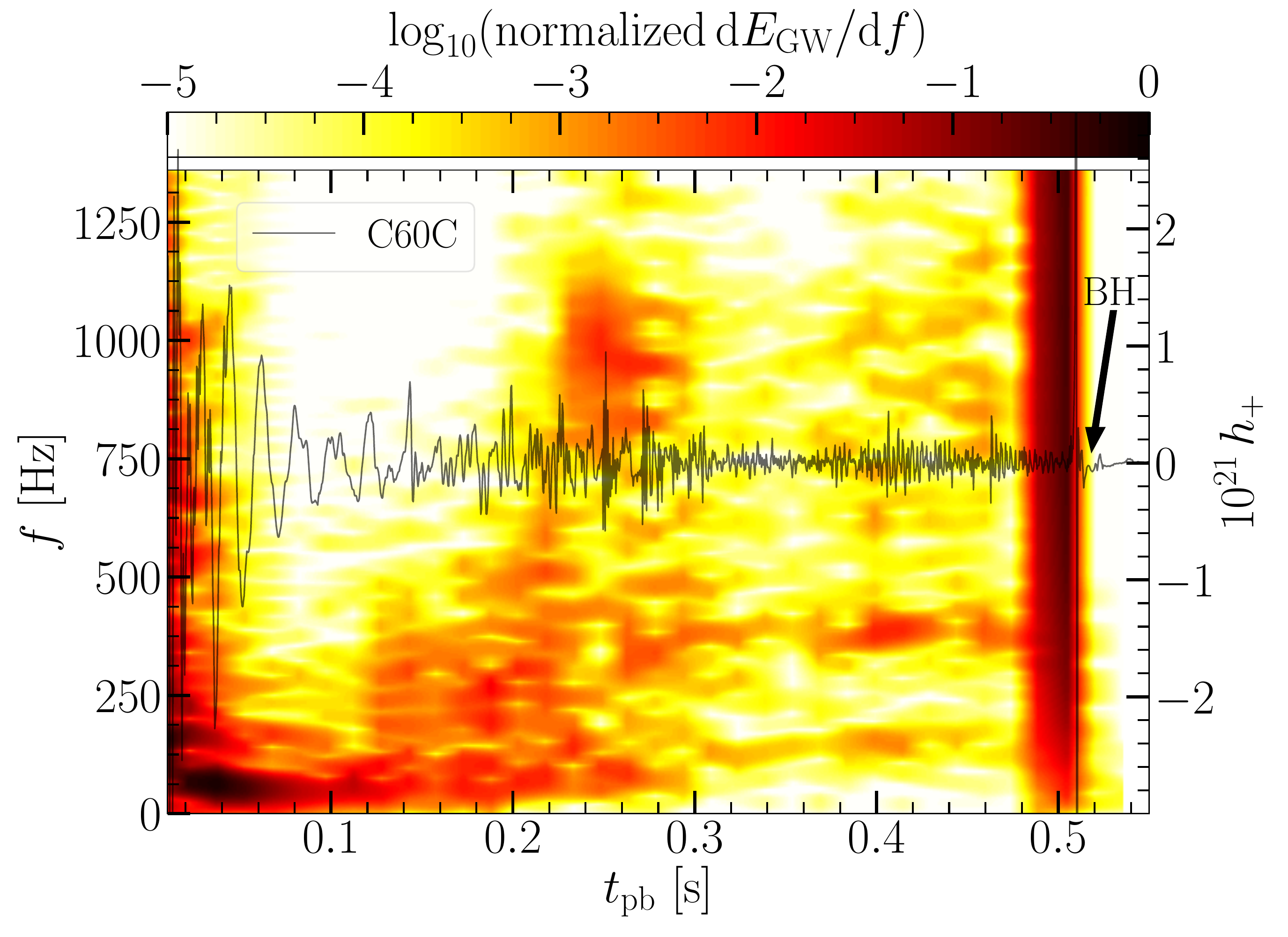}
    \vspace*{-7mm}
    \caption{Same as Fig.~30 but for the rapidly rotating Model C60C, in which shock revival does not happen.}
    \label{fig:GW_c60_rot}
\end{figure}
\begin{figure}
    \includegraphics[width=0.48\textwidth]{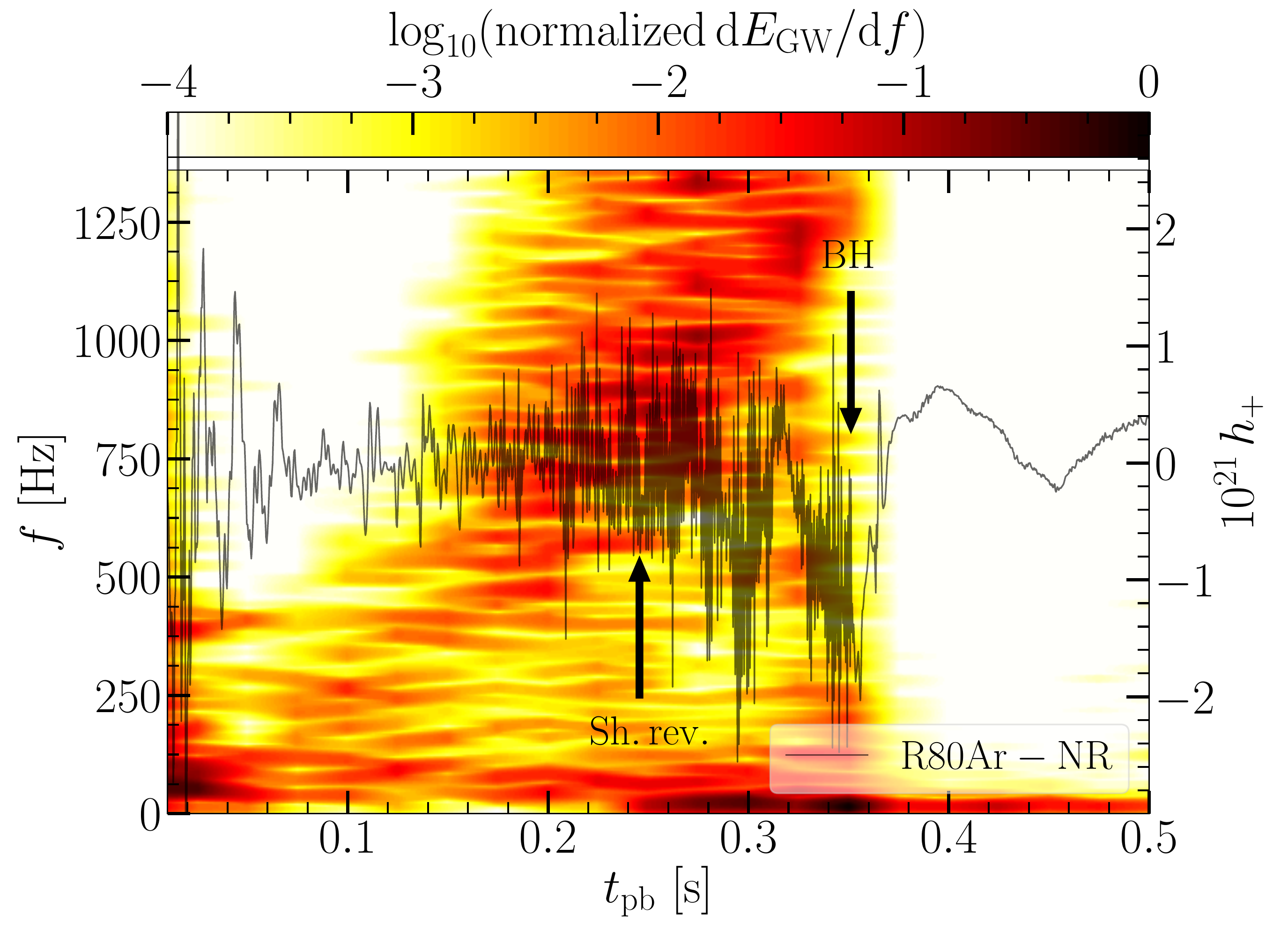}
    \vspace*{-7mm}
    \caption{Same as Fig.~30 but for the non-rotating Model R80Ar-NR.}
    \label{fig:GW_r80_norot}
\end{figure}
\begin{figure}
    \includegraphics[width=0.48\textwidth]{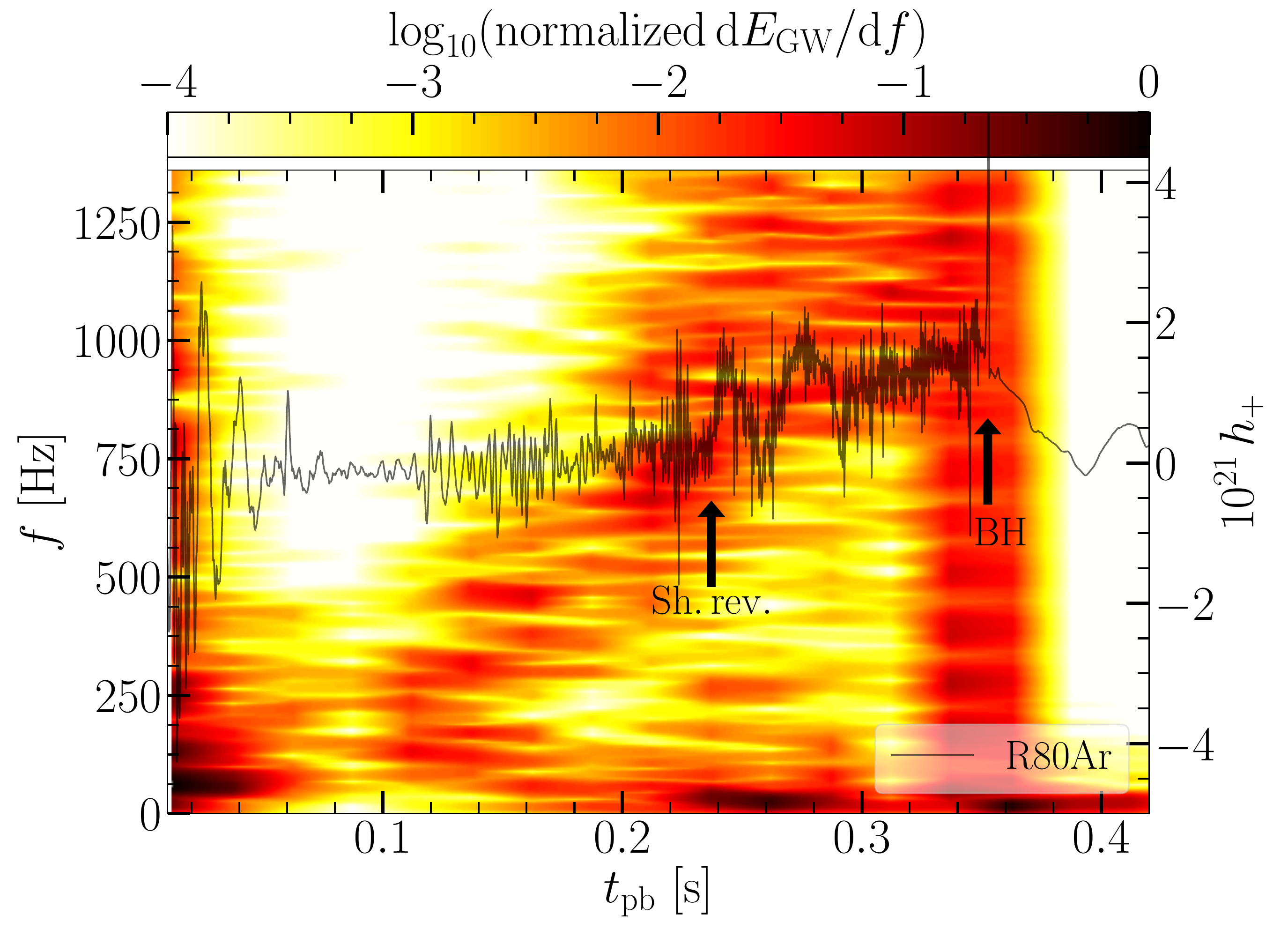}
    \vspace*{-7mm}
    \caption{Same as Fig.~30 but for the slowly rotating Model R80AR.}
    \label{fig:GW_r80_rot}
\end{figure}
\begin{figure}
    \includegraphics[width=0.48\textwidth]{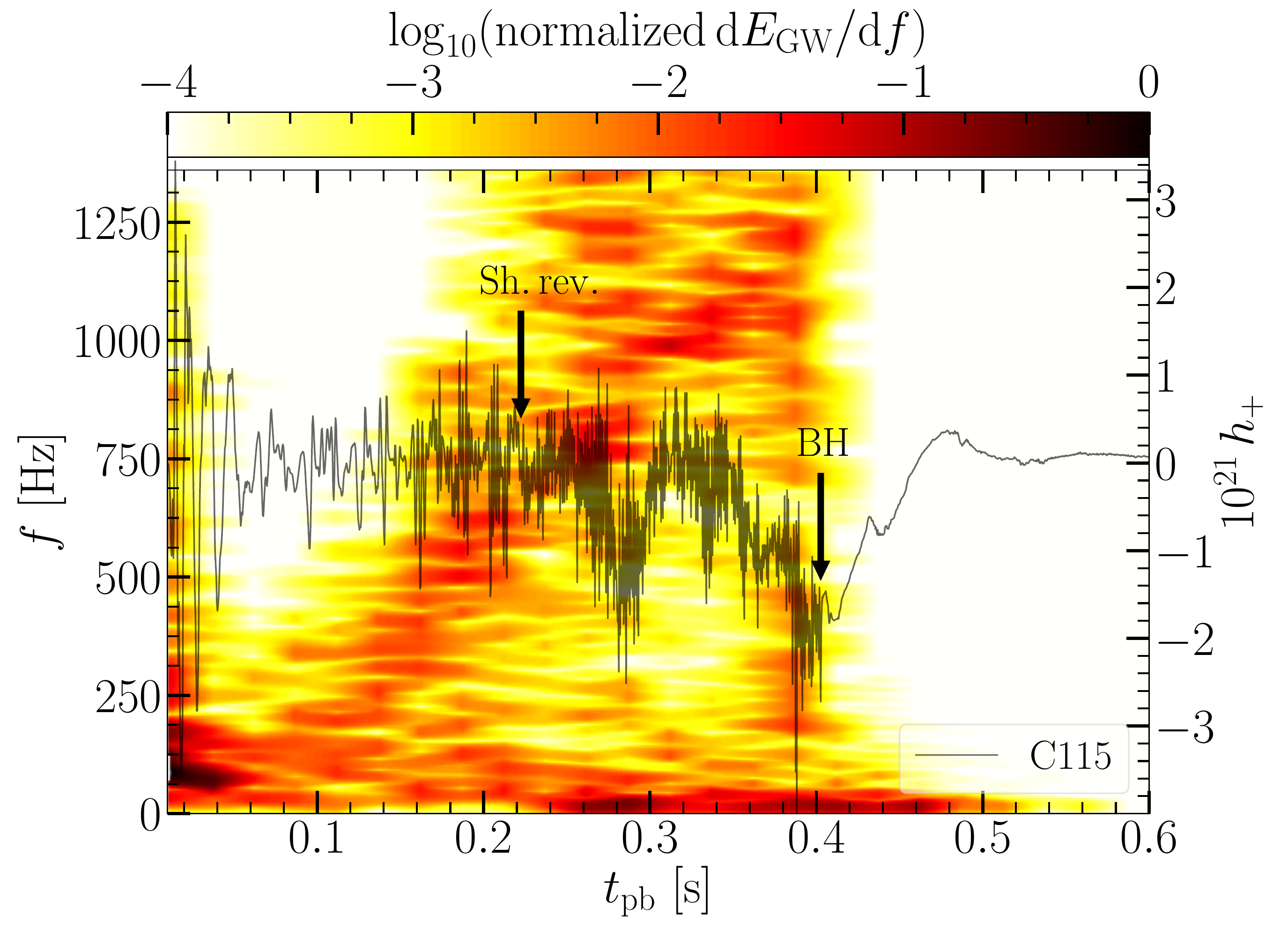}
    \vspace*{-7mm}
    \caption{Same as Fig.~30 but for the non-rotating Model C115.}
    \label{fig:GW_c115}
\end{figure}
\begin{figure}
    \includegraphics[width=0.48\textwidth]{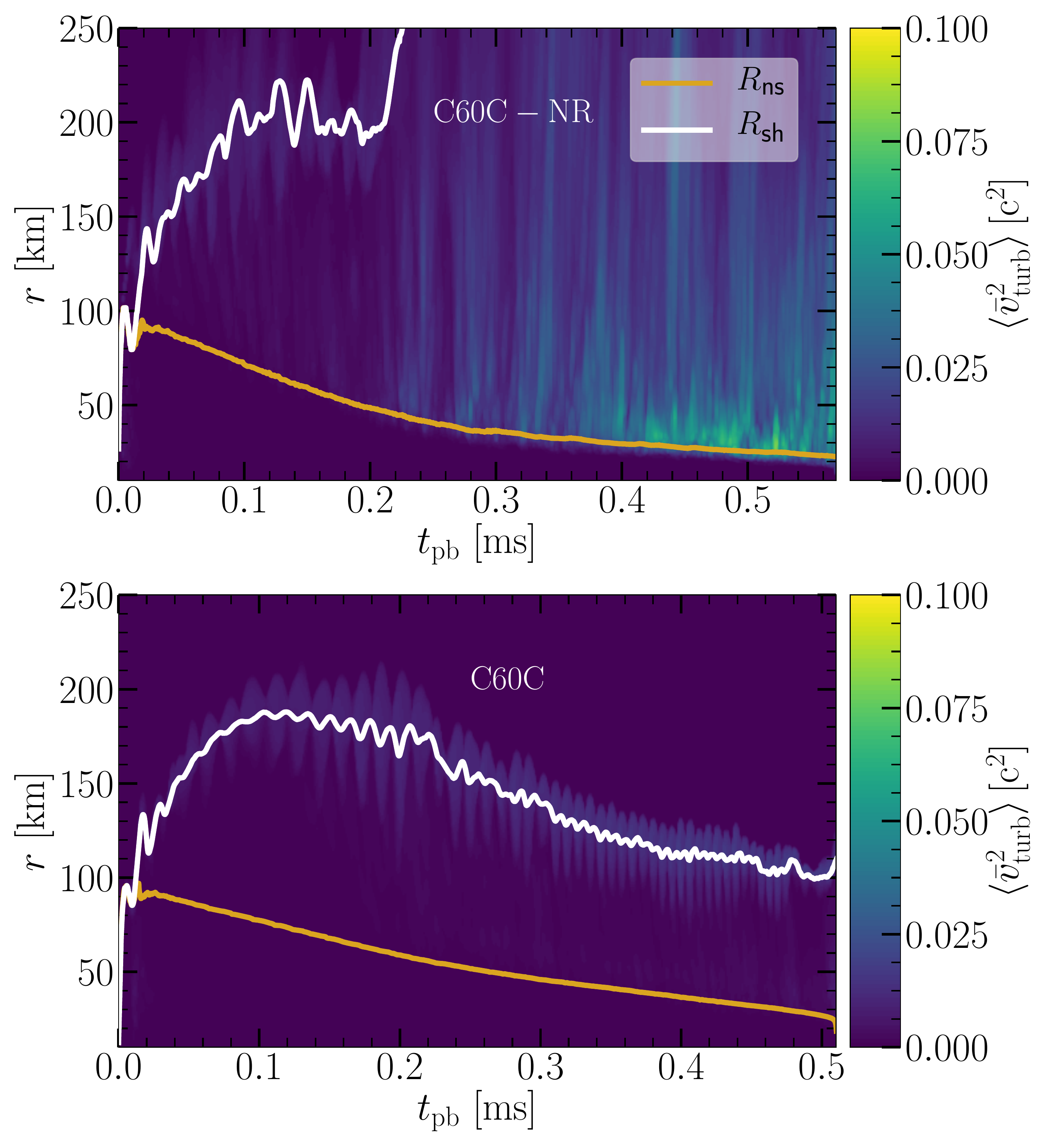}
    \vspace*{-7mm}
    \caption{Time evolution of the squared turbulent velocity for the non-rotating Model C60C-NR (top panel) and the rapidly rotating Model C60C (bottom panel). The mean PNS and shock radii are indicated by yellow and white lines, respectively.}
    \label{fig:GW_c60_vturb}
\end{figure}
\begin{figure}
	\includegraphics[width=0.48\textwidth]{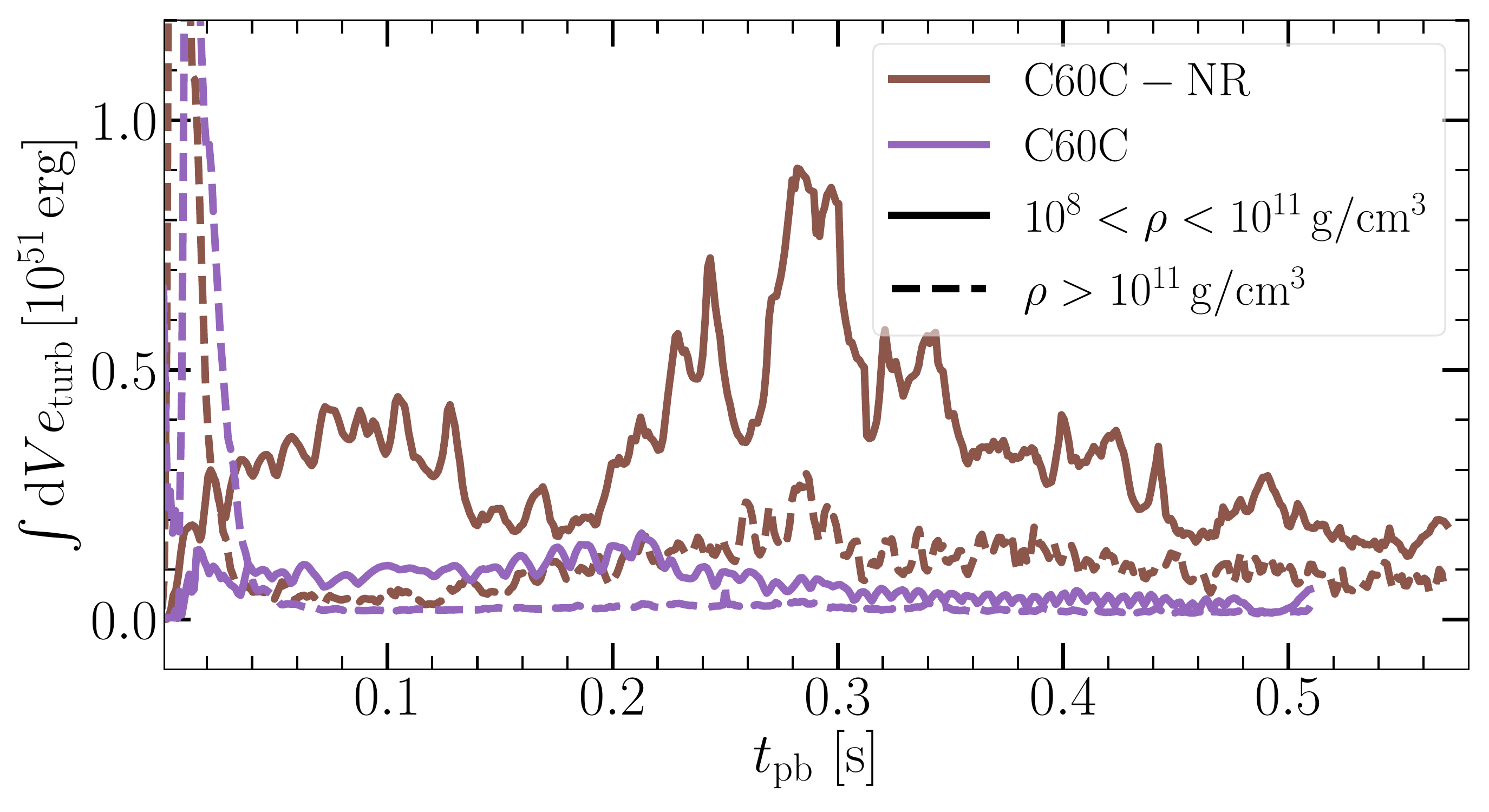}
	\vspace*{-7mm}
	\caption{Time evolution of the turbulent kinetic energy in different volumes of our rapidly rotating Model C60C (violet lines) and its non-rotating counterpart, Model C60C-NR (brown lines). The solid and dashed lines correspond to density domains inside and outside of the PNS (as specified by the labels), over which the volume integral of the quantity $e_\mathrm{turb}$ (defined in equation~\eqref{eq:turb_energy}) is performed. We assume the PNS surface at a density value of $10^{11}$\,g/cm$^3$.}
	\label{fig:GW_c60_turbulent_energy}
\end{figure}

\section{Gravitational Waves}
\label{sec:gravitational_waves}

In this section we briefly discuss the GW emission predicted for our set 
of BH-formation simulations. Since our GW calculations are only approximate
and there is already a vast recent literature on the topic, our focus 
will just be on a few essential aspects and a cursory comparison to similar
results in the literature. There are two major caveats of our models
in connection to the GW determination that we would like to point out
from the beginning. First, it is well known
from published studies (e.g., \citealt{2017MNRAS.468.2032A}) that 2D models 
overestimate the GW amplitudes typically by a factor of ten or more
compared to 3D calculations, because with the constraint of axisymmetry
all non-radial flow structures are considered to have toroidal geometry
and are unable to fragment in the azimuthal direction. We therefore refer
to the recent publication by \citet{2021MNRAS.503.2108P} for more reliable GW
predictions from 3D simulations of collapsing VMSs with masses similar
to our models and an assessment of their detectability. 
Second, although our NADA-FLD code uses general relativistic (GR)
hydrodynamics, we solve the Einstein equations in spherical symmetry
(see Section~\ref{subsec:numerical_setup}) and estimate the GW amplitudes by employing the
quadrupole formula (equations~\eqref{eq:gravitational_wave_amplitude}, \eqref{eq:gravitational_wave_strain}). In the case of rotation, in
particular for our rapidly rotating Model~C60C with its centrifugally
flattened PNS and globally aspherical accretion flow, the spherically 
symmetric approximation of the field equations may not allow to capture
deformation effects in all details and may thus limit the accuracy of
our GW calculations, too.

Figures~\ref{fig:GW_c60_norot}--\ref{fig:GW_c115} show the dimensionless GW strains $h_+$ as functions of time for all of our models along with the GW spectrograms. The calculation
of $h_+$ is based on equations \eqref{eq:gravitational_wave_amplitude}  and \eqref{eq:gravitational_wave_strain} for an inclination angle
of $\vartheta = 90^\circ$. We also adopt the assumption of a source distance
of 10\,kpc, which is standard in the CCSN literature but may be much too
optimistic for the potential detection of one of the rare collapse events
of VMSs. The spectrograms are constructed by performing a short-time
discrete Fourier transform on the dimensionless GW strain, using a 
sliding window of 50\,ms. The signals are convolved with a Kaiser function
with the shape parameter value of $\beta = 2.5$ before applying the discrete
Fourier transform. In the spectrograms, the spectral energy density,
$\mathrm{d}E_\mathrm{GW}/\mathrm{d}f$, is normalized with respect to its
maximum value.

Our GW signals exhibit all of the well known features found in most other
modern 2D and 3D CCSN simulations published over the past two decades
(e.g., \citealt{2004ApJ...603..221M,2009ApJ...707.1173M,2009A&A...496..475M,2013ApJ...766...43M,2013ApJ...779L..18C,2015PhRvD..92h4040Y,2016ApJ...829L..14K,2017hsn..book.1671K,2017ApJ...851...62K,2018ApJ...861...10M,2019ApJ...876L...9R,2020PhRvD.102b3027M,2021MNRAS.502.3066S,2018ApJ...857...13P,2021ApJ...914..140P,2017MNRAS.468.2032A,2019MNRAS.486.2238A,2021MNRAS.503.3552A,2019MNRAS.487.1178P,2021MNRAS.503.2108P,2021arXiv210501315J}).
In the following we compare some aspects of our results mostly to the works
of \citet{2013ApJ...766...43M} and \citet{2013ApJ...779L..18C},
where NS and BH forming stellar core-collapse events were simulated in 2D, and
to the recent work by \citet{2021MNRAS.503.2108P}, where the collapse of 85\,M$_\odot$
and 100\,M$_\odot$ PPISN progenitors similar to our models was investigated in
3D. All of these previous studies were based on GR hydrodynamics with
the CoCoNuT code. Although all basic features are also found in 
Newtonian models or Newtonian models with modified pseudo-GR potential,
there can be quantitative as well as minor qualitative differences in 
details (for discussions of this aspect, see, e.g.,
\citealt{2013ApJ...766...43M}; \citealt{2017MNRAS.468.2032A}). In general, the
exact properties of the computed GW signals are very sensitive to 
differences in the considered microphysics (e.g., the nuclear EOS
and the set of neutrino interaction rates), the chosen grid resolution,
and the ``noisiness'' or perturbations connected to the employed numerical
scheme and type of numerical mesh (e.g., cartesian or polar grid), 
all of which are difficult to assess in
detail without having direct access to the codes. We therefore refrain from 
a wider comparison to results in the literature listed above.

All of our models exhibit pronounced, quasi-periodic GW emission during the
first $\sim$50\,ms after bounce. This is a consequence of strong shock 
pulsations and prompt post-bounce convection, because shock expansion and
contraction phases (see Fig.~\ref{fig:shock_radius}) create negative entropy gradients that 
trigger the onset of convective mass motions. Although the GW activity
produced by such mass motions over a broad frequency range, but with a
very prominent peak below $\sim$100\,Hz, is witnessed also in the models 
of \citet{2013ApJ...766...43M} and \citet{2013ApJ...779L..18C}, the amplitudes in
our simulations ($h_+\sim (1-2)\times 10^{-21}$ at 10\,kpc source 
distance, corresponding to quadrupole amplitudes of $A_{20}^{E2} \sim
(100-200)$\,cm) are a few times bigger than those in the previous
works, even compared to the collapse of a massive (35\,M$_\odot$) progenitor
in \citet{2013ApJ...779L..18C}. This difference might partly be caused by
the very large iron cores of our models and their correspondingly
huge mass accretion rates immediately after core bounce. To an unclear
extent the amplitude might also be enhanced by an overestimation of the 
fast shock expansion and contraction during the first 15\,ms after bounce, 
which we hypothetically linked to the disregard of neutrino-electron 
scattering in our simulations (see \citealt{2019MNRAS.490.3545R}).

After the convective mass motions and strong GW emission in this early 
post-bounce period have been damped by wave dissipation, there is a
transient phase of lower activity in most of our models before GW production
with increasing amplitudes sets in when four of our five models approach 
their moments of shock revival. This enhanced GW production is a 
consequence of the growing strength of neutrino heating prior to shock
expansion, which leads to a growing violence of the hydrodynamic
instabilities in the postshock flow, boosting the GW amplitudes. Also
after the shock begins to expand, these models display vivid GW emission
with episodic outbursts when infalling matter is accreted asymmetrically onto
the PNS, a phenomenon that was described for the post-explosion phase in
a 2D simulation of an 11.2\,M$_\odot$ star by \citet{2013ApJ...766...43M}. 
The outbursts are characterized by strongly increasing high-frequency 
as well as very low-frequency ($\lesssim 50$\,Hz, though here the sampling
window for the spectrograms may have an influence) emission,
and in addition by rapid changes of overall excursions of the wave train 
away from the zero level. This indicates sudden alterations of the geometry
of the postshock flow when buoyant plumes of neutrino-heated matter and 
their separating accretion downflows rearrange themselves during the
transient phases of enhanced fallback and accretion.

The overall displacement of the wave train from the zero level is 
connected to the linear memory effect (see \citealt{2010CQGra..27h4036F} for a review),
which is a consequence of the asymmetric expansion of the shock wave and
the asymmetric mass distribution in the postshock layer. This was 
first discussed in the CCSN context
by \citet{1996PhRvL..76..352B} and \citet{2009ApJ...707.1173M}. The long-time
gradual shift occurs towards the positive side for a prolate global
deformation of the shock (as in Model R80Ar, see Figs.~\ref{fig:contour_all_model_BH} and \ref{fig:GW_r80_rot}), 
and to the negative side for more oblate deformations as in all other
cases with shock revival because of prominent, equatorially expanding high-entropy 
plumes (Figs.~\ref{fig:contour_all_model_BH}, \ref{fig:GW_c60_norot}, \ref{fig:GW_r80_norot}, \ref{fig:GW_c115}).

The spectrograms reveal the typical combination of a low-frequency
contribution around 50--100\,Hz from SASI and large-scale convective
overturn motions in the postshock layer on the one hand, and, on 
the other hand, a high-frequency component originating from the PNS 
convection layer and from gravity waves (g-mode activity)
instigated in the convectively stable
near-surface layers of the PNS mainly by the impact of accretion
downflows (for detailed discussions, see, e.g., \citealt{2009ApJ...707.1173M,2009A&A...496..475M,2013ApJ...766...43M,2013ApJ...779L..18C,2017MNRAS.468.2032A,2019MNRAS.486.2238A,2018ApJ...861...10M,2019ApJ...876L...9R}).
As found in the previous simulations, this signal component forms
a broad band in the spectrograms, whose frequency increases
continuously from initially $\sim$250\,Hz to finally well over 
1000\,Hz as the PNS contracts and becomes increasingly more compact, 
until it finally collapses to a BH (see also \citealt{2013ApJ...779L..18C}).

After the formation of the BH, the high-frequency emission
abruptly abates and the wave amplitudes, which still show some
low-frequency modulations connected to unsteady accretion flows, 
tend to swing back to the zero level, because
all initial matter asymmetries fall back to the BH and the surviving
shocks expand basically spherically symmetrically. The abrupt
disappearance of the high-frequency GW signal at the time of BH
formation is a clear confirmation of its origin from the transiently
stable PNS. Interestingly, in the case of the two rotating models
C60C and R80Ar, and considerably stronger in the fast-rotating case
of C60C, the dramatic contraction of the PNS prior to the 
BH formation (see Fig.\ref{fig:NS_property}) and the final collapse of the PNS are
accompanied by a high-amplitude GW burst that lasts some 10\,ms
and is visible as a prominent broad-band feature in the spectrograms
of Figs.~\ref{fig:GW_c60_rot} and ~\ref{fig:GW_r80_rot}. This phenomenon is absent in the other three cases without rotation (Figs.~\ref{fig:GW_c60_norot}, \ref{fig:GW_r80_norot}, \ref{fig:GW_c115}).

In contrast, during all of its evolution after the first 100\,ms post bounce 
until BH formation and beyond, the non-exploding, rapidly rotating Model
C60C exhibits weaker GW emission than all other models, in particular also
in comparison to its non-rotating counterpart, Model C60C-NR (Fig.~\ref{fig:GW_c60_rot}
compared to Fig.~\ref{fig:GW_c60_norot}). Between 
$\sim$300\,ms after bounce and the BH formation at 510\,ms, the GW 
amplitudes in C60C settle to a low level (with quadrupole amplitudes
$A_{20}^\mathrm{E2}$ around 20\,cm), fairly similar to the case simulated 
by \citealt{2013ApJ...779L..18C}. This behavior is explained by the continuous
shock contraction in Model C60C (Fig.~\ref{fig:shock_radius}), which reduces the width and the
mass in the postshock layer, thus disfavoring strong GW production in the
postshock layer. Therefore the low-frequency GW component in Model C60C
is strong only for about 100\,ms after bounce and then loses power 
gradually until it becomes invisible in the spectrogram after 300\,ms
of post-bounce evolution (Fig.~\ref{fig:GW_c60_rot}). 
Moverover, because of the rapid differential rotation, 
convection inside the PNS is suppressed (see \citealt{2001LNP...578..333J}). Since
both PNS convection and postshock accretion flows are stirring mechanisms
of g-mode activity in the PNS surface layers and are much weaker than in 
Model C60C-NR and in all other nonrotating or slowly rotating cases, also 
the high-frequency GW emission is considerably reduced in Model C60C. In
support of this reasoning, the angle-averages of the squared turbulent
velocity, $\left\langle \bar v_\mathrm{turb}^2 \right\rangle = \left\langle\sum_{i = r,\theta,\phi} ({\bar v^i}_\mathrm{turb})^2\right\rangle$ 
(see the text following equation~\eqref{eq:turb_energy} for the definition 
of ${\bar v^i}_\mathrm{turb}$) are plotted for Models C60C-NR and C60C in 
Fig.~\ref{fig:GW_c60_vturb}, and the turbulent kinetic energies 
(i.e., the volume integrals of the turbulent kinetic energy density as defined in
equation~\eqref{eq:turb_energy}) inside and outside of
the PNS are displayed for both models in Fig.~\ref{fig:GW_c60_turbulent_energy}. These
plots substantiate the much lower turbulent flow activity in Model C60C
in the postshock region as well as in the convective layer of the PNS.

Compared to the GW signals from 3D simulations of the gravitational collapse
of PPISNe performed by \citet{2021MNRAS.503.2108P}, the wave amplitudes from 
our 2D models show the typical overestimation by a factor of 10--20. Therefore the 
total GW energies emitted during the simulated accretion evolution of the
PNSs and final BHs, which are about $1.0\times 10^{47}$\,erg, $8.0\times 10^{45}$\,erg,
$7.3\times 10^{46}$\,erg, $7.7\times 10^{46}$\,erg, and $6.1\times10^{46}$\,erg for
Models C60C-NR, C60C, R80Ar-NR, R80Ar, and C115, respectively, are also
far too optimistic. We therefore refrain from a discussion of the 
detectability of our signals and refer the reader to \citet{2021MNRAS.503.2108P}
for this aspect.

We conclude this brief discussion of our results for the GW emission by repeating the caveat that we employ the pseudo-Newtonian quadrupole formula (equation~\eqref{eq:gravitational_wave_amplitude}) for approximately estimating the GW signals. This pseudo-Newtonian approximation may yield poor estimates for the GW properties after BH formation, because such an approximation cannot accurately describe the GW production at extreme space-time curvatures near the apparent horizon. Therefore, a comparison between pseudo-Newtonian GW predictions and fully relativistic results is highly desirable. Such a study requires truly multi-dimensional solutions of the Einstein equations. This, however, is beyond the scope of our current modeling approach, in which we assume spherical symmetry for solving the GR metric equations.

\section{Summary and conclusions}\label{sec:summary}

We presented results from neutrino-hydrodynamic simulations of the 
final gravitational collapse of PPISNe, considering a rapidly rotating
progenitor of 60\,M$_\odot$ ZAMS mass, a slowly rotating progenitor of
80\,M$_\odot$ ZAMS mass, and a non-rotating case of 115\,M$_\odot$ 
ZAMS mass, all of them with a metallicity of 10\% Z$_\odot$ \citep{2017ApJ...836..244W}. 
The pre-collapse stars possess gravitationally bound masses of 41.5, 47.6, 
and 45.5\,M$_\odot$, respectively, and are characterized by massive iron 
cores of $\sim$(2.4--2.7)\,M$_\odot$ with high compactness values between
0.77 and 0.89. Our calculations were performed in 2D, using the general
relativistic \textsc{NADA-FLD} code with energy-dependent, three-flavor, flux-limited
neutrino transport \citep{2019MNRAS.490.3545R}, employing the SFHo nuclear EOS
of \citep{2012ApJ...748...70H}. Our set of simulations includes non-rotating and
rotating versions of the 60\,M$_\odot$ progenitor (Models C60C-NR and
C60C), non-rotating and rotating versions of the 80\,M$_\odot$ progenitor 
(Models R80Ar-NR and R80Ar), and Model C115 of the non-rotating 
115\,M$_\odot$ progenitor.

Because of the huge mass accretion rates of more than 2\,M$_\odot$/s
over several 100\,ms after core bounce, the luminosities of $\nu_\mathrm{e}$
and $\bar\nu_\mathrm{e}$ increase up to more than $1.5\times 10^{53}$\,erg/s
and the corresponding RMS energies to $\sim$17\,MeV and $\sim$20\,MeV,
respectively. Therefore in all cases except one, strong neutrino heating 
is able to trigger shock revival after about 250\,ms of post-bounce evolution
and well before the transiently stable PNS collapses to a BH (similar to what
was found by \citealt{2018ApJ...852L..19C, 2020MNRAS.495.3751C, 2021MNRAS.503.2108P, 2018MNRAS.477L..80K, 2018ApJ...857...13P} in core-collapse simulations for other massive or 
very massive progenitors). 
The only exception is the rapidly rotating Model C60C, where the radiated
neutrino luminosities and RMS energies are considerably lower because of the 
centrifugally deformed and radially more extended PNS with a correspondingly
cooler neutrinospheric layer. In this model the
average shock radius shrinks until BH formation occurs at 510\,ms after
bounce. In the other models with shock expansion, the PNS continues to 
accrete matter until it collapses to a BH between 350\,ms and 580\,ms after 
bounce.

The diagnostic energies of postshock matter that expands in neutrino-heated
high-entropy plumes reach maxima of up to $1.6\times 10^{51}$\,erg at the 
time of BH formation, but ultimately all of this matter falls back to the 
BH because the neutrino luminosities and heating decline dramatically after 
the PNS has collapsed. Nevertheless, the SN shocks have received 
enough energy by $p\mathrm{d}V$ work of the buoyant high-entropy plumes
to propagate outward either as a shock wave or sonic pulse, despite the fact
that the diagnostic energy asymptotes to zero after some seconds latest in
all of our models. This outward propagation of the shock or acoustic pulse
is quite similar to what was found in 3D simulations by \citet{2021MNRAS.503.2108P} for a zero-metallicity 85\,M$_\odot$ PPISN model and by \citet{2018ApJ...852L..19C, 2020MNRAS.495.3751C} for a zero-metallicity 40\,M$_\odot$ progenitor, but in our models it is much 
weaker and less extreme concerning ejecta energies and estimated ejecta masses.
We tracked the evolution of the expanding SN shocks by follow-up
simulations with the \textsc{Prometheus} code, which we continued until
shock breakout from the stellar surface or at least until the shock had
converted to a weak sonic pulse. This allowed us to estimate the energies
of the outgoing waves, which are much lower than the initial diagnostic
postshock energies, because the initial diagnostic energy is only partly 
transferred by $p\mathrm{d}V$ work
before the originally neutrino-heated matter falls back and is swallowed
by the BH. We found wave energies in the range of 
(4--$7)\times 10^{49}$\,erg and estimated generous upper limits between
roughly 0.07\,M$_\odot$ and 3.5\,M$_\odot$ for the masses that can become
unbound when the shock or sonic pulse reaches the stellar surface.
Therefore our simulations imply that all of the considered PPISNe 
finally collapse to BHs in the mass range between 
$\sim$41.5\,M$_\odot$ and $\sim$46.5\,M$_\odot$.
For these numbers the neutrino mass decrement
does not play any relevant role, because it is only of the order of
$\sim$0.1\,M$_\odot$.

All of the VMSs studied in our core-collapse calculations, except the rapidly rotating Model~C60C, exhibit shock revival and the onset of an explosive expansion of the shock, despite their high compactness values of $\xi_{2.5} \ge 0.77$. Nevertheless, because of the high gravitational binding energies of the pre-collapse stars, the outgoing shocks or sonic pulses ultimately achieve little mass ejection and most of the stellar matter ends up in a BH. Shock revival for such high values of $\xi_{2.5}$ may be considered to be in conflict with a simple criterion that employs a certain value of the compactness parameter \citep{2011ApJ...730...70O} to judge the ``explodability'' of a star, assuming that progenitors with low compactness explode easily, whereas those above a certain threshold value of the compactness (typically around 0.3--0.45) do not explode. The more sophisticated and physics-based two-parameter criterion of \citet{Ertl+2016}, however, includes the possibility that also high-compactness progenitors may develop shock revival by neutrino heating. This is due to the fact that the two-parameter criterion, in contrast to the compactness threshold, accounts for the fact that high mass accretion rates of the new-born NS (as a consequence of high core compactness) can foster explosions by the neutrino-driven mechanism, because high mass accretion rates imply high accretion luminosities of neutrinos and harder radiated neutrino spectra, and a correspondingly enhanced rate of neutrino heating. Of course, the functional shape of the two-parameter criterion provided by \citet{Ertl+2016} on grounds of simplified (calibrated ``engine-driven'') 1D explosion models may not be the final answer, but it will have to be tested and possibly revised by the information obtained from large sets of detailed multi-dimensional explosion models, once all remaining uncertainties of the supernova mechanism have been settled.

The general relativistic \textsc{NADA-FLD} code also permitted us to track, for
the first time, the evolution of the PNS continuously beyond its collapse
to a BH into the subsequent aspherical accretion phase of the new-born
BH. Our \textsc{NADA-FLD} simulations were carried on until most of the initial 
explosion asymmetries had fallen back into the BH and the accretion
flow to the BH had become effectively spherical or rotationally 
deformed, depending on the progenitor properties. These simulations
enabled us to determine the neutrino emission properties not only of
the transiently stable PNS but also of the newly formed BH. Within
milliseconds after BH formation the neutrino luminosities drop by 
at least two orders of magnitude for $\nu_\mathrm{e}$ and 
$\bar\nu_\mathrm{e}$, for heavy-lepton neutrinos even by three orders
of magnitude. However, in the models with shock revival, a high 
level of extremely time-variable neutrino emission with peak 
luminosities exceeding $10^{52}$\,erg/s for electron neutrinos
and antineutrinos and of over $10^{51}$\,erg/s for 
heavy-lepton neutrinos can still be maintained for periods of 
several 100\,ms up to more than 0.5\,s,
as long as the originally neutrino-heated matter falls back and is
anisotropically accreted by the BH. Interestingly, converging
downflows that collide with each other in the close vicinity of
the BH become shock-heated to temperatures above 10\,MeV, and 
high-energy neutrinos can escape from the hot fallback gas, because
it has relatively low densities and is spread out over a rather large
volume ($>$\,100\,km in diameter). The RMS energies of the radiated
neutrinos are therefore considerably higher than during the PNS
cooling phase, namely up to more than 25\,MeV for $\nu_\mathrm{e}$,
more than 30\,MeV for $\bar\nu_\mathrm{e}$, and even more than 
50\,MeV for heavy-lepton neutrinos. This remarkable phase of neutrino emission continues
until almost all of the neutrino-heated matter has fallen back to
the BH and the accretion flow into the BH collapses effectively 
radially, at which time the neutrino luminosities plummet to very 
low values ($<$\,$10^{49}$--$10^{50}$\,erg/s). In some of our models 
this instant is reached later than one second after bounce.

The total energy loss by neutrinos in our simulations, including
the phase of significant luminosities by anisotropic accretion 
after BH formation, is less than $2.34\times 10^{53}$\,erg in all
of our models, corresponding to a neutrino mass decrement of at
most 0.13\,M$_\odot$ (see Table~\ref{tab:model_property_2}). This
is in conflict with the assumption by \citet{Belczynski+2016} that
10\% of the rest mass of the PPISN (i.e., several
solar masses) can be lost via neutrino emission (see their equation~(1)
and discussion afterwards). Our values of Table~\ref{tab:model_property_2}
correspond to 0.2--0.3\% of the pre-collapse mass of the PPISN models 
or less than 3\% of the mass reduction by neutrino emission adopted by
\citet{Belczynski+2016}.

The asymmetrically accreting PNSs and BHs emit neutrinos highly
anisotropically. This induces kick velocities to the BHs that
can transiently reach up to more than 100\,km/s. This effect is 
most extreme in cases where the BH accretes aspherically and 
high-temperature clumps or belts of shock-heated matter from
downflow collisions assemble near the BH in only one hemisphere.
However, because the outgoing shocks or sonic pulses trigger little mass ejection
from the progenitors' surface layers, the BHs ultimately swallow 
almost the entire stars. Therefore the final kick velocities are 
diminished to just a few km/s at most.

The GW signals of our models show the well known features connected
to (i) prompt post-bounce convection, (ii) a low-frequency ($<$\,200\,Hz) 
component produced by non-radial hydrodynamic flows (convective 
overturn and SASI) in the postshock layer, (iii) a high-frequency
component with growing frequency (up to over 1000\,Hz)
from g-mode activity in the near-surface layers of the contracting
PNS, and (iv) a long-time trend of the wave train away from the 
zero-level due to the linear memory effect when nonspherical
shock expansion takes place. All of these features are qualitatively
very similar to the GW predictions by \citet{2021MNRAS.503.2108P} from their
3D core-collapse simulations of PPISN models. After BH formation
only weak low-frequency GW activity continues until the infall
of stellar matter becomes quasi-spherical, at which time the 
overall excursion of the wave train is reset and the zero-level 
is restored. Our 
rapidly rotating Model C60C, which does not experience shock revival
and displays the weakest non-radial flow activity, exhibits also the 
weakest production of GWs of all of our simulations.

One caveat of these GW results as well as of our entire study is
its limitation to 2D. It is well known that 2D simulations overestimate
the GW amplitudes by factors of 10--20 compared to 3D models and, correspondingly,
the energy radiated in GWs is also massively overestimated 
(see, e.g., \citealt{2017MNRAS.468.2032A}). 
This is a consequence of the assumption of axisymmetry in 2D, where flow
structures possess toroidal geometry and are thus more massive and more 
coherent than in 3D, where fragmentation into smaller vortices and inhomogeneities is
possible. For the same reason our BH kicks by anisotropic neutrino emission are 
likely to be overestimated. Moreover, in 3D spiral SASI modes exist in
addition to SASI sloshing motions, and rotation-amplified spiral 
waves may lead to shock revival around the equatorial plane (see, e.g.,
\citealt{2018ApJ...852...28S} for an example). This could potentially affect the
evolution of the rapidly rotating Model C60C, which did not develop
shock expansion in our 2D simulations but might so in 3D. In contrast,
in the slowly rotating case of Model R80Ar we witnessed little influence 
by the angular momentum in the simulations (besides stochastic
variations of the nonradial flows in the postshock region), and therefore
we do not expect fundamentally different results in 3D. 

In the presence of rapid rotation, magnetic field amplification might 
also become relevant, even on the short time scales until BH formation.
Magnetically driven jet-like outflows may be a consequence (see, e.g.,
\citealt{2015Natur.528..376M, 2018ApJ...864..171M, 2018MNRAS.477.2366H, 2020MNRAS.492.4613O, 2020ApJ...896..102K, 2021MNRAS.500.4365A, 2021MNRAS.507..443B, 2021MNRAS.503.4942O, 2021ApJ...906..128K} for recent 3D simulations). Also AD
formation around the BH is a viable possibility, provided the rapidly 
rotating star of Model C60C continues to collapse. We estimate that 
near the equatorial plane some matter between $\sim$5\,M$_\odot$ and 
$\sim$11\,M$_\odot$ and all matter outside of $\sim$13\,M$_\odot$ in the 
pre-collapse star has more angular momentum than needed at the ISCO
around the rotating BH that is formed from the enclosed mass. In the 
slowly rotating Model R80Ar, only the outermost stellar layers fulfill
this condition, but we expect these loosely bound shells to be expelled 
when the sonic pulse reaches the surface of the star. With the possibility
of BH-AD formation and magnetic field amplification, Model C60C might
constitute an interesting case for the collapsar scenario and the
production of long-duration gamma-ray bursts \citep{1993ApJ...405..273W}.
For more discussion of possible consequences of magnetic field 
amplification and BH-AD or magnetar formation in the collapse of
rapidly spinning PPISNe, e.g.\ for superluminous SNe, see \citet{2017ApJ...836..244W}.

\section*{Acknowledgements}
We thank Thomas W. Baumgarte, Pedro J. Montero, Bernhard M\"uller, Ewald M\"uller, Robert Bollig, and Sebastiano Bernuzzi for helpful discussions, and Alexandra Kozyreva for comments
on the manuscript. We also thank Oliver Just for providing us with the source routines for the neutrino reaction rates used in the ALCAR code. At Garching, funding by the European Research Council through Grant ERC-AdG No.~341157-COCO2CASA and by the Deutsche Forschungsgemeinschaft (DFG, German Research Foundation) through Sonderforschungsbereich (Collaborative Research Centre) SFB-1258 ``Neutrinos and Dark Matter in Astro- and Particle Physics (NDM)'' and under Germany's Excellence Strategy through Cluster of Excellence ORIGINS (EXC-2094)---390783311 is acknowledged.
At Darmstadt, funding by the European Research Council (ERC) under the European Union's Horizon
2020 research and innovation programme (ERC Advanced Grant KILONOVA No.~885281) is acknowledged.

\section*{Data Availability}

The data from our simulations will be made available in the Garching Core-Collapse Supernova Archive 
(\texttt{https://wwwmpa.mpa-garching.mpg.de/ccsnarchive/}) upon reasonable request to the authors.




\bibliographystyle{mnras}
\bibliography{ref.bib} 

\appendix
\onecolumn

\section{General relativistic neutrino diffusion flux}\label{app:diff_flux_eqn}

\begin{figure*}
    \includegraphics[width=\textwidth]{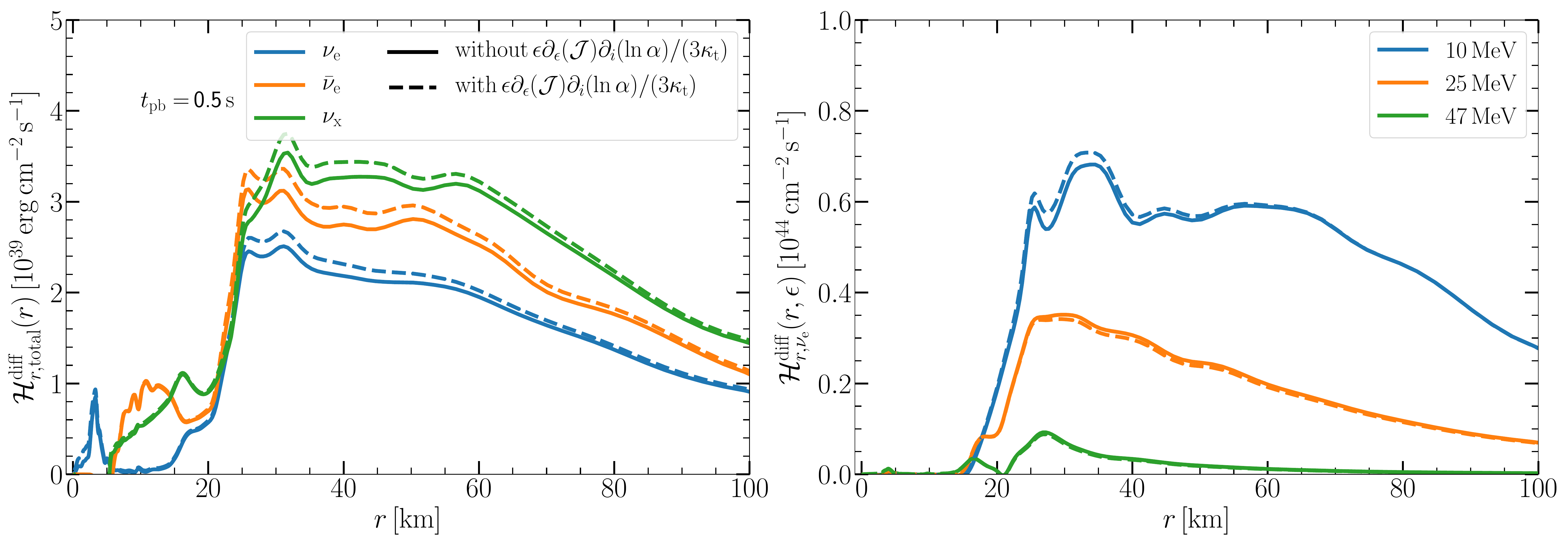}
    \vspace*{-6mm}
    \caption{{\em Left panel:} Radial profiles of the angular averages of the radial components of the energy-integrated energy-diffusion fluxes in the local comoving frame (left panel) for electron neutrinos (blue), electron anti-neutrinos (orange), and a single species of heavy-lepton neutrinos (green) in Model C60C-NR at 500\,ms after core bounce. {\em Right panel:} Corresponding radial profiles of the angular averages of the specific energy fluxes in the local comoving frame for electron neutrinos at three selected energies. The dashed (solid) lines correspond to the neutrino diffusion fluxes as given in equation~\eqref{eq:A.3}, i.e., without applying a flux limiter, with (without) the second term on the right side of this equation.}
    \label{fig:app_flux_correction}
\end{figure*}

In \citet{2019MNRAS.490.3545R} we defined the neutrino diffusion flux without taking into account a term that includes the derivatives of the zeroth angular moment of the neutrino distribution function ($\mathcal{J}$) with respect to the neutrino energy ($\epsilon$) and of the lapse function ($\alpha$) with respect to the spatial coordinates. \citet{2020MNRAS.496.2000C} pointed out that this term is needed to ensure the correct thermal structure established by energy diffusion in a (nearly) static general relativistic space-time. In this appendix we evaluate the effect of this missing term on the magnitude of the neutrino diffusion fluxes. We refer the reader to Appendices A and B of \citet{2019MNRAS.490.3545R} for the derivation of the general relativistic neutrino moment equations and the flux-limited diffusion fluxes, respectively. We begin with the neutrino momentum equation (A9) of \citet{2019MNRAS.490.3545R} (for definitions of the different neutrino moments, see Table A1 of the same paper):
\begin{eqnarray}\label{eq:A.1}
 &&\frac{1}{\alpha}\frac{\partial}{\partial t}(\hat{\mathcal{F}}_i) 
 + \frac{1}{\alpha}\frac{\partial}{\partial x^j}(\alpha {\hat{\mathcal{S}}^j}_i-{\beta}^j \hat{\mathcal{F}}_i)
 + \hat{\mathcal{E}} \frac{\partial \ln \alpha}{\partial x^i}
 - \hat{\mathcal{F}}_j \frac{1}{\alpha} \frac{\partial \beta^j}{\partial x^i} 
 - \frac{1}{2} \hat{\mathcal{S}}^{jk}\frac{\partial \gamma_{jk}}{\partial x^i}
 - \frac{\partial}{\partial \epsilon}(\epsilon \hat{S}_i^\epsilon)
 = -3\kappa_\mathrm{t}\mathcal{H}_i~,
\end{eqnarray}
where
\begin{eqnarray}\label{eq:A.2}
 S_i^\epsilon &\equiv& W\left\{P_{ij} \frac{\partial v^j}{\partial \tau} + {Q_{ij}}^k \frac{\partial v^j}{\partial x^k}
 + \frac{1}{2} {Q_i}^{jk}v^l\frac{\partial \gamma_{jk}}{\partial x^l}
 + ({P_i}^j - I_i v^j)\frac{\partial \ln\alpha}{\partial x^j}
 - {Q_i}^{jk}K_{jk} + v^j P_{ik} \frac{1}{\alpha} \frac{\partial \beta^k}{\partial x^j}\right\} \nonumber \\
 && + (P_{ik} v^k - I_i)\frac{\partial W}{\partial \tau}
 + ({Q_{ik}}^j v^k - {P_i}^j)\frac{\partial W}{\partial x^j}~.
\end{eqnarray}
Here, $\kappa_\mathrm{t}$, $\alpha$, $\beta^i$, $\gamma_{ij}$, $\gamma$ are the neutrino transport opacity, the lapse function, the shift vector, the spatial part of the metric tensor, and its determinant, respectively. Moreover, we introduce the notation $\hat X \equiv \sqrt{\gamma} X$ for any quantity $X$. As is customary in a FLD scheme, we ignore all velocity dependent terms as well as the time derivatives of the comoving flux $\mathcal{H}_i$, and apply the diffusion ansatz, $\mathcal{K}^{\hat i \hat j} = \delta^{\hat i \hat j} \mathcal{J} / 3$ in equation \eqref{eq:A.1} (see, e.g., \citealt{1981ApJ...248..321L} for a detailed description of FLD schemes). Moreover, assuming $\gamma^i_j \rightarrow \delta^i_j$ and $\beta^j \rightarrow 0$ in equation~\eqref{eq:A.1} (see Appendix B of \citealt{2019MNRAS.490.3545R} for the validity of these approximations in the context of CCSNe), we get
\begin{eqnarray}\label{eq:A.3}
 && 
 \mathcal{H}_i^{\mathrm{diff}} = - \frac{1}{3\kappa_\mathrm{t} \alpha^3} \partial_{i}(\alpha^3 \mathcal{J})
 + \frac{\epsilon}{3\kappa_\mathrm{t}} \partial_{\epsilon}(\mathcal{J}) \partial_i(\ln \alpha)\,.
\end{eqnarray}
The second term on the right side of equation~\eqref{eq:A.3} was ignored in \citet{2019MNRAS.490.3545R} and also in the simulations of the present work. However, inclusion of this term is necessary to obtain the correct thermal equilibrium between matter and radiation in a static general relativistic space-time (see, e.g., \citealt{1999ApJ...513..780P, 2020MNRAS.496.2000C}). Now, introducing the flux-limiter $\lambda$, evaluated by using $\mathcal{H}^{\mathrm{diff}}_{i}$ as given in equation~\eqref{eq:A.3} (for details, see Appendix~C of \citealt{2019MNRAS.490.3545R}) and the flux-limited diffusion coefficient $D=\lambda/\kappa_\mathrm{t}$, we obtain the FLD flux:
\begin{eqnarray}\label{eq:A.4}
 \mathcal{H}^{\hat i}
 &=& - D e^{i \hat i} \Big[ \alpha^{-3} \partial_i(\alpha^3 \mathcal{J}) - \alpha^{-1} (\partial_i \alpha) \epsilon \partial_{\epsilon}(\mathcal{J}) \Big] \,.
\end{eqnarray}

Since the logarithm of the lapse function $(\ln \alpha)$ changes most significantly only near the PNS surface, the additional correction term on the right side of equation~\eqref{eq:A.3} affects the radial neutrino-energy fluxes most visibly only in the near-surface layers of the PNS. Here, we calculate this correction term by post-processing data from our Model C60C-NR. In Fig.~\ref{fig:app_flux_correction}, the radial components of the diffusion fluxes for neutrino energy with (dashed lines) and without (solid lines) this correction term are shown for Model C60C-NR at a post-bounce time of 500\,ms, which is a time when the PNS has a large mass and a very compact structure, implying that the general relativistic effects have a strong impact on the neutrino emission. The left plot displays the energy integrated radial components of the energy-diffusion fluxes for all three neutrino species, and the right plot shows the radial components of the specific energy fluxes for electron neutrinos at three representative energies. We witness changes of a few percent in the total neutrino diffusion fluxes due to the correction term. The correction is positive for neutrinos with energies lower than the peak of the energy spectrum, and it is negative for neutrinos with energies above the spectral peak, as visible in the right panel of Fig.~\ref{fig:app_flux_correction}.

Employing the flux given in the equation~\eqref{eq:A.4}, one obtains the FLD neutrino transport equation:
\begin{eqnarray}\label{eq:A.5}
	&&\frac{1}{\alpha} \frac{\partial}{\partial t} (W \mathcal{\hat J})
    + \frac{1}{\alpha} \frac{\partial}{\partial x^j} [\alpha W (v^j-\beta^j/\alpha) \mathcal{\hat J}] \nonumber \\
    &&- \frac{1}{\alpha} \frac{\partial}{\partial x^j} \Big[\sqrt{\gamma} \Big\{ \gamma^{j i}
    + W \Big(\frac{W}{W+1}v^j-\beta^j/\alpha \Big) v^i \Big\} D \big\{ \alpha^{-2} \partial_i (\alpha^3 \mathcal{J}) - (\partial_i \alpha) \, \epsilon \partial_{\epsilon}(\mathcal{J}) \big\} \Big] \nonumber \\
    && - \frac{e^{i \hat i}}{\alpha^2} \frac{\partial}{\partial t}
    (W \sqrt{\gamma} \bar v_{\hat i}) D \big\{ \alpha^{-2} \partial_i(\alpha^3 \mathcal{J}) - (\partial_i \alpha) \, \epsilon \partial_{\epsilon}(\mathcal{J}) \big\}
    +\hat R_\epsilon - \frac{\partial}{\partial \epsilon} (\epsilon \hat R_\epsilon) \nonumber \\
    &&= \kappa_\mathrm{a} (\mathcal{\hat J}^{eq}-\mathcal{\hat J})~.
    \label{eq:tr_fld_energy_eqn_corr}
\end{eqnarray}
If we neglect the correction term for the neutrino flux $(\partial_i \alpha) \, \epsilon \partial_{\epsilon}(\mathcal{J})$ in equation \eqref{eq:A.5}, we obtain the FLD neutrino transport equation employed in the present work, which is equivalent to equation (37) of \citet{2019MNRAS.490.3545R}:
\begin{eqnarray}\label{eq:A.6}
	&&\frac{1}{\alpha} \frac{\partial}{\partial t} (W \mathcal{\hat J})
    + \frac{1}{\alpha} \frac{\partial}{\partial x^j} [\alpha W (v^j-\beta^j/\alpha) \mathcal{\hat J}] \nonumber \\
    &&- \frac{1}{\alpha} \frac{\partial}{\partial x^j} \Big[\sqrt{\gamma} \Big\{ \gamma^{j i}
    + W \Big(\frac{W}{W+1}v^j-\beta^j/\alpha \Big) v^i \Big\} D \big\{ \alpha^{-2} \partial_i (\alpha^3 \mathcal{J}) \big\} \Big] \nonumber \\
    && - \frac{e^{i \hat i}}{\alpha^2} \frac{\partial}{\partial t}
    (W \sqrt{\gamma} \bar v_{\hat i}) D \big\{ \alpha^{-2} \partial_i(\alpha^3 \mathcal{J}) \big\}
    +\hat R_\epsilon - \frac{\partial}{\partial \epsilon} (\epsilon \hat R_\epsilon) \nonumber \\
    &&= \kappa_\mathrm{a} (\mathcal{\hat J}^{eq}-\mathcal{\hat J})~.
    \label{eq:tr_fld_energy_eqn}
\end{eqnarray}

In closing, we want to point out several typos in \citet{2019MNRAS.490.3545R}. Firstly, $v^{i}$ should be replaced by $v^{k}$ and $\gamma^{ik}$ should be replaced by $\gamma^{jk}$ in the second line of equation (37) in \citet{2019MNRAS.490.3545R}. Additionally, $R_{\epsilon}$ should be replaced by $\hat{R}_{\epsilon}$ in the same equation. Moreover, plus signs are missing at the beginning of the fourth line of equations (56), (57), and (60). In Appendix B, $R_{\epsilon}$ should be $\hat{R}_{\epsilon}$ in equations (B11) and (B12).


\bsp	
\label{lastpage}
\end{document}